\documentclass[12pt,leqno]{article}
\usepackage{latexsym}
\usepackage{verbatim}
\newtheorem{theorem}{Theorem}

\newtheorem{corollary}{Corollary}
\newtheorem{lemma}{Lemma}

\newtheorem{definition}{Definition}

\newenvironment{notation}{\noindent{\bf Notation:}\em\penalty100}{}

\newcommand{\blackslug}{\mbox{\hskip 1pt \vrule width 4pt height 8pt 
depth 1.5pt \hskip 1pt}}
\newcommand{\qed}{\quad\blackslug\lower 8.5pt\null\par\noindent}
\newenvironment{proof}{\par\noindent{\bf Proof:}}{\qed \par}

\newcommand{\cC}{\mbox{${\cal C}$}}

\newcommand{\cR}{\mbox{${\cal R}$}}

\title{Concrete Foundations for Categorical Quantum Physics
\thanks{This work was partially supported 
by the Jean and Helene Alfassa fund for 
research in Artificial Intelligence}
}

\author{Daniel Lehmann\\School of Engineering
and \\ Center for Language, Logic and Cognition
\\Hebrew University, \\Jerusalem 91904, Israel 
}

\date{December 2010}

\begin{document}
\maketitle
\begin{abstract}
An original presentation of Categorical Quantum Physics, in the line of
Abramsky and Coecke~\cite{Abramsky-Coecke:catsem}, tries to introduce only
objects and assumptions that are clearly relevant to Physics and does not assume 
compact closure. 
Adjoint arrows, tensor products and biproducts are the ingredients of this presentation. 
Tensor products are defined, up to a unitary arrow, by a universal property related to
transformations of composite systems, not by assuming a monoidal structure.
Entangled states of a tensor product define mixed states on the components of the tensor 
product.
Coproducts that fit the adjoint structure are shown to be defined up to a unitary arrow and
to provide biproducts.
An abstract no-cloning result is proved.  
\end{abstract}

% If f is self-adjoint and x is an eigen vector for f, and y is orthogonal to x, then fy is orthogonal
%to x

%spectrum theorem: if f is self-adjoint it has an eigen vector

%Classical structures, for any f , g : A -> B, f* o g = g* o f

%Treat the matrices examples: objects are the natural numbers >= 1 and arrows are m x n 
%matrices on some field (commutative) or some subset of a field: GF(2), N, R+, C.
%adjoint are transpose or conjugate-transpose.

%Is A \otimes A the coproduct of A {\otimes}_{s} A and A {\otimes_{a} A ?

%s-tensor and a-tensor products for Relations

%Examples for matrices: bi-arrows

%Correct EPR in the Dieks paper bibliography

%Go over all the examples again

%Mixed states ??? Category whose arrows are Sets of arrows ????? See Selinger and Ab-%Coecke 7.2 p. 51  

%Is there 

% An arrow is right-unitary iff it preserves scalar products

\section{Introduction and plan} \label{sec:intro}
\subsection{Introduction and previous work} \label{sec:previous}
Quantum Mechanics is probably the most successful theory in the whole 
history of Physics. The admiration we feel towards QM is mixed with
surprise: Quantum Physics uses a sophisticated mathematical framework to
describe the physical world and the relation between the phenomena we see
and the mathematical objects used to describe them is never obvious or
justified on first principles. In Quantum Physics it seems one has to accept, 
for example, the fact that systems are represented by unit vectors 
in a Hilbert space and observables by self-adjoint operators, without having
much of a clue as to why Hilbert spaces and why self-adjoint operators.

The Hilbert space framework used for QM, unifying different earlier formalisms,
was presented, in 1932, by J. von Neumann in his book~\cite{vonNeumann:Quanten}
and has not been seriously challenged since.
It is nevertheless obvious that its basic ingredients have no direct physical
correlate. Hilbert spaces, for example, are vector spaces and the basic
operations of vector spaces are multiplication by a scalar and addition of
vectors. Vectors represent states of physical systems. But multiplication
of a state by a scalar (or by the physical entity represented by the scalar,
whatever that is) is not a physical operation. Addition of states is not,
either, immediately correlated to a physical operation.

This paper tries to put in evidence those mathematical structures
present in the Hilbert space formalism that are correlated to physical entities. 
It proposes a minimalist framework that is more general 
than Hilbert spaces. Doing so, it suggests interpretations for the
structures mentioned above: scalars, scalar multiplication and addition.
The framework we develop here is very similar to, 
but different from the one proposed by Abramsky and 
Coecke~\cite{Abramsky-Coecke:catsem}.
The paper provides a lazy introduction to the categorical framework for
Quantum Physics and Quantum Computation. It is intended for physicists
and tries its utmost to describe the physical meaning of the categorical
concepts presented.

\subsection{Plan of this paper} \label{sec:plan}
In Section~\ref{sec:category} we describe how objects and arrows represent
types of physical systems and their transformations.
Then, in the subsequent sections, we describe additional structure to deal
with Quantum Physics. 
Three main structures are considered.
In Section~\ref{sec:dagger} we introduce the adjoint structure and its properties.
It requires every arrow to have an adjoint arrow that flows in the opposite direction.
It formalizes a fundamental symmetry principle of QM: every quantic process 
that can be described as flowing in one direction may be equivalently described as flowing 
in the opposite direction. It is concerned with concepts such as: adjoint, self-adjoint and unitary
operations.
In Section~\ref{sec:tensor} we describe the tensor structure. It formalizes the fact that quantic
systems may be composed of sub-systems. 
One may notice that tensor product is conspicuously absent from the Quantum Logic program 
initiated in~\cite{BirkvonNeu:36}. 
The tensor structure is concerned with notions such as: scalar products, 
tensor products, entanglement and mixed states. 
The present treatment of tensor products differs from standard treatments since it involves no 
additive structure. 
It also differs from previous categorical treatments since tensor products are 
defined up to a unitary arrow by universal properties and not by a monoidal structure.
In Section~\ref{sec:biprod} we describe the additive structure: if $A$ and
$B$ are types of systems, there is a type \mbox{$A \oplus B$} for systems
that are, either of type $A$ or of type $B$, or, equivalently 
(because of the adjoint structure), both of type $A$ and of type $B$. It is concerned with 
orthogonality, biproducts, superpositions and bases.
Section~\ref{sec:BT-cat} deals with the relation between the mutiplicative and additive structures.
Its main result is that multiplication distributes over addition.
Section~\ref{sec:quantic} defines those categories that exhibit truly quantic properties and
about which a no-cloning result is proved in Section~\ref{sec:nocloning}.
Section~\ref{sec:further} discusses open questions and conclusions.

The plan of this paper has been chosen in order to minimize the need for definition, 
justification and discussion of categorical properties early in the paper. Therefore it
deals first with properties of very general structures and moves on towards more and more
specialized structures. The drawback of such a choice is that the same QM concept will be discussed at different stages of this paper: mixed states, for example, are introduced, in a very
general framework, in Section~\ref{sec:mixed}, and meaningful results on mixed states are 
presented in Section~\ref{sec:mixed2} under more stringent categorical assumptions.

\section{The categorical structure} \label{sec:category}
\subsection{Operations} \label{sec:operations}
The fundamental notion to be studied is that of a quantic operation. 
Intuitively an operation is some transformation of a physical system,
quantic or classical. 
Examples of such operations are: the evolution in time of a quantic system,
the change brought about by a measurement in a quantic system, a step
in a quantic computation, or a classical computation.
We shall represent such operations by letters such as: f, g, a, b, and so on.
The topic of this paper is the algebraic structure of such operations.

\subsection{Types} \label{sec:types}
Operations represent transformations of physical systems, represented, 
for example, by phase spaces. What do those transformations operate upon?
One should not assume that all operations operate on the same phase space. 
Consider, for example, the absorption of a photon by an atom. 
We need to assume that the transformation acts on a state of a space that describes 
an atom and an incoming photon and produces a state of a space that represents an atom.
Those two spaces are different.
Similarly, a computation step such as \mbox{$z := x + y$} can be seen as operating on 
the type {\em two numbers} and into the type {\em three numbers}.

As noticed by Abramsky and Coecke~\cite{Abramsky-Coecke:catsem}, 
then, such operations have types: 
they transform an input type, the {\em domain}, into an output type, 
the {\em co-domain}.
We shall write: \mbox{$f :$} \mbox{$A \rightarrow B$} 
and say that $f$ is an arrow
from type $A$ to type $B$. 
Such a point of view may be surprising for physicists, accustomed to the idea they are 
studying transformations of a system of a fixed type, i.e., an operation from a type $A$ 
to itself, but it has proved extremely useful in many areas of computer science and logic.

\subsection{Sequential composition} \label{sec:sequential}
Our first remark is that operations may be composed sequentially, if they have
proper types. 
If \mbox{$f :$} \mbox{$A \rightarrow B$} 
is an operation that transforms any system 
of type $A$ into a system of type $B$, and \mbox{$g :$}
\mbox{$B \rightarrow C$}
transforms systems of type $B$ into systems of type $C$, then there is an
operation that transforms systems of type $A$ into systems of type $C$,
denoted \mbox{$g \circ f$} that consists in the application of $f$, first,
and then $g$. Another remark is that, for any type $A$, 
there is an operation: ${id}_{A}$ of type
\mbox{$A \rightarrow A$} that represents the {\em do nothing} operation.

We shall assume the following, for any types $A$, $B$, $C$, and $D$ and for any
operations \mbox{$f :$} \mbox{$A \rightarrow B$}, 
\mbox{$g :$} \mbox{$B \rightarrow C$} and \mbox{$h :$}
\mbox{$C \rightarrow D$}
\begin{equation} \label{eq:assoc_comp}
(h \circ g) \circ f = h \circ (g \circ f)
\end{equation}
and
\begin{equation} \label{eq:id_comp}
{id}_{B} \circ f = f = f \circ {id}_{A}.
\end{equation}
We, in a nutshell, require that sequential composition of operations
be associative and that identities behave as neutral elements for sequential
composition. Those requirements are precisely those that define 
the mathematical notion of a category,
the {\em objects} of which are types and the {\em arrows} of which are operations.
We shall, from now on use freely {\em objects} for {\em types} and 
{\em arrows} for {\em operations}.
We shall assume that, for any two types $A$ and $B$, the arrows $f$ with domain $A$ and
codomain $B$ (\mbox{$f : A \rightarrow B$}) form a set denoted \mbox{$Hom(A , B)$}. 
The most important techniques and results in category theory are concerned with universal 
characterizations. We shall show, on an example, how physically meaningful concepts
can be described by universal properties.

A word of caution is in place here. Contrary to the impression given above that we equate
quantic operations with arrows, we shall see in Section~\ref{sec:mixed} that there may be
quantic operations that are not represented by arrows. Two reasons may be given for this
situation:
\begin{itemize}
\item arrows are linear operations as will be required by part~\ref{global} of 
Definition~\ref{def:unit-object}, whereas some quantic operations are antilinear, 
as will be explained in Section~\ref{sec:mixed}. 
A presentation involving both linear and antilinear arrows 
may be possible, but, in such a presentation, technical problems may blur the motivation,
\item to ensure that all quantic operations are represented by an arrow requires some closure
property that the author does not know how to formulate yet. It should probably require that 
any formula built out of $\circ$, $\star$, $\otimes$, $\oplus$ and $+$ be represented by arrows.
\end{itemize}
Therefore the arrows \mbox{$f : A \rightarrow B$} should be seen as {\em basic quantic 
operations} from which more complex quantic operations may be defined.

\subsection{Products and coproducts} \label{sec:prod&coprod}
We shall now show how categorical universal characterizations can express basic concepts 
related to Logic and QM, on the example of the dual notions of product and coproduct 
in category theory.
Let us begin by coproducts. If $A$ and $B$ are types of system, one may ask how would look
a type that would be the type of all systems that are either of type $A$ or of type $B$.
Note that we are not claiming that such a type should exist, we shall claim that in 
Section~\ref{sec:biprod}, but we are asking what properties should such a type enjoy, 
if it were to exist.
If $X$ is the type {\em $A$ or $B$}, then any system of type $A$ (resp. $B$) 
should be associated with a system of type $X$, 
and this association should be, in a sense, uniform, i.e., expressed by an arrow. 
Therefore, there should be arrows 
\mbox{$u_{1} : A \rightarrow X$} and \mbox{$u_{2} : B \rightarrow X$}
representing this association. 
The property we are going to describe now is a fine example of 
the {\em universal} characterizations that are the hallmark of category theory.
Suppose we have two arrows \mbox{$f : A \rightarrow Y$} and
\mbox{$g : B \rightarrow Y$}. 
They represent transformations that associate a system of type $Y$ 
with systems of type $A$ and $B$ respectively. 
Then, by using $f$ and $g$, we have a way of associating a system of type $Y$ with any 
system of type {\em $A$ or $B$}: test if it is $A$ or $B$ and, in accordance, 
act with $f$ or with $g$ , and, therefore, there should be an arrow \mbox{$X \rightarrow Y$}, 
that we shall denote, following the notations of~\cite{Abramsky-Coecke:catsem}, 
\mbox{$[ f \  g ]$} that is
the unique arrow \mbox{$x : X \rightarrow Y$} such that 
\mbox{$f =$} \mbox{$x \circ u_{1}$} and \mbox{$g =$} \mbox{$x \circ u_{2}$}.
Such an object $X$ and arrows $u_{1} , u_{2}$ are called, in category theory, 
a {\em coproduct} for $A$ and $B$. The coproduct object, $X$, of $A$ and $B$ is often
denoted $A + B$.

The dual notion of a {\em product} for $A$ and $B$ consists of an object
$A \times B$ together with two arrows 
\mbox{$p_{1} : A \times B \rightarrow A$} and
\mbox{$p_{2} : A \times B \rightarrow B$} such that, for any arrows
\mbox{$f : Y \rightarrow A$} and \mbox{$g : Y \rightarrow B$} there exists a unique
arrow \mbox{$(f \ g) :$} \mbox{$Y \rightarrow A \times B$} such that \mbox{$f =$}
\mbox{$p_{1} \circ (f \ g)$} and \mbox{$g =$}
\mbox{$p_{2} \circ (f \ g)$} . 
Our notation is traditional, but differs here from~\cite{Abramsky-Coecke:catsem} which uses
\mbox{$\langle \ldots \rangle$}.
The type $A \times B$ represents the type of all systems that can be understood
as having both an $A$ aspect and a $B$ aspect, i.e., in short, that are {\em both
$A$ and $B$}. This should {\em not} be confused with the notion of a system that
has two parts, a part of type $A$ and a part of type $B$.
In Section~\ref{sec:biprod}, we shall develop the topic of composite systems.

\section{The adjoint structure} \label{sec:dagger}
\subsection{Introduction} \label{sec:intro_dagger}
The first structure we want to study formalizes an essential symmetry
principle ubiquitous in Quantum Physics. In short, every physical system
may be described in one of any two dual ways: 
what can be described as flowing in one
direction can as well be described as flowing in the opposite direction.
It fits the spirit of Asher Peres'~\cite{Peres:QuantumTheory} p. 35
{\em Law of Reciprocity}. In the Hilbert space framework it is embedded
in the fact that any operator has an adjoint. This fact is certainly one
of the central features of Hilbert spaces 
that qualify operators in Hilbert space
for representing quantum transformations adequately.
Our formal requirements, to appear in Section~\ref{sec:req_dagger},
form the definition of a {\em dagger category} in Abramsky and 
Coecke's~\cite{Abramsky-Coecke:catsem}, Definition 10 p. 285.
We do not follow their typography or their terminology.

\subsection{Requirements} \label{sec:req_dagger}
\begin{definition} \label{def:A-cat}
An $A-category$ is a category in which every arrow 
\mbox{$f : A \rightarrow B$} is associated with an adjoint arrow
\mbox{$f^{\star} : B \rightarrow A$}, such that for any such $f$ and any
\mbox{$g : B \rightarrow C$} one has:
\begin{equation} \label{eq:adj}
{(f^{\star})}^{\star} \ = \ f \ , 
\ {(g \circ f)}^{\star} \ = \ {f}^{\star} \circ {g}^{\star}.
\end{equation}
\end{definition}
The operation $\star$ can be described, as in~\cite{Abramsky-Coecke:catsem}, as a
strictly involutive, contravariant endo-functor that is the identity on objects.
The $\star$ operation provides a bijection between 
\mbox{$Hom(A , B)$} and \mbox{$Hom(B , A)$}. 
We shall see, in Section~\ref{sec:unit-object}, once we have introduced unit objects, 
that this bijection formalizes, in particular, the following fundamental principle of QM: 
any state in space $A$ is also a linear measure on $A$ 
(giving a scalar value to any state of $A$, by the scalar product) 
and any linear measure on $A$ is characterized in this way by some state of $A$.
States always have a dual aspect: {\em states the system can be put in}
and {\em tests that can be applied to the system}.
Definition~\ref{def:A-cat} requires that this bijection behave as expected
with respect to composition.

\subsection{Consequences} \label{sec:cons_dagger}
The adjoint structure alone has already noticeable consequences, that we shall
develop now.

\subsubsection{Limits and colimits} \label{sec:lim_cons}
In a nutshell, for category theory literate readers, 
the basic property of A-categories is that adjoints provide a tight fit 
between categorical limits and colimits.
\begin{lemma} \label{le:A-lim-colim}
In an A-category, a diagram $L$ is a limit (resp. colimit) 
for a diagram $D$ iff
${L}^{\star}$ is a colimit (resp. limit) for ${D}^{\star}$.
\end{lemma}
The proof is obvious.
Theorem~\ref{the:products}, proved in Section~\ref{sec:ucoprod},
includes a special case of Lemma~\ref{le:A-lim-colim}: products and coproducts 
coincide in an A-category.

\subsubsection{Unitary arrows} \label{sec:unitary}
In an A-category, it is only natural to guess that arrows that are left or right 
inverses of their adjoint have an important role to play.
\begin{definition} \label{def:unitary}
In an A-category, an arrow  \mbox{$f : A \rightarrow B$} is {\em left-unitary} iff
\mbox{$f \circ f^{\star} =$} ${id}_{B}$. It is {\em right-unitary} iff
\mbox{$f^{\star} \circ f =$} ${id}_{A}$. It is {\em unitary} iff it is both left-unitary and
right-unitary.
\end{definition}
Clearly an arrow $f$ is right-unitary (resp. left-unitary, resp. unitary) iff ${f}^{\star}$ is 
left-unitary (resp. right-unitary, resp. unitary).
Left-unitary, right-unitary and unitary arrows are all closed under composition.
Note that a unitary arrow is an isomorphism, but not all isomorphisms are
unitary: in Hilbert spaces, multiplication by $2$ is an isomorphism but not unitary. 
In an A-category, objects that are related by a unitary
arrow have the same categorical properties, but isomorphic objects do not.
Unitary arrows will play the role usually played by isomorphisms: we are interested
in categorical notions that are defined up to a unitary arrow. 
Right-unitary arrows are monos that preserve scalar products, as will be shown in
Theorem~\ref{the:unitary-scalar}. They correspond to subspace injections.
Left-unitary arrows correspond to projections on subspaces. 
Notice that, if \mbox{$p : A \rightarrow B$} is left-unitary, the arrow 
\mbox{${p}^{\star} \circ p : A \rightarrow A$} is self-adjoint 
(see Definition~\ref{def:self-adjoint} below) and idempotent, i.e., what is called a projection
in the framework of Hilbert spaces.

\subsubsection{Self-adjoint arrows} \label{sec:self-adjoint}
In an A-category self-adjoint arrows are salient. 
Note that a self-adjoint arrow must be an inner arrow
\mbox{$A \rightarrow A$}.
 \begin{definition} \label{def:self-adjoint}
 An arrow \mbox{$f : A \rightarrow A$} is {\em self-adjoint} iff
 \mbox{${f}^{\star} =$} $f$.
 \end{definition}
 Note that, if \mbox{$f : A \rightarrow B$},  the arrows \mbox{${f}^{\star} \circ f$} and
 \mbox{$f \circ {f}^{\star}$} are self-adjoint arrows.
We shall now prove that identities are self-adjoint and unitary.
\begin{lemma} \label{le:id}
In an A-category, for any object $A$, ${id}_{A}$ is self-adjoint and unitary.
\end{lemma}
\begin{proof}
\mbox{${({id}_{A})}^{\star} =$}
\mbox{${({id}_{A})}^{\star}  \circ {id}_{A} =$}
\mbox{${({({id}_{A})}^{\star}  \circ {id}_{A})}^{\star} =$}
\mbox{${({({id}_{A})}^{\star})}^{\star} =$}
\mbox{${id}_{A}$}.
We have shown that ${id}_{A}$ is self-adjoint, and it follows that it is unitary.
\end{proof}
\begin{definition} \label{def:eigen}
Let \mbox{$f : A \rightarrow A$}, \mbox{$x : B \rightarrow A$} and
\mbox{$s : B \rightarrow B$}. We shall say that $x$ is an eigenvector for $f$,
with eigenvalue $s$ iff $x$ is right-unitary and we have
\mbox{$f \circ x =$} \mbox{$x \circ s$}.
\end{definition}
\begin{lemma} \label{le:eigen1}
If \mbox{$f : A \rightarrow A$} is self-adjoint and
\mbox{$x : B \rightarrow A$} is  an eigenvector for $f$ with eigenvalue 
\mbox{$s : B \rightarrow B$}, then $s$ is self-adjoint
\end{lemma}
\begin{proof}
\[
s \ = \ {x}^{\star} \circ x \circ s \ = \ {x}^{\star} \circ f \circ x \ = \ 
{({f}^{\star} \circ x)}^{\star} \circ x \ = \ 
\]
\[
{(f \circ x)}^{\star} \circ x \ = \ 
{(x \circ s)}^{\star} \circ x \ = \ {s}^{\star} \circ {x}^{\star} \circ x \ = \ {s}^{\star}  
\]
\end{proof}
\begin{lemma} \label{le:self-prod}
If \mbox{$f , g : A \rightarrow A$} are self-adjoint, then \mbox{$g \circ f$}
is self-adjoint iff $f$ and $g$ commute, i.e., \mbox{$g \circ f =$}
\mbox{$f \circ g$}.
\end{lemma}
\begin{proof}
Assume $f$ and $g$ are self-adjoint.
If \mbox{$g \circ f$} is self adjoint we have:
\mbox{$g \circ f =$} 
\mbox{${(g \circ f)}^{\star} =$}
\mbox{${f}^{\star} \circ {g}^{\star} =$}
\mbox{$f \circ g$}.
If $f$ and $g$ commute we have:
\mbox{${(g \circ f)}^{\star} =$}
\mbox{${(f \circ g)}^{\star} =$}
\mbox{${g}^{\star} \circ {f}^{\star} =$}
\mbox{$g \circ f$}.
\end{proof}
Examples of A-categories are provided in Appendix~\ref{sec:examples1}.

\section{The tensor structure} \label{sec:tensor}
\subsection{Introduction} \label{sec:intro_tensor}
After treating the first fundamental aspect of the categorical treatment of QM, 
the adjoint structure, 
we are now set to treat the second one: the tensor structure. 
Systems of QM may be {\em composite}, i.e., composed of a number of parts. 
In this paper we shall limit ourselves to systems composed of two parts. 
It is indeed a fundamental feature of QM, and of Quantum Computation that
one deals with systems that may have different parts 
that are isolated enough to be clearly identifiable: 
e.g., a pair of of photons with opposite spins in well-separated locations. 
Many of the challenging aspects of QM and Quantum Computation (QC) come from this feature: 
we are dealing with a system of entangled subsystems, 
each of which has some kind of individual identity.

This aspect of QM has been treated so far in the literature by describing the
tensor structure as a symmetric monoidal structure, a well studied topic in
category theory. 
This paper proposes a different, original, approach, closer to the method of 
universal characterizations so pervasive in category theory, 
but some aspects of which are at odds with what is generally considered
good taste in category theory.

Here is a summary of our proposal. To every pair of types $A$, $B$ 
we shall associate a type, denoted \mbox{$A \otimes B$}, the tensor product of
$A$ and $B$, that will represent the type of all systems that can be described as
composed of two subsystems, one of type $A$ and one of type $B$.
Tensor product will be characterized, as expected in
a categorical approach, by what it does, not just on objects, but on arrows.

The basic intuition is that operations
\mbox{$A \otimes B \rightarrow X$} that operate on composite systems
made of subsystems of type $A$ and $B$ to give a system of type $X$ are
in one-to-one correspondence with {\em arrows of two domains} of sort
\mbox{$A , B \rightarrow X$}. 
This intuition is formalized in Section~\ref{sec:T-cat}, Definition~\ref{def:tensor-product}.

A prerequisite to this definition is a proper formalization of the notion of an {\em arrow of two
variables}. This is achieved by the definition of a {\em bi-arrow}
in Section~\ref{sec:bi-arrows}, Definition~\ref{def:bi-arrow}.
Bi-arrows are to arrows what functions of two variables are to functions.
But bi-arrows are defined relative to a specific object. In other terms bi-arrows
are defined only in a pointed category, a category with a designated object.
The properties of the bi-arrows depend on the choice of this object.

For a categorical treatment of QM this object must enjoy certain properties.
Those properties are encapsulated in the definition of a 
{\em unit object} in an A-category, Definition~\ref{def:unit-object}, 
Section~\ref{sec:unit-object}. A unit object represents the type of trivial systems,
i.e., systems that carry no information whatsoever.
As mentioned just above, the definition of a unit object requires an A-category,
it refers to the adjoint structure. 
A remarkable aspect of our approach is that the categorical concepts we will be dealing with: 
unit objects and tensor products are defined, 
not up to an isomorphism as is usually the case in category theory, 
but up to a unitary arrow, a special class of isomorphisms.

\subsection{Previous work: closure properties} \label{sec:closure}
Before we proceed to describe our novel approach, let us point out a significant difference
with the treatment proposed by Abramsky and Coecke in~\cite{Abramsky-Coecke:catsem}.
They view the closure properties of their base category as central. The closure property 
essentially says that, for any two objects (i.e., types) $A$ and $B$ there is an object 
(\mbox{${A}^{\star} \otimes B$}) that
can be understood as the type of all arrows (i.e., quantic operations) from $A$ into $B$.
In other words they require and they consider as a fundamental property the fact that
the quantic operations from type $A$ to type $B$ form a type. 

This paper proposes a different approach, in which we do not require such a closure property
and this seems more in line with the practice of physicists. For physicists transformations
of systems are not a type of systems. 
There may be operations transforming a system of two particles into a system of three particles 
but the set of all such operations does not provide 
a type of systems of transformations: transformations are not states.

\subsection{Unit objects} \label{sec:unit-object}
\subsubsection{Definition of unit objects} \label{sec:unit-def}
We are interested in A-categories in which a designated object $I$ represents
the type of all trivial systems, i.e., systems that carry no information. 
We shall now try to characterize, up to a unitary arrow, the type, $I$, of systems
that contain no information.
 
An arrow such as \mbox{$a : I \rightarrow A$} takes, as input, a trivial system
and produces, as output, a system of type $A$. It is fair to say that such an operation creates, 
or prepares, a system of type $A$ ({\em from nothing}).
An operation \mbox{$b :$} \mbox{$A \rightarrow I$}, on the 
contrary, transforms a system of type $A$ into a system that 
carries no information, therefore it is a method to destroy systems of type $A$.
\begin{definition} \label{def:prep-destruc}
An arrow of sort \mbox{$I \rightarrow A$} is called a {\em preparation}.
An arrow of sort \mbox{$A \rightarrow I$} is called a {\em destruction}.
An arrow of sort \mbox{$I \rightarrow I$} is both a preparation and
a destruction and is called a {\em scalar}.
\end{definition}
In an A-category, the adjoint of a preparation is a destruction, the adjoint of
a destruction is a preparation and the adjoint of a scalar is a scalar.
Let \mbox{$a : I \rightarrow A$} be a preparation. The destruction
${a}^{\star}$ is the destruction that corresponds to the preparation $a$,
in a sense the destruction of $a$. 
The scalar \mbox{${a}^{\star} \circ a$} represents
the effect of preparing $a$ and then destroying it. We cannot hope that,
in general, this be equal to the identity on $I$, but it is reasonable to expect that
those preparations $a$ for which this is the case, i.e., those
preparations that are right-unitary, are more central than others.
Physicists will recognize here {\em normalized states}. 
\begin{definition} \label{def:normalized}
A preparation is said to be {\em normalized} iff it is right-unitary. 
\end{definition}
We shall indeed take the view that the bona fide preparations are normalized.
In other terms, only normalized preparations are first-class citizens.
We shall see in Section~\ref{sec:trivial}, Theorem~\ref{the:norm-trivial} that the 
additive structure forces on us scalars that are normalized only in trivial situations and therefore 
forces on us preparations that are not normalized, in any non-trivial situation.
 
Let us think about the meaning of scalars.
A scalar represents a quantic operation transforming trivial systems into trivial
systems. It may seem that there should be no such meaningful transformation
other than the identity. 
Lemma~\ref{le:trivial} will show that the only categories 
that support an additive structure and satisfy this property are trivial. 
In fact, QM live in categories with scalars other than zero and identity, even many
normalized scalars different from the identity and a host of non-normalized scalars.
A tentative understanding of this fact may be that scalars represent fundamental 
symmetries in physical systems, symmetries that cannot be apprehended
by an observer in any way. 
No information about which of the many symmetric states a system is in, can,
even in principle, be obtained. 
Property~\ref{global} precisely requires that those symmetries are global, 
i.e., they appear in any type and are preserved by preparations and destructions. 
\begin{definition} \label{def:unit-object}
An object $I$ in an A-category is said to be a {\em unit object} iff
\begin{enumerate}
\item \label{atleast}
for any object $A$ there is at least one normalized preparation of sort \mbox{$I \rightarrow A$},
\item \label{prep-arrows}
for any objects $A$, $B$ and any arrows \mbox{$f , g : A \rightarrow B$},
if for any preparation \mbox{$a : I \rightarrow A$} one has
\mbox{$f \circ a =$} \mbox{$g \circ a$} then one has \mbox{$f = g$},
\item \label{global}
for every object $A$ and every scalar \mbox{$s : I \rightarrow I$} there
is an arrow \mbox{$s_{A} :$} \mbox{$A \rightarrow A$} such that 
\begin{itemize}
\item 
for any preparation \mbox{$a : I \rightarrow A$} one has
\mbox{$a \circ s =$} \mbox{$s_{A} \circ a$}, and
\item
for any destruction \mbox{$b : A \rightarrow I$} one has
\mbox{$s \circ b =$} \mbox{$b \circ s_{A}$}.
\end{itemize}
Note that property~\ref{prep-arrows} implies that this characterizes the arrow \mbox{${s}_{A}$}
in a unique way.
\end{enumerate}
\end{definition}
Note that property~\ref{atleast} is the only condition that refers to the adjoint structure.
Note that in systems in which the only scalar is the identity, Property~\ref{global}
is trivially satisfied.
We shall now justify the requirements of Definition~\ref{def:unit-object}.
Property~\ref{atleast} expresses the fact that, if no system of type $A$ can be prepared, then, 
$A$ is not really a type of systems to be considered. Since only normalized preparations 
are bona fide preparations, we require the existence of at least one normalized preparation
for any type $A$.
Property~\ref{prep-arrows} expresses our point of view
that only systems that can be prepared from $I$, i.e., from no information, are really interesting.
If two parallel arrows $f$ and $g$ operate in the same way on all preparations,
i.e., define the same generalized quantic operation,
they must be the same arrow, or at least we shall not distinguish between them.
Property~\ref{global} expresses the {\em global} character of scalars: the symmetries that
exist in $I$, exist in every type of systems. Every preparation preserves those symmetries,
a requirement that parallels the {\em linearity} of quantic transformations. If one wanted to 
consider {\em antilinear} arrows, one would have to modify this requirement.

If we indeed think of scalars as symmetries, it would be reasonable to expect that every
scalar be invertible (with respect to composition). 
Theorem~\ref{the:scalar-com} will show that scalar composition is commutative and therefore
normalized scalars are invertible.
Section~\ref{sec:zero} will force on us a scalar, 
namely \mbox{${0}_{I , I}$}, the zero scalar, that is not, except in trivial situations, invertible. 
In QM, all scalars, except zero are invertible. We shall therefore make this assumption, 
but only in Definition~\ref{def:Q-cat}, when we shall need it.

Building on the intuition that quantic operations realize a correspondence between preparations
of type \mbox{$I \rightarrow A$} and preparations of type \mbox{$I \rightarrow B$}, one considers
a generalized quantic operation from type $A$ to type $B$ to be a transformation associating a
preparation \mbox{$I \rightarrow B$} with every preparation \mbox{$I \rightarrow A$}. 
Certain such generalized quantic operations are defined by an arrow
\mbox{$f : A \rightarrow B$} by: \mbox{$a : I \rightarrow A \mapsto$} 
\mbox{$f \circ a$}, but, in the sequel, we shall consider generalized operations that cannot be
defined by an arrow and, in particular, antilinear generalized quantic operations.

\subsubsection{Properties of unit objects and scalars} \label{sec:unit-prop}
We shall now prove that, if $I$ is a unit object, 
the global character of the scalars expressed in part~\ref{global} of 
Definition~\ref{def:unit-object} implies that they 
behave as expected with respect to the categorical
structure (composition and identities) and the adjoint structure.

\begin{lemma} \label{le:global-cat-adj}
Let $I$ be a unit object in an A-category. For any object $A$ and any scalars 
\mbox{$s , t : I \rightarrow I$}, one has:
\begin{enumerate}
\item  \label{comp}
\mbox{${(t \circ s)}_{A} =$} \mbox{$t_{A} \circ s_{A}$}, 
\item \label{identity}
\mbox{${(id_{I})}_{A} =$} \mbox{${id}_{A}$}, and
\item \label{star}
\mbox{${(s_{A})}^{\star} =$}
\mbox{${(s^{\star})}_{A}$}.
\end{enumerate}
\end{lemma}
\begin{proof}
For item~\ref{comp}, note that,
by item~\ref{global} of Definition~\ref{def:unit-object}, 
for any object $A$, any scalars $s$, $t$ and any
arrow \mbox{$a : I \rightarrow A$} one has 
\[
{(t \circ s)}_{A} \circ a \ = \  a \circ t \circ s \ = \ 
t_{A} \circ a \circ s \ = \  (t_{A} \circ s_{A}) \circ a
\]
and therefore, by
item~\ref{prep-arrows} of Definition~\ref{def:unit-object},
\mbox{${(t \circ s)}_{A}  =$}  \mbox{$t_{A} \circ s_{A}$}.
For item~\ref{identity}, note that, for any preparation \mbox{$a : I \rightarrow A$},
\mbox{${({id}_{I})}_{A} \circ a =$} \mbox{$a \circ {id}_{I} =$}
\mbox{$a =$} \mbox{${id}_{A} \circ a$}. By item~\ref{prep-arrows} of Definition~\ref{def:unit-object}, we conclude that \mbox{${({id}_{I})}_{A} =$} \mbox{${id}_{A}$}.
For item~\ref{star}, note that 
\[
{({s}^{\star})}_{A} \circ a \ = \ 
a \circ {s}^{\star} \ = \ {(s \circ {a}^{\star})}^{\star} \ = \ 
{({a}^{\star} \circ s_{A})}^{\star}  \ = \ 
{({s}_{A})}^{\star} \circ a.
\]
We see that \mbox{${({s}^{\star})}_{A} =$}
\mbox{${({s}_{A})}^{\star}$}.
\end{proof}

\begin{lemma} \label{le:forgotten}
In an A-category with unit object $I$, for any \mbox{$f :$}
\mbox{$A \rightarrow B$} and any scalar \mbox{$s :$}
\mbox{$I \rightarrow I$} one has:
\mbox{$f \circ s_{A} =$} \mbox{$s_{B} \circ f$}.
\end{lemma}
\begin{proof}
For any \mbox{$a : I \rightarrow A$} we have, by item~\ref{global} of 
Definition~\ref{def:unit-object},
\mbox{$f \circ s_{A} \circ a =$}
\mbox{$f \circ a \circ s =$}
\mbox{$(s_{B} \circ f) \circ a$}.
By item~\ref{prep-arrows}, we conclude that
\mbox{$f \circ s_{A} =$} \mbox{$s_{B} \circ f$}.
\end{proof}
Another consequence of property~\ref{global} of Definition~\ref{def:unit-object} is that
composition of scalars is commutative.

\begin{theorem} \label{the:scalar-com}
Assume $I$ is a unit object in an A-category. 
For any scalars \mbox{$s , t: I \rightarrow I$}, one has \mbox{${s}_{I} =$} $s$ and
\mbox{$t \circ s =$} \mbox{$s \circ t$}.
\end{theorem}
\begin{proof}
In Definition~\ref{def:unit-object}, item~\ref{global}, take 
\mbox{$A = I$}.
For any scalars $s , t$ we have \mbox{$t \circ s =$}
\mbox{$s_{I} \circ t$}. Taking \mbox{$t = {id}_{I}$} we see that,
for every scalar $s$ one has \mbox{$s_{I} =$} $s$. Therefore 
\mbox{$t \circ s =$} \mbox{$s \circ t$}.
\end{proof}
Theorem~\ref{the:scalar-com} is a fundamental insight into the structure
of scalars in QM.
The fact that the commutativity of scalar multiplication is a such a basic
property of our framework begs the question of the status of Hilbert spaces 
on the field of quaternions in Quantum Physics. We can say that either a more general framework
in which scalar multiplication is not necessarily commutative would be 
preferable, or the non-commutativity of scalar multiplication has no physical
meaning for those systems described by Hilbert spaces on the quaternions.

Now, we shall ask how precisely is a unit object defined, if one exists.
We note that if $I$ and $J$ are isomorphic objects, the fact that $I$ is a unit 
object does not imply that $J$ is a unit object. 
We show that unit objects are defined up to a unitary transformation.
\begin{theorem} \label{the:unit-objects}
In an A-category, if $I$ is a unit object then $J$ is a unit object iff there is some unitary arrow
\mbox{$u : I \rightarrow J$}.
\end{theorem}
\begin{proof}
Assume $I$ is a unit object.
For the {\em if} part, assume \mbox{$u : I \rightarrow J$} is unitary. 
\begin{enumerate}
\item
For any object $A$ there is a normalized
preparation \mbox{$a : I \rightarrow A$}. The arrow \mbox{$a \circ {u}^{\star} :$} 
\mbox{$J \rightarrow A$} is a normalized $J$-preparation.
\item
Assume, now, that \mbox{$f \circ a' = g \circ a'$} for any 
\mbox{$a' : J \rightarrow A$}.
Then, for any \mbox{$a : I \rightarrow A$} we have 
\mbox{$f \circ a \circ {u}^{\star} =$}
\mbox{$g \circ a \circ {u}^{\star}$}. Compose with $u$ on the right to find that:
\mbox{$f \circ a =$} \mbox{$g \circ a$} for any \mbox{$a : I \rightarrow A$}.
Therefore \mbox{$f =$} $g$.
\item
Let \mbox{$t : J \rightarrow J$}, then $t^{A} =$
\mbox{${({u}^{\star} \circ t \circ u)}_{A}$} is easily seen to 
satisfy Property~\ref{global} of Definition~\ref{def:unit-object}.
\end{enumerate}

For the {\em only if} part, assume that $J$ is a unit object. 
Since $I$ is a unit object, by item~\ref{atleast} of Definition~\ref{def:unit-object},
there is some right-unitary arrow \mbox{$k : I \rightarrow J$}.
We shall show that \mbox{$k \circ {k}^{\star} =$} ${id}_{J}$ and therefore $k$ is unitary.
Since $J$ is a unit object, and \mbox{$k \circ {k}^{\star}$} is a J-scalar, 
we let \mbox{$i =$} \mbox{${(k \circ {k}^{\star})}_{I} :$}
\mbox{$I \rightarrow I$} and,
for any \mbox{$m : J \rightarrow I$} one has \mbox{$m \circ k \circ {k}^{\star} =$}
\mbox{$i \circ m$}.
Therefore, taking \mbox{$m = {k}^{\star}$}, one has
 \mbox{${k}^{\star} \circ k \circ {k}^{\star} =$}
\mbox{$i \circ {k}^{\star}$} and
 \mbox{${k}^{\star} =$}
\mbox{$i \circ {k}^{\star}$}.
Composing by $k$ on the right, we have \mbox{${id}_{I} =$} \mbox{$i =$}
\mbox{${(k \circ {k}^{\star})}_{I}$}. By Lemma~\ref{le:global-cat-adj}, then, we have:
\mbox{${({id}_{J})}_{I} =$} \mbox{${(k \circ {k}^{\star})}_{I}$} and we conclude that
\mbox{${id}_{J} =$} \mbox{$k \circ {k}^{\star}$}.
\end{proof}
The following will be used in Section~\ref{sec:nocloning}.
\begin{theorem} \label{the:characterization}
In an A-category, if $I$ is a unit object and \mbox{$a : I \rightarrow A$} is a normalized 
preparation, then the following propositions are equivalent:
\begin{enumerate}
\item \label{unitary}
$a$ is unitary,
\item \label{unit}
$A$ is a unit object,
\item \label{every}
for every preparation \mbox{$b : I \rightarrow A$} there is some scalar $s$ such that
\mbox{$b =$} \mbox{$a \circ s$}.
\end{enumerate}
\end{theorem}
\begin{proof}
\mbox{$\ref{unit} \Rightarrow \ref{unitary}$}. Assume $A$ is a unit object. By Theorem~\ref{the:unit-objects}
there is some unitary \mbox{$u : I \rightarrow A$}. We have:
\mbox{$a =$} \mbox{${id}_{A} \circ a =$} \mbox{$u \circ {u}^{\star} \circ a$}.
Therefore, by Theorem~\ref{the:scalar-com}, 
\[
a \circ {a}^{\star} \ = \ u \circ ({u}^{\star} \circ a) \circ ({a}^{\star} \circ u) \circ {u}^{\star} \ = \ 
\]
\[
u \circ ({a}^{\star} \circ u) \circ ({u}^{\star} \circ a) \circ {u}^{\star} \ = \ 
u \circ {a}^{\star} \circ a \circ {u}^{\star} \ = \ 
u \circ {u}^{\star} \ = \ {id}_{A}
\]
and preparation $a$, that is normalized, is unitary.

\mbox{$\ref{unitary} \Rightarrow \ref{unit}$}. By Theorem~\ref{the:unit-objects}.

\mbox{$\ref{unitary} \Rightarrow \ref{every}$}. One has:
\mbox{$b =$} \mbox{${id}_{A} \circ b =$} \mbox{$a \circ ({a}^{\star} \circ b)$}.

\mbox{$\ref{every} \Rightarrow \ref{unitary}$}.
Since $a$ is normalized, it is enough to show that \mbox{$a \circ {a}^{\star} =$}
\mbox{${id}_{A}$}.
For this, it is enough to show that for every preparation
\mbox{$b : I \rightarrow A$} one has:
\mbox{$a \circ {a}^{\star} \circ b =$} \mbox{${id}_{A} \circ b$}.
But, for every such $b$ there is some scalar $s$ such that \mbox{$b =$} \mbox{$a \circ s$}
and we have:
\[
a \circ {a}^{\star} \circ b \ = \ a \circ {a}^{\star} \circ a \circ s \ = \ 
a \circ s \ = \ b.
\]
\end{proof}

\subsubsection{Scalar products} \label{sec:scalar-products}
In an A-category with a unit object one may define a scalar product for any two
preparations of the same type, and a norm for any preparation. We shall study arrows that 
preserve those quantities.
\begin{definition} \label{def:scalar-product}
Assume an A-category and a unit object $I$. For any object $A$ and any preparations
\mbox{$a , b : I \rightarrow A$} we define their scalar product 
\mbox{$\langle b \mid a \rangle$} to be the scalar \mbox{${b}^{\star} \circ a$}.
The sqnorm of $a$, $sq(a)$ is defined to be the scalar product of $a$ with itself:
\mbox{${a}^{\star} \circ a$}.
\end{definition}
Note that, by Theorem~\ref{the:unit-objects}, the scalar product and sqnorm do not
depend on the choice of the unit object.
The proof of the next lemma is left to the reader.
\begin{lemma} \label{le:scalar1}
In an A-category with unit object, for any object $A$ and any preparations
\mbox{$a , b : I \rightarrow A$}, one has:
\begin{itemize}
\item scalar product is adjoint-symmetric 
\mbox{$\langle b \mid a \rangle =$} \mbox{${\langle a \mid b \rangle}^{\star}$},
\item sqnorm is self-adjoint: \mbox{${sq^{\star}(a)} =$} \mbox{$sq(a)$},
\item for any \mbox{$f : A \rightarrow B$}, \mbox{$a : I \rightarrow A$},
\mbox{$b : I \rightarrow B$} one has:
\mbox{$\langle b \mid f \circ a \rangle =$} \mbox{$\langle {f}^{\star} \circ b \mid a \rangle$}.
\end{itemize}
\end{lemma}
\begin{definition} \label{def:preserve}
In an A-category with unit object $I$, an arrow \mbox{$f : A \rightarrow B$} is
said to {\em preserve scalar products} iff for any preparations
\mbox{$a , b : I \rightarrow A$}, one has
\mbox{$\langle f \circ b \mid f \circ a \rangle =$} \mbox{$\langle b \mid a \rangle$}.
It is said to be an {\em isometry} iff \mbox{$sq(f \circ a) =$}
\mbox{$sq(a)$}.
\end{definition}
Clearly, any arrow that preserves scalar products is an isometry.
In Hilbert spaces the converse holds:
any linear isometry preserves scalar products. 
Since the proof of this involves heavily the linear structure
of the Hilbert space and the group properties of addition, 
it seems improbable that such a result holds in our more general framework.
Our next result characterizes right-unitary arrows as those arrows that preserve scalar products.
\begin{theorem} \label{the:unitary-scalar}
In an A-category with unit object, an arrow \mbox{$f : A \rightarrow B$} preserves scalar
products iff it is right-unitary.
\end{theorem}
\begin{proof}
If \mbox{${f}^{\star} \circ f = {id}_{A}$}, then, for any \mbox{$a , b : I \rightarrow A$},
\[
\langle f \circ b \mid f \circ a \rangle \ = \ 
{b}^{\star} \circ {f}^{\star} \circ f \circ a \ = \ 
{b}^{\star} \circ a \ =\  \langle b \mid a \rangle.
\]
If, for any \mbox{$a , b : I \rightarrow A$}, 
\mbox{$\langle f \circ b \mid f \circ a \rangle =$}
\mbox{$\langle b \mid a \rangle$}, then, by Definition~\ref{def:unit-object}, item~\ref{prep-arrows}
we have, for any \mbox{$b : I \rightarrow A$},
\mbox{${b}^{\star} \circ {f}^{\star} \circ f =$}
\mbox{${b}^{\star}$} and therefore
\mbox{${f}^{\star} \circ f \circ b =$} $b$.
By  Definition~\ref{def:unit-object}, item~\ref{prep-arrows} again, we conclude that
\mbox{${f}^{\star} \circ f =$} ${id}_{A}$.
\end{proof}
Our next result strengthens Property~\ref{prep-arrows} of Definition~\ref{def:unit-object}.
\begin{theorem} \label{the:unique-scalar-prod}
In an A-category with unit object $I$, for any arrows \mbox{$f , g : A \rightarrow B$} one has
\mbox{$f = g$} iff \mbox{$\langle f \circ a \mid b \rangle =$} 
\mbox{$\langle g \circ a \mid b \rangle$}
for any preparations \mbox{$a : I \rightarrow A$} and \mbox{$b : I \rightarrow B$}.
In particular, for any \mbox{$f : A \rightarrow B$} and \mbox{$g : B \rightarrow A$}
one has \mbox{$g =$} \mbox{${f}^{\star}$} iff
\mbox{$\langle g \circ b \mid a \rangle =$} \mbox{$\langle b \mid f \circ a \rangle$} for any 
preparations \mbox{$a : I \rightarrow A$} and \mbox{$b : I \rightarrow B$}.
\end{theorem}
\begin{proof}
Assume that for any preparations \mbox{$a : I \rightarrow A$} and \mbox{$b : I \rightarrow B$}
one has \mbox{$\langle f \circ a \mid b \rangle =$} \mbox{$\langle g \circ a \mid b \rangle$}.
For any preparation $b$ one has:
\mbox{${a}^{\star} \circ {f}^{\star} \circ b =$}  \mbox{${a}^{\star} \circ {g}^{\star} \circ b$} and,
by item~\ref{prep-arrows} of Definition~\ref{def:unit-object} one has:
\mbox{${a}^{\star} \circ {f}^{\star}  =$}  \mbox{${a}^{\star} \circ {g}^{\star} $}  and therefore
\mbox{$f \circ a =$} \mbox{$g \circ a$} for any preparation $a$. We conclude that 
\mbox{$f = g$}. Our last claim follows from Lemma~\ref{le:scalar1}.
\end{proof}
Examples of unit objects in A-categories are presented in Appendix~\ref{sec:examples2}.

\subsection{Bi-arrows} \label{sec:bi-arrows}
\subsubsection{Systems and sub-systems} \label{sec:biarrows-intro}
We assume an A-category with a unit object $I$.
An arrow \mbox{$f : A \rightarrow B$} can be seen as a transformation that 
transforms every preparation \mbox{$a : I \rightarrow A$} into the preparation
\mbox{$f \circ a : I \rightarrow B$}.
We shall now try to define an {\em arrow of two domains}, generalizing the notion of
a function of two variables.
In the category of sets and functions, any function that associates an element of 
$X$ with any pair of elements from $A$ and $B$ is a proper function of two variables,
i.e., a proper arrow of two domains $A$ and $B$ into $X$.
We generalize this idea by defining an arrow of two domains by associating 
a preparation \mbox{$I \rightarrow X$} to any pair of preparations
\mbox{$I \rightarrow A$} and \mbox{$I \rightarrow B$}. But to respect the structure
of the base category we require a commutation property. 
The preparation associated with \mbox{$a : I \rightarrow A$} and \mbox{$b : I \rightarrow B$}
must be obtained by composing $a$ with some arrow of sort 
\mbox{$A \rightarrow X$} that depends only on $b$ and also by composing $b$
with some arrow of sort 
\mbox{$B \rightarrow X$} that depends only on $a$.

\subsubsection{Definition of bi-arrows} \label{sec:biarrows-def}
Bi-arrows of sort \mbox{$A , B \rightarrow X$} describe 
the ways one can operate on systems composed of two subsystems, 
one of type $A$ and one of type $B$.
\begin{definition} \label{def:bi-arrow}
\begin{sloppypar}
Assume $I$ is a unit object in an A-category \cC\ and let
\mbox{$A , B , C$} be objects of \cC.
A {\em bi-arrow} of sort \mbox{$A , B \rightarrow C$} is a function 
\mbox{$\alpha :$} 
\mbox{$Hom(I , A) \times Hom(I , B)$} \mbox{$\: \rightarrow \:$}
\mbox{$Hom(I , C)$}
such that:
\begin{itemize}
\item for any arrow \mbox{$a :$} \mbox{$I \rightarrow A$} there is an arrow
\mbox{$\alpha^{1}(a) :$} \mbox{$B \rightarrow C$},
\item for any arrow \mbox{$b :$} \mbox{$I \rightarrow B$} there is an arrow
\mbox{$\alpha^{2}(b) :$} \mbox{$A \rightarrow C$}, such that
\item for any \mbox{$a :$} \mbox{$I \rightarrow A$} and any
\mbox{$b :$} \mbox{$I \rightarrow B$} one has:
\begin{equation} \label{eq:bi-arrow}
\alpha(a , b) \ = \ \alpha^{2}(b) \circ a \ = \ \alpha^{1}(a) \circ b.
\end{equation}
\end{itemize}
\end{sloppypar}
\end{definition}
Note that we have defined a bi-arrow $\alpha$ by the function of two arguments
$\alpha(- , -)$ and {\em not} by the pair of functions of one argument 
$\alpha^{1}$ and $\alpha^{2}$. The distinction will prove important in
the interpretation of the uniqueness property of 
Definition~\ref{def:T-cat} since a bi-arrow $\alpha$ does not, in
general, define uniquely the functions $\alpha^{i}$ for \mbox{$i = 1 , 2$}. 

\subsubsection{A family of bi-arrows} \label{sec:kappa}
We shall now present an example of a bi-arrow. Assume $I$ is a unit object in an
A-category and $A$ is an object. One may define a bi-arrow 
\mbox{$\kappa : I , A \rightarrow A$} by setting, for any scalar
\mbox{$s : I \rightarrow I$} and any preparation \mbox{$a : I \rightarrow A$},
\mbox{$\kappa(s , a) =$} \mbox{$a \circ s$}.
Indeed if one sets \mbox{$\kappa^{1}(s) =$} \mbox{${s}_{A}$} and
\mbox{$\kappa^{2}(a) =$} \mbox{$a$} one indeed has
\mbox{$a \circ s =$} \mbox{$\kappa^{1}(s) \circ a =$} \mbox{$\kappa^{2}(a) \circ s$}, 
and $\kappa$ is a bi-arrow.
Examples of bi-arrows in A-categories with unit object are presented 
in Appendix~\ref{sec:examples3}.

\subsubsection{Composing bi-arrows} \label{sec:composing-bi}
The next lemmas show that bi-arrows may be composed with arrows, on both sides.
First, consider a bi-arrow followed by an arrow: 
the result is a bi-arrow.
\begin{lemma} \label{le:comp_bi}
Let \mbox{$\alpha :$} \mbox{$A , B \rightarrow C$} be a bi-arrow and let
\mbox{$f : C \rightarrow D$} be an arrow.
The function \mbox{$\beta :$} 
\mbox{$Hom(I , A) \times Hom(I , B) \rightarrow D$} defined by
\mbox{$\beta(a , b) = f \circ \alpha(a , b)$} for every 
\mbox{$a :$} \mbox{$I \rightarrow A$} and every
\mbox{$b :$} \mbox{$I \rightarrow B$} is a bi-arrow. It will be denoted
\mbox{$f \circ \alpha$}.
Moreover, if \mbox{$g : D \rightarrow E$}, one has
\mbox{$(g \circ f) \circ \alpha =$}
\mbox{$g \circ (f \circ \alpha)$}.
\end{lemma}
\begin{proof}
Since $\alpha$ is a bi-arrow, for any arrows $a , b$ we have
\mbox{$\alpha^{2}(b) \circ a =$} \mbox{$\alpha^{1}(a) \circ b$}.
Therefore \mbox{$(f \circ \alpha^{2}(b)) \circ a =$}
\mbox{$(f \circ \alpha^{1}(a)) \circ b$}.
We can take \mbox{$\beta^{1}(a) =$} \mbox{$f \circ \alpha^{1}(a)$} and
\mbox{$\beta^{2}(b) =$} \mbox{$f \circ \alpha^{2}(b)$} and we have
\[
\beta(a , b) \: = \: \beta^{2}(b) \circ a \: = \: \beta^{1}(a) \circ b
\]
showing that $\beta$ is indeed a bi-arrow. The last part of our claim is obvious.
\end{proof}
Then, consider two arrows into $A$ and $B$ respectively
followed by a bi-arrow \mbox{$\alpha :$}
\mbox{$A , B \rightarrow C$}.
The composition is a bi-arrow.
\begin{lemma} \label{le:comp-bi2}
Let \mbox{$\alpha :$} \mbox{$A , B \rightarrow C$} be a bi-arrow and let
\mbox{$f :$} \mbox{$A' \rightarrow A$} and \mbox{$g :$}
\mbox{$B' \rightarrow B$} be arrows.
The function \mbox{$\beta :$} 
\mbox{$Hom(I , A') \times Hom(I , B') \rightarrow C$} defined by
\mbox{$\beta(a , b) =$} \mbox{$\alpha(f \circ a , g \circ b)$} for every 
\mbox{$a :$} \mbox{$I \rightarrow A'$} and every
\mbox{$b :$} \mbox{$I \rightarrow B'$} is a bi-arrow. It will be denoted
\mbox{$\alpha \circ (f , g)$}.
Moreover, for any \mbox{$f' :$} \mbox{$A'' \rightarrow A'$} and 
\mbox{$g' :$} \mbox{$B'' \rightarrow B'$}, one has:
\mbox{$\alpha \circ (f \circ f' , g \circ g') =$} 
\mbox{$(\alpha \circ (f , g)) \circ (f' , g')$}.
\end{lemma}
\begin{proof}
Take \mbox{$\beta^{1}(a) =$} \mbox{$\alpha^{1}(f \circ a) \circ g$} and
\mbox{$\beta^{2}(b) = $} \mbox{$\alpha^{2}(g \circ b) \circ f$}.
We need to check that
\mbox{$\beta^{1}(a) \circ b =$} \mbox{$\beta^{2}(b) \circ a$},
which holds since $\alpha$ is a bi-arrow.
The last claim is obvious.
\end{proof}

\subsection{Tensor products in A-categories} \label{sec:T-cat}
We may now define tensor products. The definition will be followed by an explanation.
\begin{definition} \label{def:tensor-product}
We assume an A-category with unit object $I$.
A bi-arrow \mbox{$\kappa : A , B \rightarrow X$} is said to be a {\em tensor
product} (for $A$ and $B$) iff:
\begin{enumerate}
\item \label{tensorprod-unique}
for any object $Y$ and any bi-arrow \mbox{$\alpha : A , B \rightarrow Y$} there
is a unique arrow \mbox{$x : X \rightarrow Y$} such that
\mbox{$\alpha =$} \mbox{$x \circ \kappa$}, and
\item \label{tensorprod-star}
for any preparations \mbox{$a , a' : I \rightarrow A$} and \mbox{$b , b' : I \rightarrow B$}
one has 
\begin{equation} \label{eq:cond2}
\kappa^{\star}(a , b) \circ \kappa(a' , b') \ = \ 
{a}^{\star} \circ a' \circ {b}^{\star} \circ b'.
\end{equation}
\end{enumerate}
In such a case we shall write $A \otimes B$ for $X$.
\end{definition}
Definition~\ref{def:tensor-product} defines the notion of a composite system. A system composed
of a part of type $A$ and a part of type $B$ has type $A \otimes B$. There is a canonical way,
$\kappa$, to build a system out of a preparation $a$ of type $A$ and a preparation $b$ of type 
$B$. Condition~\ref{tensorprod-unique} expresses the fact that quantic operations of sort 
\mbox{$A \otimes B \rightarrow X$} are in one-to-one correspondence with bi-arrows of sort
\mbox{$A , B \rightarrow X$}, as expected since bi-arrows have been introduced to describe 
operations on composite systems. Condition~\ref{tensorprod-star} requires that 
the tensor product behave properly with respect to the adjoint structure:
the adjoint, i.e., the destruction, associated with the preparation of a composite system 
composed of two parts \mbox{$(a , b)$} applied to a system with parts \mbox{$(a' , b')$}
behaves as the composition of the destruction associated to $a$ on $a'$ and the destruction
associated to $b$ on $b'$.
It may be better understood in terms of scalar products. Condition~\ref{tensorprod-star} is
equivalent to:
\begin{equation} \label{eq:kappastar1}
\langle \kappa(a , b) \mid \kappa(a' , b') \rangle \ = \ 
\langle a \mid a' \rangle \circ \langle b \mid b' \rangle.
\end{equation}
In Lemma~\ref{le:kappa-tensor}, below, we shall show that, equivalently, 
using the notion of a tensor product of arrows defined in Definition~\ref{def:prodarrows}, 
we can write Condition~\ref{tensorprod-star} as:
\begin{equation} \label{eq:kappastar2}
\langle a \otimes b \mid a' \otimes b' \rangle \ = \ 
\langle a \mid a' \rangle \circ \langle b \mid b' \rangle,
\end{equation}
which is familiar to physicists.
As for the case of unit objects, tensor products are defined up to a unitary arrow, but
the proof, that will be presented in Theorem~\ref{the:tensor-products} 
requires some preliminary work.

It is a fundamental property of QM that if $A$ and $B$ are types, then
there is a type for composite systems that have a part of type $A$ and
a part of type $B$. We shall therefore assume that all pairs of objects have 
a tensor product.
\begin{definition} \label{def:T-cat}
An A-category with unit object is said to be a {\em T-category} iff
for every pair of objects $A$, $B$ there exists a tensor product
\mbox{$\kappa_{A , B} : A , B \rightarrow$} \mbox{$A \otimes B$}.
\end{definition}
Examples of T-categories are presented in Appendix~\ref{sec:examples4}.

\subsection{Bifunctorial character of the tensor product} \label{sec:bifunctorial}
In the framework of monoidal categories proposed 
in~\cite{Abramsky-Coecke:catsem} the bifunctorial character of tensor product
is assumed. We shall now show that, in every T-category, tensor products have this character..
We must, first, define the tensor product of arrows.
\begin{definition} \label{def:prodarrows}
Let \mbox{$f :$} \mbox{$A' \rightarrow A$} and \mbox{$g :$}
\mbox{$B' \rightarrow B$} be arrows in a T-category.
Their tensor product \mbox{$f \otimes g :$} 
\mbox{$A' \otimes B' \rightarrow A \otimes B$} is defined as the unique
arrow such that
\mbox{$(f \otimes g) \circ \kappa_{A' , B'} =$}
\mbox{$\kappa_{A , B} \circ (f , g)$}.
\end{definition} 
Let us check now that this definition makes tensor product a bi-functor.
\begin{lemma} \label{le:otimes-comp}
Let \mbox{$f':$} \mbox{$A'' \rightarrow A'$}, 
\mbox{$f :$} \mbox{$A' \rightarrow A$}, 
\mbox{$g' ;$}\mbox{$B'' \rightarrow B'$} and \mbox{$g :$}
\mbox{$B' \rightarrow B$}.
We have: \mbox{$(f \circ f') \otimes (g \circ g') =$}
\mbox{$(f \otimes g) \circ (f' \otimes g')$}.
\end{lemma}
\begin{proof}
By Definition~\ref{def:prodarrows} and Lemma~\ref{le:comp-bi2} we have
\mbox{$(f \otimes g) \circ (f' \otimes g') \circ \kappa_{A'' , B''} =$}.
\mbox{$\kappa_{A , B} \circ (f \circ f' , g \circ g') =$}
\mbox{$((f \circ f') \otimes (g \circ g')) \circ \kappa_{A'' , B''}$}.
Our claim now follows from the uniqueness property of 
Definition~\ref{def:tensor-product}.
\end{proof}
The tensor product of arrows represents quantic operations on composite systems: 
those operations that can be seen as operating separately on the parts. The tensor
product enjoys many properties that we shall describe now.
The following is an obvious consequence of Lemma~\ref{le:otimes-comp}.
We leave the proof to the reader.
\begin{corollary} \label{co:co-funct}
We have:
\begin{itemize}
\item
\mbox{$(g \circ f) \otimes id_{X} =$} 
\mbox{$(g \otimes id_{X}) \circ (f \otimes id_{X})$},
\item
\mbox{$id_{A} \otimes id_{B} =$} \mbox{$id_{A \otimes B}$}.
\end{itemize}
\end{corollary}
We shall now prove deeper results.

\subsection{Fundamental properties of tensor products} \label{sec:fund-tensprod}
\subsubsection{Coherence and density properties} \label{sec:coherence}
We shall first look for coherence (see~\cite{MacLane:working}) properties.
Coherence is the property that says that all diagrams (that should
commute) commute.
In abstract approaches, such as that of Abramsky and 
Coecke~\cite{Abramsky-Coecke:catsem}, coherence has to be assumed.
More precisely, it is proved out of assumptions about the commutation of 
certain diagrams. In T-categories coherence is less of a problem
since we have a very powerful tools to prove the commutation of diagrams.
It is the uniqueness property included in 
Definition~\ref{def:tensor-product}: in many cases a diagram commutes because both
paths are the unique arrow that mediates between two bi-arrows.

Consider preparations of states in a tensor product, i.e. arrows of sort 
\mbox{$I \rightarrow A \otimes B$} 
into a tensor product. For any \mbox{$a :$} \mbox{$I \rightarrow A$} and
\mbox{$b :$} \mbox{$I \rightarrow B$} the tensor product arrow
\mbox{$\kappa_{A , B}(a , b)$} is such an arrow, but, in general, not all arrows
of sort \mbox{$I \rightarrow A \otimes B$} are tensor products of arrows.
Our first fundamental result is that, in a T-category, 
such tensor products of arrows are {\em dense} in \mbox{$A \otimes B$}.
\begin{lemma} \label{le:dense}
Assume a T-category.
Let \mbox{$\kappa : A , B \rightarrow A \otimes B$} be a tensor product and
let \mbox{$f , g :$} \mbox{$A \otimes B \rightarrow X$}.
If, for any \mbox{$a :$} \mbox{$I \rightarrow A$} and
\mbox{$b :$} \mbox{$I \rightarrow B$} one has
\mbox{$ f \circ \kappa(a , b) =$}
\mbox{$g \circ \kappa(a , b)$}, then \mbox{$f = g$}.
\end{lemma}
\begin{proof}
There is a unique
arrow $h$ such that \mbox{$h \circ \kappa =$} 
\mbox{$f \circ \kappa$}. We conclude that \mbox{$f = g$}.
\end{proof}

\subsubsection{Tensor products respect the adjoint structure} \label{sec:tensor-adjoint}
Our next result is that tensor products behave as expected with respect to
the adjoint structure: the adjoint of a tensor is the tensor of the adjoints.

\begin{lemma} \label{le:star-tens}
In a T-category, for any arrows \mbox{$f : A' \rightarrow A$} and \mbox{$g : B' \rightarrow B$},
one has: \mbox{${f}^{\star} \otimes {g}^{\star} =$} \mbox{${(f \otimes g)}^{\star}$}.
\end{lemma}
\begin{proof}
The result essentially follows from Condition~\ref{tensorprod-star} of 
Definition~\ref{def:tensor-product}.
Let \mbox{$a : I \rightarrow A$}, \mbox{$a' : I \rightarrow A'$}, 
\mbox{$b : I \rightarrow B$} and \mbox{$b' : I \rightarrow B'$}.
We have:
\[
{\kappa}^{\star}_{A , B}(f \circ a' , g \circ b') \circ {\kappa}_{A , B}(a , b) \ = \ 
{(f \circ a')}^{\star} \circ a \circ {(g \circ b')}^{\star} \circ b \ = \ 
\]
\[
{a'}^{\star} \circ ({f}^{\star} \circ a) \circ {b'}^{\star} \circ ({g}^{\star} \circ b) \ = \ 
{\kappa}^{\star}_{A' , B'}(a' , b') \circ \kappa_{A' , B'}({f}^{\star} \circ a , {g}^{\star} \circ b).
\]
But \mbox{$(f \otimes g) \circ \kappa_{A' , B'}(a' , b') =$}
\mbox{$\kappa_{A , B}(f \circ a' , g \circ b')$}.
We see that:
\[
{\kappa}^{\star}_{A' B'}(a' , b') \circ {(f \otimes g)}^{\star} \circ \kappa_{A , B}(a , b) \ = \ 
{\kappa}^{\star}_{A' , B'}(a' , b') \circ \kappa_{A' , B'}({f}^{\star} \circ a , {g}^{\star} \circ b).
\]
Therefore, by Lemma~\ref{le:dense}, after considering the adjoint of both sides, we have:
\[
{(f \otimes g)}^{\star} \circ \kappa_{A , B}(a , b) \ = \ 
\kappa_{A' , B'}({f}^{\star} \circ a , {g}^{\star} \circ b)
\]
which characterizes \mbox{${(f \otimes g)}^{\star}$} as the tensor product
\mbox{${f}^{\star} \otimes {g}^{\star}$}.
\end{proof}

\subsubsection{Tensor products are defined up to a unitary arrow} \label{sec:tensor-unitary}
The third of our results is that tensor products are defined up to a unitary arrow.
\begin{theorem} \label{the:tensor-products}
In a T-category, if \mbox{$\kappa :$} \mbox{$A , B \rightarrow A \otimes B$}
is a tensor product, then \mbox{$\lambda : $} \mbox{$A , B \rightarrow X$} is a tensor product iff
there is some unitary arrow \mbox{$u :$} \mbox{$A \otimes B \rightarrow X$} such that
\mbox{$\lambda =$} \mbox{$u \circ \kappa$}. 
\end{theorem}
\begin{proof}
We leave the proof of the {\em if} part to the reader.
Assume, now, that both $\kappa$ and $\lambda$ are tensor products.
Since $\kappa$ is a tensor product, there exists a unique \mbox{$i : A \otimes B \rightarrow X$}
such that \mbox{$\lambda =$} \mbox{$i \circ \kappa$}.
But, now, Condition~\ref{tensorprod-star} of Definition~\ref{def:tensor-product},
for any preparations 
\mbox{$a , a' : I \rightarrow A$} and \mbox{$b , b' : I \rightarrow B$}, we have
\[
{\lambda}^{\star}(a , b) \circ \lambda(a' , b') \ = \ 
{a}^{\star} \circ a' \circ {b}^{\star} \circ b' \ = \ 
{\kappa}^{\star}(a , b) \circ \kappa(a' , b').
\]
Since \mbox{$\lambda =$} \mbox{$i \circ \kappa$} we see that
\[
{\kappa}^{\star}(a , b) \circ {i}^{\star} \circ i \circ \kappa(a' , b') \ = \ 
{\kappa}^{\star}(a , b) \circ \kappa(a' , b').
\]
By Lemma~\ref{le:dense}, we conclude that
\[
{\kappa}^{\star}(a , b) \circ {i}^{\star} \circ i \ = \ 
{\kappa}^{\star}(a , b) \ , \ 
{i}^{\star} \circ i \circ \kappa(a ,b) \ = \  \kappa(a ,b) 
\]
and therefore \mbox{${i}^{\star} \circ i =$} \mbox{${id}_{A \otimes B}$}.
We have:
\[
i \circ {i}^{\star} \circ \lambda  \ = \ i \circ {i}^{\star} \circ i \circ \kappa \ = \ 
i \circ \kappa \ = \ \lambda.
\]
Therefore \mbox{$i \circ {i}^{\star} =$} \mbox{${id}_{X}$}
and $i$ is unitary.
\end{proof}

\subsection{Monoidal properties} \label{sec:monoidal}
\subsubsection{Introduction} \label{sec:intro-monoid}
In the symmetric monoidal structures studied in~\cite{Abramsky-Coecke:catsem}:
\begin{itemize}
\item
\mbox{$(A \otimes B) \otimes C$} is naturally isomorphic to 
\mbox{$A \otimes (B \otimes C)$},
\item
\mbox{$A \otimes B$} is naturally isomorphic to \mbox{$B \otimes A$}, and
\item
\mbox{$I \otimes A$} and
\mbox{$A \otimes I$} are naturally isomorphic to $A$.
\end{itemize}
In our framework can we expect those properties to hold when {\em naturally isomorphic}
is interpreted as {\em unitarily equivalent}, i.e., are there unitary arrows
\mbox{$u : (A \otimes B) \otimes C \rightarrow A \otimes (B \otimes C)$},
\mbox{$u : A \otimes B \rightarrow B \otimes A$},
\mbox{$u : I \otimes A \rightarrow A$} and \mbox{$u : A \otimes I \rightarrow A$}?
We shall consider those questions in (the opposite) order.

\subsubsection{Tensorial properties of a unit object} \label{sec:tens-unit} 
We shall now prove that the bi-arrow presented in Section~\ref{sec:kappa} 
is a tensor product, proving the last one the three properties above.
\begin{lemma} \label{le:I-times-A}
If $I$ is a unit object in an A-category, the bi-arrow \mbox{$\sigma : I , A \rightarrow A$},
defined by \mbox{$\sigma(s , a) =$} \mbox{$a \circ s$}, is a tensor product.
From now on we shall take \mbox{$I \otimes A =$} \mbox{$A =$} \mbox{$A \otimes I$}, 
\mbox{$\kappa_{I , A}(s , a) =$} \mbox{$a \circ s$} and \mbox{$\kappa_{A , I}(a , s) =$}
\mbox{$a \circ s$}.
\end{lemma}
\begin{proof}
First one sees that $\sigma$ is a bi-arrow with \mbox{$\sigma^{1}(s) =$} \mbox{${s}_{A}$}
and \mbox{$\sigma^{2}(a) =$} $a$, by Definition~\ref{def:unit-object}.
For property~\ref{tensorprod-unique} of Definition~\ref{def:tensor-product},
let \mbox{$\alpha : I , A \rightarrow X$} be any bi-arrow.
If \mbox{$\alpha =$} \mbox{$f \circ \sigma$} for some
\mbox{$f : A \rightarrow X$}, we must have 
\mbox{$\alpha(s , a) =$} \mbox{$f \circ a \circ s$} for any 
\mbox{$s : I \rightarrow I$} and any \mbox{$a : I \rightarrow A$},
and in particular \mbox{$f \circ a =$} \mbox{$\alpha({id}_{I} , a)$}.
By Definition~\ref{def:unit-object}, Property~\ref{prep-arrows}, we conclude
that there can be at most one such $f$.
But, if we take \mbox{$f = \alpha^{1}({id}_{I}) : A \rightarrow X $}, we have
\mbox{$f \circ a =$} \mbox{$\alpha({id}_{I} , a) =$}
\mbox{$\alpha^{2}(a)$} and therefore
\mbox{$f \circ \sigma(s , a) =$}
\mbox{$f \circ a \circ s =$}
\mbox{$\alpha^{2}(a) \circ s =$}
\mbox{$\alpha(s , a)$}.

For property~\ref{tensorprod-star}, notice that 
\[
{\sigma}^{\star}(s , a) \circ \sigma(s' , a') \ = \ 
{s}^{\star} \circ {a}^{\star} \circ  a' \circ s' \ = \ 
{s}^{\star} \circ s' \circ {a}^{\star} \circ a'
\]
by the commutativity of scalar composition (see Theorem~\ref{the:scalar-com}).   
\end{proof}
Theorem~\ref{the:tensor-products} implies the existence of a unitary arrow 
\mbox{$u_{A} : I \otimes A \rightarrow A$}. Our convention 
\mbox{$I \otimes A =$} \mbox{$A$} implies that \mbox{$u_{A} =$}
\mbox{${id}_{A}$}. No harm can ensue from identifying \mbox{$I \otimes A$} 
and \mbox{$A \otimes I$} and our convention is harmless because composition of scalars
is commutative.
We now describe the notational consequences of this convention.
The next lemma has been announced in Equation~\ref{eq:kappastar2}.
\begin{lemma} \label{le:kappa-tensor}
In a T-category, for any \mbox{$a : I \rightarrow A$} and any
\mbox{$b : I \rightarrow B$} one has \mbox{$\kappa_{A , B}(a , b) =$}
\mbox{$a \otimes b$} and one can take
\mbox{$\kappa_{A , B}^{1}(a) =$} \mbox{$a \otimes id_{B}$} and
\mbox{$\kappa_{A , B}^{2}(b) =$} \mbox{$id_{A} \otimes b$}.
\end{lemma}
\begin{proof}
By Lemma~\ref{le:I-times-A},
\mbox{${u}_{I} \circ \kappa_{I , I}(id_{I} , id_{I}) = id_{I}$}.
Our convention to identify \mbox{$I \otimes I$} to $I$ and ${u}_{I}$ to the identity ${id}_{I}$
implies \mbox{$\kappa_{I , I}(id_{I} , id_{I}) = id_{I}$}.
By Lemma~\ref{le:comp-bi2}, we have
\mbox{$(\kappa_{A , B} \circ (a , b))(id_{I} , id_{I}) =$}
\mbox{$\kappa_{A , B}(a , b)$}.
By Definition~\ref{def:prodarrows},
\mbox{$(a \otimes b) \circ \kappa_{I , I}(id_{I} , id_{I}) =$}
\mbox{$(\kappa_{A , B} \circ (a , b))(id_{I} , id_{I})$}.
We conclude that \mbox{$a \otimes b =$}
\mbox{$\kappa_{A , B}(a , b)$}.
By Lemma~\ref{le:otimes-comp} we see that
\mbox{$\kappa_{A , B}(a , b) =$}
\mbox{$(a \otimes id_{B}) \circ b =$}
\mbox{$(id_{A} \otimes b) \circ a$}.
\end{proof}
We can now rephrase Lemma~\ref{le:dense}.
\begin{corollary} \label{co:dense}
Assume a T-category.
Let \mbox{$f , g :$} \mbox{$A \otimes B \rightarrow X$}.
If, for any \mbox{$a :$} \mbox{$I \rightarrow A$} and
\mbox{$b :$} \mbox{$I \rightarrow B$} one has
\mbox{$ f \circ (a \otimes b) =$}
\mbox{$g \circ (a \otimes b)$}, then \mbox{$f = g$}.
\end{corollary}
Our next lemma deals with tensor products of scalars and preparations.
\begin{lemma} \label{le:global}
In a T-category, for any object $A$, any scalar $s$ and 
any \mbox{$a :$} \mbox{$I \rightarrow A$},
one has \mbox{$a \circ s =$} 
\mbox{$s \otimes a =$} \mbox{$a \otimes s$}.
\end{lemma}
\begin{proof}
By Lemmas~\ref{le:I-times-A} and~\ref{le:kappa-tensor}
\[
a \circ s \ = \ \kappa_{I , A}(s , a) \ = \ s \otimes a
\]
and
\[
a \circ s \ = \ \kappa_{A , I}(a , s) \ = \ 
a \otimes s.
\]
\end{proof}
Lemma~\ref{le:global} can be applied to scalars and one obtains:
\begin{corollary} \label{co:tensor-scalar}
In a T-category, for any scalars $s$, $t$, one has 
\mbox{$s \otimes t =$} \mbox{$s \circ t$}.
\end{corollary}
Our last result follows.
\begin{corollary} \label{co:sA}
In a T-category, for any object $A$ and any scalar \mbox{$s : I \rightarrow I$}
one has \mbox{$s_{A} =$} \mbox{${id}_{A} \otimes s =$}
\mbox{$s \otimes {id}_{A}$}.
\end{corollary}
\begin{proof}
By Lemmas~\ref{le:otimes-comp} and~\ref{le:global},
for any \mbox{$a : I \rightarrow A$}, one has:
\mbox{$({id}_{A} \otimes s) \circ a =$}
\mbox{$({id}_{A} \otimes s) \circ (a \otimes {id}_{I}) =$}
\mbox{$({id}_{A} \circ a) \otimes (s \circ {id}_{I}) =$}
\mbox{$a \otimes s =$}
\mbox{$a \circ s =$}
\mbox{${s}_{A} \circ a$}.
By Property~\ref{prep-arrows} of Definition~\ref{def:unit-object}, we conclude that
 \mbox{$s_{A} =$} \mbox{${id}_{A} \otimes s$}. The last equality follows from
 Lemma~\ref{le:global}.
\end{proof}

\subsubsection{Symmetry} \label{sec:symmetry}
We shall now show that the tensor products are symmetric: 
there is a unitary arrow \mbox{$u:$}
\mbox{$A \otimes B \rightarrow B \otimes A$}, proving the second one of
the three properties above.
\begin{lemma} \label{le:sym}
In a $T$-category, If \mbox{$\kappa :$} \mbox{$A , B \rightarrow A \otimes B$} 
is a tensor product, then
the bi-arrow \mbox{$\lambda :$} \mbox{$B , A \rightarrow A \otimes B$} defined by
\mbox{$\lambda(b , a) =$} \mbox{$\kappa(a , b)$} for any preparations
\mbox{$a : I \rightarrow A$}, \mbox{$b : I \rightarrow B$} is also a tensor product.
There is, therefore, a unitary arrow \mbox{$u :$}
\mbox{$B \otimes A \rightarrow A \otimes B$} such that
\mbox{$u \circ \kappa_{B , A} =$} \mbox{$\lambda$}.
\end{lemma}
\begin{proof}
One checks that $\lambda$ is a bi-arrow with \mbox{$\lambda^{1}(b) =$}
\mbox{$\kappa^{2}(b)$} and \mbox{$\lambda^{2}(a) =$}
\mbox{$\kappa^{1}(a)$}. Let \mbox{$\alpha : B , A \rightarrow X$} and
\mbox{$f : A \otimes B \rightarrow X$}. One sees that
\mbox{$\alpha =$} \mbox{$f \circ \lambda$} iff 
\mbox{$\beta =$} \mbox{$f \circ \kappa$} with 
\mbox{$\beta(a , b) = \alpha(b , a)$}.
We conclude that $\lambda$ is a tensor product.
The last claim follows from Theorem~\ref{the:tensor-products}.
\end{proof}
One may assume that \mbox{$B \otimes A =$} \mbox{$A \otimes B$} and that 
the unitary arrow $u$ is the identity, i.e., that 
\mbox{$\kappa_{B , A}(b , a) =$} \mbox{$\kappa_{A , B}(a , b)$} for any preparations
\mbox{$a :$} \mbox{$I \rightarrow A$} and \mbox{$b :$} \mbox{$I \rightarrow B$}.

\subsubsection{Associativity} \label{sec:associativity}
I have not been able to prove, in the very general framework of $T$-categories, the existence 
of a unitary arrow from \mbox{$(A \otimes B) \otimes C$} 
to \mbox{$A \otimes (B \otimes C)$}, and I doubt this holds but lack a counter example.
If one assumes an additive structure, in the more restricted framework of BT-categories
that will be presented in Section~\ref{sec:B-cat}, one should be able to develop a the notion
of basis and dimension for an object along lines similar to what has been done 
in~\cite{SP:IJTP}. I conjecture that associativity of tensor product holds for finite dimensional
objects in BT-categories. 

It should be noticed that none of the results in this paper require associativity of tensor products,
which comes as a surprise since associativity is the fundamental property of the monoidal
structures advocated in the literature.
One may ask whether the associativity of the tensor product is a fundamental principle of QM.
Is it obviously the case that systems composed of two parts: 
one a composite system with a part of type $A$ and a part of type $B$ 
and the other of type $C$ are the same as systems composed of a part of type $A$ 
and a part of composite type \mbox{$B \otimes C$}? 
Is a system composed of an hydrogen atom and a neutron really (or naturally) the same as 
as a system composed of a deuterium nucleus and an electron?

\subsection{Tensor products of indistinguishable  subsystems} \label{sec:identical}
The tensor product \mbox{$\kappa : A , B \rightarrow A \otimes B$} represents the way
a composite system is built out of two systems, one in $A$ and one in $B$. 
In QM, a special case of fundamental importance occurs when one considers tensor products 
of the type \mbox{$A \otimes A$}.
Systems composed of two parts of type $A$ may be of three different types. 
If the two parts can be, even only in principle, distinguished, 
then the construction described above is the right one, 
but if the two subsystems cannot, even in principle, be distinguished, 
the ways one can operate on such a composite system are different. 
The bi-arrows to be considered should be the {\em symmetric} ones (for bosons) or the 
anti-symmetric ones (for fermions) and the tensor construction should be different. 
This construction will be presented in 
Section~\ref{sec:tensor-sym}, once we have defined the additive structure.

\subsection{Mixed states, I} \label{sec:mixed}
We shall now study more carefully the structure of preparations 
\mbox{$ c : I \rightarrow A \otimes B$} of entangled states and define the {\em partial
traces} defined by any such $c$ on the component types $A$ and $B$.
One should note that this treatment of mixed states does not make use of an additive structure,
it is purely (tensor) multiplicative.

A preparation \mbox{$ c : I \rightarrow A \otimes B$} defines naturally a generalized quantic
operation that sends any preparation \mbox{$a : I \rightarrow A$} into
the preparation \mbox{$x_{c}(a) : I \rightarrow B$} defined by
\begin{equation} \label{eq:x-anti}
x_{c}(a) \ = \ ({a}^{\star} \otimes {id}_{B}) \circ c.
\end{equation}
Similarly it defines an operation that sends any preparation \mbox{$b : I \rightarrow B$} into
the preparation \mbox{$y_{c}(b) : I \rightarrow A$} defined by
\begin{equation} \label{eq:y-anti}
y_{c}(b) \ = \ ({id}_{A} \otimes {b}^{\star}) \circ c.
\end{equation}
Note that $x_{c}$ and $y_{c}$ are antilinear operations in $a$ and $b$ respectively, i.e.,
for any \mbox{$c : I \rightarrow A \otimes B$}, any \mbox{$a : I \rightarrow A$} and any
\mbox{$s : I \rightarrow I$}, one has \mbox{$x_{c}(a \circ s) =$} \mbox{$x_{c}(a) \circ {s}^{\star}$}.
The operations $x_{c}$ and $y_{b}$ are legitimate quantic operations. For example, $x_{c}$
can be performed by preparing the entangled state $c$ and measuring the local $A$-part
of the entangled system, by projecting it onto $a$. If one indeed finds the $A$-part in state 
$a$, the resulting state (in $A \otimes B$) is a product state \mbox{$a \otimes b$} for some
state $b$ of $B$. In other terms, measuring $a$ on the state $c$ is a legitimate preparation
of the state \mbox{$b =$} \mbox{$x_{c}(a)$}.
We shall now show that the (generalized, antilinear) quantic operations $x_{c}$ and $y_{c}$
are {\em adjoint}. Note the definition of adjointness has to be edited for antilinear operations.
\begin{theorem} \label{the:x-y-adjoint}
For any \mbox{$ c : I \rightarrow A \otimes B$}, \mbox{$a : I \rightarrow A$} and 
 \mbox{$b : I \rightarrow B$}, one has:
 \[
 \langle b \mid x_{c}(a) \rangle \ = \ \langle a \mid y_{c}(b) \rangle.
 \]
\end{theorem}
\begin{proof}
Indeed
\mbox{$\langle b \mid x_{c}(a) \rangle =$} 
\mbox{${b}^{\star} \circ ({a}^{\star} \otimes {id}_{B}) \circ c$}
and 
\[ 
{b}^{\star} \circ ({a}^{\star} \otimes {id}_{B}) \ = \ 
({id}_{I} \otimes {b}^{\star}) \circ ({a}^{\star} \otimes {id}_{B}) \ = \ 
({id}_{I} \circ {a}^{\star}) \otimes ({b}^{\star} \circ {id}_{B}) \ = \ 
{a}^{\star} \otimes {b}^{\star}
\]
by Lemmas~\ref{le:global} and~\ref{le:otimes-comp}.
Similarly, one shows that 
\mbox{${a}^{\star} \circ ({id}_{A} \otimes {b}^{\star}) =$}
\mbox{${a}^{\star} \otimes {b}^{\star}$}.
\end{proof}
If we think of the system $c$ as being an entangled system the A-part and B-part of which are
at a distance, it is natural to consider the partial traces of $c$ on $A$ and $B$ respectively as
being the generalized quantic operations
\mbox{$d_{c}^{A}$} and \mbox{$d_{c}^{B}$} defined by:
for any \mbox{$a : I \rightarrow A$}, \mbox{$d_{c}^{A}(a) =$} \mbox{$y_{c}(x_{c}(a))$}
and for any \mbox{$b : I \rightarrow B$}, \mbox{$d_{c}^{B}(b) =$} \mbox{$x_{c}(y_{c}(b))$}.
Note that $d_{c}^{A}$ and $d_{c}^{B}$ that are compositions of two antilinear operations are 
linear: \mbox{$d_{c}^{A}(a \circ s) =$} \mbox{$d_{c}^{A}(a) \circ s$}.
In Corollary~\ref{co:complete-positive} below, the term {\em completely  } 
must be understood loosely: in T-categories,
the arrows of the form \mbox{${f}^{\star} \circ f$} or even the scalars of the form
\mbox{${s}^{\star} \circ s$} are not necessarily positive in any sense, as explained in 
Section~\ref{sec:normalizable} following Lemma~\ref{le:fstarfzero}.
\begin{corollary} \label{co:complete-positive}
The generalized quantic operation \mbox{$d_{c}^{A}$} (resp. \mbox{$d_{c}^{B}$}) is
self-adjoint, i.e., \mbox{$\langle a' \mid d_{c}^{A}(a) \rangle =$}
\mbox{$\langle d_{c}^{A}(a') \mid a \rangle$}
 and {\em completely positive}, i.e.,
\mbox{$\langle a \mid d_{c}^{A}(a) \rangle =$} \mbox{$x_{c}^{\star}(a) \circ x_{c}(a)$},
for any \mbox{$a : I \rightarrow A$}.
\end{corollary}
\begin{proof}
By Theorem~\ref{the:x-y-adjoint},
for any \mbox{$a , a' : I \rightarrow A$},
\[
\langle a' \mid d_{c}^{A}(a) \rangle \ = \ 
\langle a' \mid y_{c}(x_{c}(a)) \rangle \ = \ 
\langle x_{c}(a) \mid x_{c}(a') \rangle
\]
Therefore, by Lemma~\ref{le:scalar1}
\[
\langle d_{c}^{A}(a') \mid a \rangle \ = \ 
\langle x_{c}(a) \mid x_{c}(a') \rangle.
\]
Therefore  \mbox{$\langle a' \mid d_{c}^{A}(a) \rangle =$}
\mbox{$\langle d_{c}^{A}(a') \mid a \rangle$}
 and \mbox{$\langle a \mid d_{c}^{A}(a) \rangle =$} \mbox{$x_{c}^{\star}(a) \circ x_{c}(a)$}.
\end{proof}
This treatment should be compared with that of B. Coecke's~\cite{Coecke:complete-positivity}
who also noticed that the notion of {\em complete-positivity} does not presuppose any notion
of {\em positivity}.
In the case $c$ is a product state, one obtains the expected results:
\[
x_{a' \otimes b'}(a) \ = \  ({a}^{\star} \otimes {id}_{B}) \circ (a' \otimes b') \ = \ 
(a^{\star} \circ a') \otimes b' \ = \  b' \circ (a^{\star} \circ a')
\]
and
\[
d_{a' \otimes b'}^{A}(a) \ = \  
({id}_{A} \otimes (a'^{\star} \circ a \circ b'^{\star})) \circ (a' \otimes b') \ = \ 
a' \circ (a'^{\star} \circ a) \circ (b'^{\star} \circ b').
\]
As a consequence, we shall prove, in Theorem~\ref{the:equal-eigen}, that in a restricted family
of T-categories, the mixed states $d_{c}^{A}$ and $d_{c}^{B}$ have the same eigenvalues, 
a result that is part of the background of, but not proved in, the study of two-party entanglement 
found in~\cite{Nielsen:LOCC}.

\section{The additive structure} \label{sec:biprod}
\subsection{Introduction} \label{sec:additive-intro}
In Section~\ref{sec:prod&coprod} we presented the notions of a product 
and the dual notion of a coproduct, 
and introduced the notations 
\mbox{$( f \ g)$} and \mbox{$[ f \  g ]$}. 
For us, the product \mbox{$A \times B$} , of types $A$ and $B$
is the type of all systems that are both $A$ and $B$, 
or have both an $A$ aspect and a $B$ aspect. 
The coproduct \mbox{$A + B$}, of types
$A$ and $B$ is the type of all systems that are either $A$ or $B$.

Our goal is to present a theory of biproducts in A-categories that fits QM.
We shall assume that, in our base category, there are coproducts that respect the adjoint structure
and prove that such coproducts are biproducts and are defined up to a unitary arrow.
A number of steps are necessary.
We shall now, first, in Section~\ref{sec:add-gen} describe the functorial character 
of products and coproducts in arbitrary categories and present some notations.
We shall explain why we require that every pair of objects have a coproduct.
In Section~\ref{sec:coprod-A-cat} we study coproducts and products in A-categories.
Then, in Section~\ref{sec:zero} we shall explain that we need a fixed family of zero arrows
to express the notion of orthogonality, a notion
that is central to QM: the different injections of a coproduct are orthogonal. 
In Section~\ref{sec:ucoprod}, we shall define a special kind of coproducts in A-categories, 
called u-coproducts, that fit the adjoint structure.
We shall claim, on first principles, that it is reasonable to assume that, in a
T-category for QM, every pair of objects has a u-coproduct, i.e., a coproduct 
with right-unitary injections that are orthogonal.
We shall show that u-coproducts are defined up to a unitary arrow and
that we have biproducts.
Then, we shall define a family of A-categories 
that have u-coproducts for every pair of objects, B-categories. 
We shall show, in the remainder of Section~\ref{sec:biprod},  
that B-categories possess an almost abelian structure.

A word of explanation is in order: we consider only products and coproducts of pairs of objects,
and consider neither limits or colimits of other diagrams nor products or coproducts of sets other
than pairs.
We do not want to require terminal (product of an empty set of objects) 
or initial (coproduct of an empty set of objects) objects. We could, easily, have considered
products and coproducts of arbitrary finite non-empty sets of objects. More interestingly, it seems,
at first sight, that products and coproducts of infinite sets pose no special problems, but we 
leave this for further study.

\subsection{Products and coproducts in arbitrary categories} \label{sec:add-gen}
There is nothing original in this Section.
Products and coproducts have a functorial character. 
If \mbox{$p^{1}_{i} : A_{i} \times B_{i} \rightarrow A_{i}$}
and \mbox{$p^{2}_{i} : A_{i} \times B_{i} \rightarrow B_{i}$} are products, for 
\mbox{$i = 1 , 2$}, and if one has \mbox{$f : A_{1} \rightarrow A_{2}$} and 
\mbox{$g : B_{1} \rightarrow B_{2}$}, there is a unique arrow
\mbox{$ f \times g : A_{1} \times B_{1} \rightarrow A_{2} \times B_{2}$} such that
\mbox{$ p^{1}_{2} \circ (f \times g) =$} \mbox{$f \circ p^{1}_{1}$} and
\mbox{$ p^{2}_{2} \circ (f \times g) =$} \mbox{$g \circ p^{2}_{1}$}.
In fact, one has \mbox{$f \times g =$} \mbox{$(f \circ p^{1}_{1} \ \ g \circ p^{2}_{1})$}.
The product of arrows behaves as expected with respect to composition of arrows:
\mbox{$(f_{1} \times g_{1}) \circ (f_{2} \times g_{2}) =$}
\mbox{$(f_{1} \circ f_{2}) \times (g_{1} \circ g_{2})$}.
Dual results hold for coproducts.
\begin{comment}
Assume \mbox{$p^{A , B} : A \times B \rightarrow A$} and 
\mbox{$q^{A , B} : A \times B \rightarrow B$} is a product. 
One sees that the pair \mbox{$( q , p$} is a product and therefore
one can take \mbox{$B \times A =$} \mbox{$A \times B$}, 
\mbox{$p^{B , A} =$} \mbox{$q^{A , B}$}, \mbox{$q^{B , A} =$} \mbox{$p^{A , B}$}
 and also 
\mbox{$f \times g =$} \mbox{$g \times f$}. Similarly one can take
\mbox{$(A \times B) \times C =$} \mbox{$A \times (B \times C)$} and
\mbox{$(f \times g) \times h =$} \mbox{$f \times (g \times h)$}. Dual results hold for coproducts.
\end{comment}

\subsection{Products and coproducts in A-categories} \label{sec:coprod-A-cat}
We shall now assume that our base category is an A-category that possesses coproducts
for every pair of objects. The coproduct \mbox{$A \oplus B$} of $A$ and $B$ represents 
the type of systems that are either of type $A$ or of type $B$. We shall then study the interplay 
between the adjoint structure and products and coproducts: in an A-category products and 
coproducts coincide in a very strong sense.
\begin{theorem} \label{the:products}
In an A-category, for any objects $A$, $B$, if \mbox{$a : A \rightarrow A \oplus B$},
\mbox{$b : B \rightarrow A \oplus B$} is a coproduct, then
\begin{enumerate}
\item \label{prod-coprod-objects}
\mbox{${a}^{\star} : A \oplus B \rightarrow A$}, 
\mbox{${b}^{\star} : A \oplus B \rightarrow B$} is a product.
\item \label{hor-ver}
For any \mbox{$f : X \rightarrow A$}, \mbox{$g : X \rightarrow B$} one has:
\mbox{${(f \ g)}^{\star} =$}  \mbox{$[ {f}^{\star} \  {g}^{\star} ]$}.
Similarly, for any \mbox{$f : A \rightarrow X$}, \mbox{$g : B \rightarrow X$} one has:
\mbox{${({f}^{\star}  {g}^{\star})} =$}  \mbox{${[f \  g ]}^{\star}$}.
\item \label{prod=coprod}
Moreover, for any objects $A_{i}$, $B_{i}$ , \mbox{$i = 1 , 2$}, if 
\mbox{$a_{i} :$} \mbox{$A_{i} \rightarrow A_{i} \oplus B_{i}$} and
\mbox{$b_{i} :$} \mbox{$B_{i} \rightarrow A_{i} \oplus B_{i}$}  are coproducts
for \mbox{$i = 1 , 2$},
then, for any \mbox{$f :$} \mbox{$A_{1} \rightarrow A_{2}$} and any 
\mbox{$g :$} \mbox{$B_{1} \rightarrow B_{2}$}
the coproduct of $f$ and $g$
\mbox{$f \oplus g :$} \mbox{$A_{1} \oplus B_{1} \rightarrow A_{2} \oplus B_{2}$} 
is also their product, i.e., 
\mbox{${a}_{2}^{\star} \circ (f \oplus g) =$} 
\mbox{$f \circ {a}_{1}^{\star}$} and
\mbox{${b}_{2}^{\star} \circ (f \oplus g) =$} 
\mbox{$g \circ {b}_{1}^{\star}$}.
\item \label{star-prod}
For any $f$, $g$ as above, \mbox{${(f \oplus g)}^{\star} =$}
\mbox{${f}^{\star} \oplus {g}^{\star}$}. 
\end{enumerate}
\end{theorem}
\begin{proof}
\begin{enumerate}
\item 
Let \mbox{$f : X \rightarrow A$} and \mbox{$g : X \rightarrow B$} and 
\mbox{$x : X \rightarrow A \oplus B$}. We have:
\begin{equation} \label{eq:prodbi}
{a}^{\star} \circ x \ = \ f  \ {\rm and} \  {b}^{\star} \circ x \ = \  g 
\end{equation}
iff
\[
{f}^{\star}  \  = \ {x}^{\star} \circ a \ {\rm and} \  {g}^{\star} \ = \  {x}^{\star} \circ b.
\]
But there is a unique arrow \mbox{$k : A \oplus B \rightarrow X$} such that
\[
{f}^{\star} \ = \ k \circ a \  {\rm and} \  {g}^{\star} \ = \  k \circ b. 
\]
We conclude that there is a unique $x$ that satisfies Equation~\ref{eq:prodbi}, 
namely ${k}^{\star}$.
\item
Suppose \mbox{$f : X \rightarrow A$} and \mbox{$g : X \rightarrow B$}.
Let \mbox{$x : X \rightarrow A \oplus B$} be the arrow
\mbox{$( f \ g )$}, i.e., the only arrow such that satisfies Equation~\ref{eq:prodbi}.
The arrow $k$ described just above is the arrow
\mbox{$[ {f}^{\star} \  {g}^{\star} ]$}.
We have seen that \mbox{$x =$} ${k}^{\star}$ and therefore
\mbox{${x}^{\star} =$} $k$.
\item
With the notations of Theorem~\ref{the:products}, part~\ref{prod=coprod}:
\mbox{$f \oplus g =$}
\mbox{$[ (a_{2} \circ f) \ (b_{2} \circ g) ]$}.
Therefore, by part~\ref{hor-ver} just above, we have:
\mbox{${(f \oplus g)}^{\star} =$}
\mbox{$\left({f}^{\star} \circ {a}^{\star}_{2} \ \  {g}^{\star} \circ b^{\star}_{2}  \right) $} 
which is the categorical product of ${f}^{\star}$ and ${g}^{\star}$.
We conclude that categorical products and coproducts of arrows coincide.
\item
We have seen that \mbox{${(f \oplus g)}^{\star}$} is the product of ${f}^{\star}$ and ${g}^{\star}$
and that products are coproducts. It is, therefore, their coproduct 
\mbox{${f}^{\star} \oplus {g}^{\star}$}.
\end{enumerate}
\end{proof}
We have just seen that the adjoint structure forces on us the identification of products and coproducts. This is an essential feature of QM. If one considers the coproduct type 
\mbox{$A \oplus B$},
i.e., the type of systems that are either of type $A$ or of type $B$ it is the product type
\mbox{$A \times B$}, i.e., the type of systems that have both an $A$ aspect and a $B$ aspect.
This does not, though, mean that our category possesses biproducts as defined, for example, 
in~\cite{Mitchell:Categories} or~\cite{MacLane:working}. We need some more assumptions
for that.

\subsection{Zero arrows and orthogonal arrows} \label{sec:zero}
We assume that our base category is an A-category. If \mbox{$a : A \rightarrow A \oplus B$}
and \mbox{$b : B \rightarrow A \oplus B$} form a coproduct, 
a first, fundamental, interaction between the adjoint structure and the coproduct structure reveals
itself when considering the arrow
\mbox{${b}^{\star} \circ a : A \rightarrow B$}. Intuitively, $a$ is the injection into the A-part
of \mbox{$A \oplus B$} and ${b}^{\star}$ is the projection onto its B-part. 
Since an A-object has no B-part, i.e., $a$ and $b$ are {\em orthogonal},
this arrow should represent this fact. 
We shall therefore assume that, for any objects $A$, $B$, 
there is a designated arrow \mbox{${0}_{A , B}: A \rightarrow B$}. 
This arrow will represent the transformation of an A-object that has no B-part into a B-object.
Any arrow such as 
\mbox{${b}^{\star} \circ a : A \rightarrow B$} coming from a proper coproduct should be equal to 
\mbox{${0}_{A , B}$}.

We shall now give a categorical characterization of the family of zero arrows.
This is done without any assumption on the base category.
\begin{definition} \label{def:zero}
A family of arrows \mbox{${0}_{A , B} : A \rightarrow B$}, for each pair of objects $A$, $B$, 
is said to be a {\em z-family} iff for any objects $A$, $A'$, $B'$ and $B$ and any arrows
\mbox{$f : A \rightarrow A'$} and \mbox{$g : B' \rightarrow B$} one has:
\mbox{$g \circ {0}_{A' , B'} \circ f =$} \mbox{${0}_{A , B}$}.
Any category that admits a z-family will be called a z-category.
An A-category that is a z-category is called an Az-category.
A T-category that is a z-category is a Tz-category.
\end{definition}
\begin{lemma} \label{le:z-family}
If a category admits a z-family this family is uniquely defined.
\end{lemma}
\begin{proof}
Suppose both \mbox{${x}_{A , B}$} and \mbox{${y}_{A , B}$} are z-families.
We have \mbox{${x}_{A , B} \circ y_{A , A} =$} \mbox{${x}_{A , B}$} because $x$ is a
z-family. But \mbox{${x}_{A , B} \circ y_{A , A} =$} \mbox{${y}_{A , B}$} because $y$ is a
z-family.
\end{proof}
Note that, contrary to what happens for abelian categories~\cite{Freyd:abelian}, there
is no zero object. The remarks introducing Definition~\ref{def:zero} explain 
the nature of zero arrows in QM better
than the definition involving a zero object that does not represent any physical system.
When no confusion can arise we shall drop the lower index and write $0$ for
${0}_{A , B}$.
We shall now prove that zero arrows behave as expected with respect to adjoints.
\begin{lemma} \label{le:zero-star}
In an Az-category, for any objects $A$, $B$,
\mbox{${({0}_{A , B})}^{\star} =$} \mbox{$0_{B , A}$}.
\end{lemma}
\begin{proof}
Note, first, that we have:
\[
{0}_{A , A} \ = \ {({0}_{A , B})}^{\star} \circ {0}_{A , B} \ = \  
{({({0}_{A , B})}^{\star} \circ {0}_{A , B})}^{\star} \ = \  
{({0}_{A , A})}^{\star}.
\]
Therefore:
\[
{({0}_{A , B})}^{\star} \ = \ 
{({0}_{A , B} \circ {0}_{A , A})}^{\star} \ = \ 
{({0}_{A , A})}^{\star} \circ {({0}_{A , B})}^{\star} \ = \ 
{0}_{A , A} \circ {({0}_{A , B})}^{\star} \ = \ 
{0}_{B , A}.
\]
\end{proof}
We shall now study the behavior of zero arrows with respect to the unit object and
the tensor structure of a $T$-category.
\begin{lemma} \label{le:zsA}
In an Az-category with unit object $I$,
for any object $A$, one has  \mbox{${({0}_{I , I})}_{A} =$} \mbox{${0}_{A , A}$}.
\end{lemma}
\begin{proof}
For any preparation \mbox{$a : I \rightarrow A$}, one has 
\mbox{$a \circ {0}_{I , I} =$} \mbox{${0}_{I , A} =$}
\mbox{${0}_{A , A} \circ a$}.
\end{proof}
\begin{lemma} \label{le:zero-tens}
In a Tz-category for any objects $A$, $B$, $C$, $D$ and any arrow 
\mbox{$f : A \rightarrow B$} one has:
\mbox{$f \otimes {0}_{C , D} =$} \mbox{${0}_{A \otimes C , B \otimes D}$} and
\mbox{${0}_{C , D} \otimes f =$}
\mbox{${0}_{C \otimes A , D \otimes B}$}.
\end{lemma}
\begin{proof}
For any \mbox{$a : I \rightarrow A$}, \mbox{$c : I \rightarrow C$}
we have: \mbox{$(f \otimes {0}_{C , D}) \circ \kappa_{A , C}(a , c) =$}
\mbox{$\kappa_{B , D}(f \circ a , {0}_{C , D} \circ c) =$}
\mbox{$\kappa_{B , D}(f \circ a , {0}_{I , D}) =$}
\mbox{$\kappa_{B , D}^{1}(f \circ a) \circ {0}_{I , D} =$}
\mbox{${0}_{I , B \otimes D}$}.
We note that 
\mbox{${0}_{A \otimes C , B \otimes D} \circ \kappa_{A , C}(a , c) =$}
\mbox{${0}_{I , B \otimes D}$} and therefore
\mbox{${0}_{A \otimes C , B \otimes D} \circ \kappa_{A , C} =$}
\mbox{$(f \otimes {0}_{C , D}) \circ \kappa_{A , C}$}
and we conclude that \mbox{$f \otimes {0}_{C , D} =$}
\mbox{${0}_{A \otimes C , B \otimes D}$}.
\end{proof}
We shall now define orthogonal arrows, they are the A-category version of the exact sequences
of abelian categories. 
\begin{definition} \label{def:ortho}
In an Az-category, 
\begin{enumerate}
\item \label{right}
arrows \mbox{$f : A \rightarrow C$} and \mbox{$g : B \rightarrow C$} with common co-domain 
are said to be {\em orthogonal} iff the two
equivalent conditions below hold:
\begin{itemize}
\item \mbox{${g}^{\star} \circ f =$} \mbox{${0}_{A , B}$},
\item \mbox{${f}^{\star} \circ g =$} \mbox{${0}_{B , A}$},
\end{itemize}
and
\item \label{left}
arrows \mbox{$f : C \rightarrow A$} and \mbox{$g : C \rightarrow B$} with common domain
are said to be {\em orthogonal} iff the two equivalent conditions below hold:
\begin{itemize}
\item \mbox{$g \circ {f}^{\star} =$} \mbox{${0}_{A , B}$},
\item \mbox{$f \circ {g}^{\star} =$} \mbox{${0}_{B , A}$},
\end{itemize}
\end{enumerate}
The equivalence of both conditions follows from Lemma~\ref{le:zero-star}.
Note that for parallel arrows \mbox{$f , g : A \rightarrow B$} both cases~\ref{right} and~\ref{left}
give the same answer, so that no confusion can arise.
In fact, we shall use case~\ref{right} only when $f$ and $g$ are right-unitary, i.e., subspace 
injections, and case~\ref{left} only when $f$ and $g$ are left-unitary, i.e., projections on subspaces, but the definition is more general.
\end{definition}
We can now show that, if an arrow $f$ is self-adjoint, $x$ is an eigenvector of $f$ 
(see Definition~\ref{def:eigen}) and $y$
is orthogonal to $x$, then the image of $y$ by $f$, \mbox{$f \circ y$} is also orthogonal to
$x$.
\begin{lemma} \label{le:spec2}
In an Az-category, if \mbox{$f : A \rightarrow A$} is self-adjoint, \mbox{$x : B \rightarrow A$}
is an eigenvector of $f$ and \mbox{$y : B \rightarrow A$} is orthogonal to $x$, then
\mbox{$f \circ y$} is also orthogonal to $x$.
\end{lemma}
\begin{proof}
Suppose \mbox{$s : B \rightarrow B$} and \mbox{$f \circ x =$} \mbox{$x \circ s$}.
We have:
\[
{x}^{\star} \circ (f \circ y) \ = \ {x}^{\star} \circ {f}^{\star} \circ y \ = \ 
{(f \circ x)}^{\star} \circ y \ = \ 
\]
\[
{(x \circ s)}^{\star} \circ y \ = \ 
{s}^{\star} \circ {x}^{\star} \circ y \ = \
{s}^{\star} \circ {0}_{B , B} \ = \ {0}_{B , B}
\]
\end{proof}
Examples of families of zero arrows are presented in Appendix~\ref{sec:examples5}.

\subsection{U-coproducts} \label{sec:ucoprod}
Consider two types of quantic systems, $A$ and $B$, corresponding, say, to classes of systems 
on which some observable takes on definite, different values. 
In QM, one assumes that there is a
type $A \oplus B$ that is the type of all systems that are either of type $A$ or of type $B$.
There are arrows \mbox{$u : A \rightarrow A \oplus B$} 
(resp.  \mbox{$v : B \rightarrow A \oplus B$})
that represents the transformation that transforms any system of type $A$ (resp. $B$) 
into a system of type $A \oplus B$. 
Quantic transformations of sort \mbox{$A \oplus B \rightarrow X$} 
should be in one-to-one correspondence with pairs of transformations of sort 
\mbox{$A \rightarrow X$} and \mbox{$B \rightarrow X$}, since acting on a system of type 
{\em $A$ or $B$} should be given by a recipe on how to act if the system is of type $A$ and
another recipe if it is of type $B$ and it is therefore reasonable to require $u$ and $v$ to
be a coproduct.

We think that it is reasonable to require the existence of products and coproducts of pairs of types
in QM, but that it is not reasonable to require the existence of an initial, terminal, or zero object.
The zero object of Hilbert spaces does {\em not} represent the type of any physical system.
In a sense we shall develop a theory of almost abelian categories 
that do not necessarily possess a zero object.
But the notion of a coproduct does not fully fit the requirements of QM, and this can be sensed
when one notices the following. 
First, coproducts are defined up to any isomorphism, but
we explained in Section~\ref{sec:unitary} that we are looking for categorical properties 
defined up to a unitary arrow. 
Secondly, the arrows \mbox{$i : A \rightarrow A \oplus B$} and
\mbox{$j : B \rightarrow A \oplus B$} that form a coproduct should preserve scalar products, i.e., 
by Theorem~\ref{the:unitary-scalar},
they should be right-unitary, which is not necessarily the case for coproducts.
Thirdly, the arrows $i$ and $j$ should be orthogonal in the sense of Definition~\ref{def:ortho}.
We shall therefore request the existence of coproducts that satisfy those properties, i.e.,
that are compatible with the adjoint structure.
We shall call such coproducts u-coproducts and they shall be defined in 
Definition~\ref{def:u-coproducts}.
We shall then show that u-coproducts are biproducts.

We require that the coproduct arrows
\mbox{$u : A \rightarrow A \oplus B$} and  \mbox{$v : B \rightarrow A \oplus B$} be
{\em right-unitary}.
This is a most reasonable requirement: 
the transformations $u$ and $v$ represent, in a sense, only a change of point of view: 
viewing a system as being of type $A$ or of type {\em $A$ or $B$} and such a change of
point of view should not alter the relations between different systems of type $A$. 
In particular the scalar product of preparations of type $A$ should be equal 
to the scalar product of their images by $u$ (resp.$v$).
Then, we shall require that the arrows $u$ and $v$ be orthogonal, 
as in Definition~\ref{def:ortho}:
\mbox{${v}^{\star} \circ u =$} \mbox{${0}_{A , B}$} and
\mbox{${u}^{\star} \circ v =$} \mbox{${0}_{B , A}$}.
\begin{definition} \label{def:u-coproducts}
Let $A$, $B$ and $C$ be objects in an Az-category.
The pair of arrows \mbox{$u : A \rightarrow C$}, \mbox{$v : B \rightarrow C$} is said to be
a {\em u-coproduct} for $A$ and $B$ iff:
\begin{enumerate}
\item \label{coproduct}
they are a coproduct for $A$ and $B$,
\item \label{right-inj}
the arrows $u$ and $v$ are right-unitary and orthogonal (see Definition~\ref{def:ortho}).
\end{enumerate}
The definition of a u-product is dual: a product \mbox{$p : C \rightarrow A$},
\mbox{$q : C \rightarrow B$} such that $p$ and $q$ are left-unitary and orthogonal.
\end{definition}
The notion of a u-product is dual: a product and such the 
unique arrow corresponding to two left-unitary orthogonal arrows is left-unitary.
We shall now prove that u-coproducts are defined up a unitary arrow.
\begin{theorem} \label{the:biprod-unique}
In an Az-category, if \mbox{$u : A \rightarrow A \oplus B$},
\mbox{$v : B \rightarrow A \oplus B$} provide a u-coproduct, then 
\mbox{$u' : A \rightarrow X$},
\mbox{$v' : B \rightarrow X$} provide a u-coproduct iff there exists a unitary arrow
\mbox{$w : A \oplus B \rightarrow X$} such that \mbox{$u' = $} \mbox{$w \circ u$} and 
\mbox{$v' =$} \mbox{$w \circ v$}.
\end{theorem}
\begin{proof}
Suppose $w$ is such a unitary arrow. The arrows $u'$ and $v'$ provide a coproduct because
$w$ is an isomorphism. They are right-unitary because $w$ is right-unitary and the composition
of right-unitary arrows is right-unitary. 
It is easy to see they are orthogonal.

Assume, now, that $u'$ and $v'$ provide a u-coproduct.
There is a unique arrow \mbox{$w : A \oplus B \rightarrow X$} such that
\mbox{$w \circ u =$} \mbox{$u'$} and \mbox{$w \circ v =$} \mbox{$v'$}.
In our notation \mbox{$w =$} \mbox{$[ u' \ v' ]$}.
By Theorem~\ref{the:products}, we have:
\mbox{${w}^{\star} =$} \mbox{$\left( {u'}^{\star} \ \ {v'}^{\star} \right)$}.
Therefore \mbox{${w}^{\star} \circ w =$}
\mbox{$[ ({u'}^{\star} \circ u \ \ {v'}^{\star} \circ u ) \ \ \ \  
( {u'}^{\star} \circ v \ \ {v'}^{\star} \circ v ) ] =$} 
\mbox{$[ ( {id}_{A} \ \ {0}_{A , B} ) \ \ \ \  ( {0}_{B , A} \ \ {id}_{B} ) ]$}, 
i.e., the only arrow 
\mbox{$x : A \oplus B \rightarrow A \oplus B$} such that
\mbox{${u'}^{\star} \circ x \circ u =$} \mbox{${id}_{A}$},
\mbox{${v'}^{\star} \circ x \circ u =$} \mbox{${0}_{A , B}$},
\mbox{${u'}^{\star} \circ x \circ v =$} \mbox{${0}_{B , A}$} and
\mbox{${v'}^{\star} \circ x \circ v =$} \mbox{${id}_{B}$}.
We conclude that  \mbox{${w}^{\star} \circ w =$} \mbox{${id}_{A \oplus B}$}, i.e., $w$ is right-unitary.
But, then,
\[
w \circ {w}^{\star} \circ u' \ = \ w \circ {w}^{\star} \circ w \circ u \ = \ 
w \circ u \ = \ u'
\]
and
\[
w \circ {w}^{\star} \circ v' \ = \ w \circ {w}^{\star} \circ w \circ v \ = \ 
w \circ v \ = \ v'.
\]
We conclude that \mbox{$w \circ {w}^{\star} =$} \mbox{${id}_{A \oplus B}$}, i.e.,
$w$ is left-unitary and therefore unitary.
\end{proof}
The notion of a biproduct is defined in~\cite{Mitchell:Categories} 
or in~\cite{MacLane:working}.
We shall now show that, in an Az-category, coproducts, or products, are biproducts..
A proof parallel to ours in the monoidal framework can be found in~\cite{Houston:2008}. 
Note, though, that the zero arrows are defined here without the need to consider a zero object.
 \begin{theorem} \label{the:biproduct}
 In an Az-category, arrows 
 \mbox{$u : A \rightarrow A \oplus B$} and \mbox{$v : B \rightarrow A \oplus B$} 
 form a u-coproduct iff the arrows \mbox{${u}^{\star}$} and \mbox{${v}^{\star}$} form a  
 u-product. 
 For any arrows \mbox{$a : A \rightarrow A$}, \mbox{$b : B \rightarrow B$},
 \mbox{$f : A \rightarrow B$} and \mbox{$g : B \rightarrow A$}, there is a unique arrow
 \mbox{$x : A \oplus B \rightarrow A \oplus B$}, denoted
 \mbox{$\left( \begin{array}{c} a \  \ f \\ 
g \  \ b \end{array} \right)$} such that \mbox{${u}^{\star} \circ x \circ u =$} 
$a$, \mbox{${v}^{\star} \circ x \circ u =$} \mbox{$f$},
\mbox{${u}^{\star} \circ x \circ v =$} \mbox{$g$} and
\mbox{${v}^{\star} \circ x \circ v =$} $b$.
We have \mbox{$x =$} \mbox{$( [a \ g ] \ \ [ f \ b ] ) =$} \mbox{$[ ( a \ f ) \ \ ( g \ b ) ]$}.
For \mbox{$a =$} \mbox{${id}_{A}$}, \mbox{$b =$} \mbox{${id}_{B}$},
\mbox{$f =$} \mbox{${0}_{A , B}$} and \mbox{$g =$} \mbox{${0}_{B , A}$}
the arrow $x$ is the identity
\mbox{${id}_{A \oplus B}$}. Any product (resp. coproduct) is a biproduct.
 \end{theorem}
 \begin{proof}
 By Theorem~\ref{the:products} coproducts correspond to products. The arrows
 $u$ and $v$ are left-unitary iff their adjoints are right-unitary. We have proved our first claim.
 For any $x$ satisfying the four equations, we have \mbox{${u}^{\star} \circ x =$}
 \mbox{$[ a \ g ]$}, \mbox{${v}^{\star} \circ x =$} \mbox{$[ f \ b ]$},
 \mbox{$x \circ u =$} \mbox{$( a \ f )$} and \mbox{$x \circ v =$} \mbox{$( g \ b )$}.
 Therefore \mbox{$x =$} \mbox{$( [a \ g ] \ \ [ f \ b ] ) =$} \mbox{$[ ( a \ f ) \ \ ( g \ b ) ]$}.
 One now sees easily that any one of those last two expressions is a suitable $x$.
 If one takes \mbox{$a =$} \mbox{${id}_{A}$}, \mbox{$b =$} \mbox{${id}_{B}$},
\mbox{$f =$} \mbox{${0}_{A , B}$} and \mbox{$g =$} \mbox{${0}_{B , A}$},
one sees that \mbox{${id}_{A \oplus B}$} satisfies the four equations.equations,
Therefore \mbox{$x =$} \mbox{${id}_{A \oplus B}$}.
Since \mbox{$\left( \begin{array}{c} {id}_{A} \  \ {0}_{A , B} \\ 
{0}_{B , A} \  \ {id}_{B} \end{array} \right)$} is an isomorphism, we have a biproduct as defined 
in~\cite{Mitchell:Categories} or~\cite{MacLane:working}.
 \end{proof}
\begin{lemma} \label{le:biprod-arrows}
In an Az-category, if  
\mbox{$u_{i} : A_{i} \rightarrow A_{i} \oplus B_{i}$} and 
\mbox{$v_{i} : B_{i} \rightarrow A_{i} \oplus B_{i}$} are u-coproducts for \mbox{$i = 1 , 2$},
then, for any \mbox{$f : A_{1} \rightarrow A_{2}$} and \mbox{$g : B_{1} \rightarrow B_{2}$},
one has
\[
(f \oplus g) \circ u_{1} \ = \ ( f \ {0}_{A_{1} , B_{2}} ) \ , \ 
(f \oplus g) \circ v_{1} \ = \ ({0}_{B{1} , A_{2}} \ g ) \ , \ 
\]
\[
u_{2}^{\star} \circ (f \oplus g) \ = \ [ f \ {0}_{B{1} , A_{2}} ] \ , \ 
v_{2}^{\star} \circ (f \oplus g) \ = \ [ {0}_{A_{1} , B_{2}} \ g ].
\]
\end{lemma}
\begin{proof}
We shall prove the first equality.
By Theorem~\ref{the:products}, item~\ref{prod=coprod}, we have:
\mbox{${u}_{2}^{\star} \circ (f \oplus g) =$} \mbox{$f \circ {u}_{1}^{\star}$} and
\mbox{${v}_{2}^{\star} \circ (f \oplus g) =$} \mbox{$g \circ {v}_{1}^{\star}$}.
Therefore
\mbox{${u}_{2}^{\star} \circ (f \oplus g) \circ {u}_{1} =$} 
\mbox{$f \circ {u}_{1}^{\star} \circ {u}_{1} =$} $f$ and
\mbox{${v}_{2}^{\star} \circ (f \oplus g) \circ {u}_{1} =$} 
\mbox{$g \circ {v}_{1}^{\star} \circ {u}_{1} =$} \mbox{${0}_{A_{1} , B_{2}}$}.
This proves our claim.
\end{proof}

\subsection{B-categories} \label{sec:B-cat}
We shall require a u-coproduct for every pair of objects.
In the sequel we shall always assume our base category is a B-category.
Some of the results below hold without this assumption, if one assumes the existence
of the u-coproducts and arrows that appear explicitly in the claim.
\begin{definition} \label{def:B-cat}
An Az-category is said to be a {\em B-category} iff
every pair of objects admits a u-coproduct.
\begin{comment}
\item \label{negation}
If \mbox{$u : A \rightarrow X$} is right-unitary, then, either
\begin{enumerate}
\item  \label{T+-unitary-case}
$u$ is unitary, or
\item \label{T+-decomp}
 there is an object $B$ and a right-unitary
arrow \mbox{$v : B \rightarrow X$} such that $u$ and $v$ provide a u-coproduct.
\end{enumerate}
\end{comment}
\end{definition}
In a B-category, we shall use the following notation.

\begin{notation}
For any objects $A$, $B$ the arrows \mbox{${u}_{A , B} :$} \mbox{$A \rightarrow A \oplus B$} and
\mbox{${v}_{A , B} :$} \mbox{$B \rightarrow A \oplus B$} will denote a u-coproduct.
\end{notation}

In Appendix~\ref{sec:examples6}, examples of B-categories are discussed.

\subsection{Trivial categories} \label{sec:trivial}
We shall now answer the question raised in Section~\ref{sec:unit-def}: could it be that
all preparations are normalized? We shall show that there is essentially only one 
category with a unit object that is a B-category in which all preparations are normalized.
\begin{definition} \label{def:trivial}
A category \cC\ is said to be trivial iff, for any objects $A$, $B$ the set $hom(A , B)$ is a
singleton.
\end{definition}
One easily sees that any trivial category is, in a unique way, a B-category, admits a
unit object and that every preparation is normalized. We shall show the converse.
We begin by a lemma.
\begin{lemma} \label{le:trivial}
Any B-category with a unit object $I$, in which \mbox{${id}_{I} =$} \mbox{${0}_{I , I}$} is trivial.
\end{lemma}
\begin{proof}
By Lemmas~\ref{le:global-cat-adj} and~\ref{le:zsA}, the fact that \mbox{${id}_{I} =$} 
\mbox{${0}_{I , I}$}
implies that, for any object $A$, one has \mbox{${id}_{A} =$} \mbox{${0}_{A , A}$}.
Any arrow \mbox{$f : A \rightarrow B$} satisfies
\mbox{$f =$} \mbox{$f \circ {id}_{A} =$} \mbox{$f \circ {0}_{A , A} =$} \mbox{${0}_{A , B}$}.
\end{proof}
\begin{theorem} \label{the:norm-trivial}
Any B-category with a unit object in which every scalar is normalized is trivial.
If \mbox{$a : I \rightarrow A$} is a normalized preparation, for any scalar $s$, the preparation
\mbox{$a \circ s$} is normalized iff $s$ is normalized.
\end{theorem}
\begin{proof}
Assume that the arrow \mbox{${0}_{I , I}$} is normalized.
We have \mbox{${({0}_{I , I})}^{\star} \circ {0}_{I , I}= $}
\mbox{${id}_{I}$}. But 
\mbox{${({0}_{I , I})}^{\star} \circ {0}_{I , I}= $}
\mbox{${0}_{I , I}$}.
We conclude by Lemma~\ref{le:trivial} that the category is trivial.
If $a$ and $s$ are normalized, the preparation \mbox{$a \circ s$} is easily seen to be normalized.
If $a$ and \mbox{$a \circ s$} are normalized, then \mbox{${id}_{I} =$}
\mbox{${s}^{\star} \circ {a}^{\star} \circ a \circ s =$} \mbox{${s}^{\star} \circ s$} and therefore
$s$ is normalized.
\end{proof}

\subsection{Additive structure on Hom sets} \label{sec:additive-Hom}
We shall, now, show that B-categories are semi-additive
in the sense of~\cite{Mitchell:Categories}. 
Our treatment is close to Freyd's~\cite{Freyd:abelian}, p.45 and following.
For any two parallel arrows \mbox{$f , g : A \rightarrow B$} we shall define their
{\em superposition}, \mbox{$ f + g : A \rightarrow B$}, that corresponds to the quantic operation
that performs both $f$ and $g$, so to speak in parallel.
First, we shall define a diagonal and a codiagonal for every object.
\begin{definition} \label{def:diagonal}
Let $A$ be an object of a B-category. Let \mbox{$u : A \rightarrow A \oplus A$} and
\mbox{$v : A \rightarrow A \oplus A$} be a coproduct.
The diagonal \mbox{$\Delta_{A}$} of $A$ is the arrow \mbox{$A \rightarrow A \oplus A$} defined
by: \mbox{$\Delta_{A} =$} \mbox{$\left({id}_{A} \  {id}_{A}\right)$}, i.e.,
the unique arrow such that 
\mbox{${u}^{\star} \circ \Delta_{A} =$} \mbox{${id}_{A}$} and
\mbox{${v}^{\star} \circ \Delta_{A} =$} \mbox{${id}_{A}$}.
The codiagonal \mbox{$\nabla_{A}$} of $A$ is the arrow \mbox{$A \oplus A \rightarrow A$} 
defined by: \mbox{$\nabla_{A} =$} \mbox{$[  {id}_{A}  \ {id}_{A} ]$}, i.e.,
the unique arrow such that 
\mbox{$\nabla_{A} \circ u =$} \mbox{${id}_{A}$} and
\mbox{$\nabla_{A} \circ v =$} \mbox{${id}_{A}$}.
\end{definition}
One checks that $\Delta$ is well-defined because $A \oplus B$ is a product and $\nabla$ is
well-defined because it is a coproduct. 
Let us reflect now on the meaning of diagonals and codiagonals. The diagonal 
\mbox{$\Delta_{A}$} is the quantic transformation that transforms any system $a$ of type $A$
into the system of type \mbox{$A \oplus A$} that is {\em either $a$ or $a$}, or, equivalently into
the system that has both an $A$ aspect and an $A$ aspect and such that both aspects are $a$.
The following lemma shows that the codiagonal $\nabla_{A}$ is the adjoint of the diagonal
$\Delta_{A}$.
\begin{lemma} \label{le:delta}
Let $A$ be an object of a B-category. One has \mbox{${\Delta}^{\star}_{A} =$}
\mbox{$\nabla_{A}$} and \mbox{${\nabla}^{\star}_{A} =$} \mbox{$\Delta_{A}$}. 
\end{lemma}
\begin{proof}
By Theorem~\ref{the:products} and Lemma~\ref{le:id}.
\end{proof}
\begin{lemma} \label{le:fgA1}
In a B-category, let \mbox{$f , g : A \rightarrow B$}. Then
\mbox{$(f \oplus g) \circ \Delta_{A} =$} \mbox{$( f \ g )$} and
\mbox{$\nabla_{B} \circ (f \oplus g) =$} \mbox{$[f \ g ]$}.
\end{lemma}
\begin{proof}
By Theorem~\ref{the:products}, item~\ref{prod=coprod}, we have:
\[
{u_{B}^{\star} \circ (f \oplus g) \circ \Delta_{A} \ = \ f \circ {u}_{A}^{\star} \circ ({id}_A} \ {id}_{A} ) \ = \ 
f \circ {id}_{A} \ = \ f
\]
and
\[
{v_{B}^{\star} \circ (f \oplus g) \circ \Delta_{A} \ = \ g \circ {v}_{A}^{\star} \circ ({id}_A} \ {id}_{A} ) \ = \ 
g \circ {id}_{A} \ = \ g.
\]
This proves our first claim. For the second one, note that:
\[
\nabla_{B} \circ (f \oplus g) \circ {u}_{A} \ = \ [{id}_{B} \ {id}_{B} ] \circ {u}_{B} \circ f \ = \ 
{id}_{B} \circ f \ = \ f
\]
and
\[
\nabla_{B} \circ (f \oplus g) \circ {v}_{A} \ = \ [{id}_{B} \ {id}_{B} ] \circ {v}_{B} \circ g \ = \ 
{id}_{B} \circ g \ = \ g.
\]
\end{proof}
\begin{lemma} \label{le:delta+}
Let \mbox{$f : A \rightarrow B$} be an arrow in a B-category.
We have:
\begin{equation} \label{eq:delta}
 (f \oplus f) \circ \Delta_{A} \ = \  \Delta_{B} \circ f
 \end{equation}
 and similarly:
 \begin{equation} \label{eq:nabla}
\nabla_{B} \circ (f \oplus f) \ = \  f \circ \nabla_{A}.
 \end{equation}
\end{lemma}
\begin{proof}
Let \mbox{$u_{A} , v_{A} : A \rightarrow A \oplus A$} and 
\mbox{$u_{B} , v_{B} : B \rightarrow B \oplus B$} be u-coproducts.
By Definition~\ref{def:diagonal}, we have
\[
{u}_{B}^{\star} \circ \Delta_{B} \ = \ f \ = \ {v}_{B}^{\star} \circ \Delta_{B}.
\]
By Theorem~\ref{the:products}, we have
\[
{u}_{B}^{\star} \circ (f \oplus f) \ = \ f \circ {u}_{A}^{\star} \ {\rm and} \ 
v_{B}^{\star} \circ (f \oplus f) \ = \ f \circ {v}_{A}^{\star}.
\]
We see that 
\[
{u}_{B}^{\star} \circ (f \oplus f) \circ \Delta_{A} \ = \ 
f \circ {u}_{A}^{\star} \circ \Delta_{A} \ = \  f \ = \ 
{u}_{B}^{\star} \circ \Delta_{B} \circ f
\]
and similarly
\[
{v}_{B}^{\star} \circ (f \oplus f) \circ \Delta_{A} \ = \ 
f \circ {v}_{A}^{\star} \circ \Delta_{A} \ = \ f \ = \ 
{v}_{B}^{\star} \circ \Delta_{B} \circ f.
\]
But there is a unique arrow \mbox{$x =$} \mbox{$(f \ f) :$}
\mbox{$A \rightarrow B \oplus B$} such that
\mbox{${u}_{B}^{\star} \circ x =$} $f$ and
\mbox{${v}_{B}^{\star} \circ x =$} $f$,
We conclude that Equation~\ref{eq:delta} holds.
For Equation~\ref{eq:nabla}, use Equation~\ref{eq:delta} with ${f}^{\star}$, 
Theorem~\ref{the:products} and Lemma~\ref{le:delta}.
\end{proof}
We now define a binary operation \mbox{${+}_{A , B}$} on any hom-set Hom($A$ , $B$).
\begin{definition} \label{def:addition}
Let $A$, $B$ be objects in a B-category.
If \mbox{$f , g : A \rightarrow B$} then we define the arrow
\mbox{$f + g : A \rightarrow B$} by
\mbox{$f + g =$} \mbox{$\nabla_{B} \circ (f \oplus g) \circ \Delta_{A}$}.
\end{definition}
Note that the definition of addition of arrows in Definition~\ref{def:addition} requires both a product 
(for $\Delta$) and a coproduct (for $\nabla$) and that the properties of addition depend on their 
coincidence, i.e., on the existence of biproducts, but is not at all connected to the tensor structure.
\begin{lemma} \label{le:fgA2}
In a B-category, let \mbox{$f , g : A \rightarrow B$}. Then
\mbox{$f + g =$} \mbox{$[ f \ g ] \circ \Delta_{A} =$} \mbox{$\nabla_{B} \circ ( f \ g )$}.
\end{lemma}
\begin{proof}
Obvious from Definition~\ref{def:addition} and Lemma~\ref{le:fgA1}.
\end{proof}
\begin{lemma} \label{le:zero+}
In a B-category, for any \mbox{$f : A \rightarrow B$},
\mbox{$f + {0}_{A , B} =$} \mbox{${0}_{A , B} + f =$} $f$.
\end{lemma}
\begin{proof}
Let \mbox{$u : A \rightarrow A \oplus A$} and \mbox{$v : A \rightarrow A \oplus A$} be 
a u-coproduct.
The arrow \mbox{$[ f \ {0}_{A , B} ]$} is the only arrow 
\mbox{$x : A \oplus A \rightarrow B$} such that
\mbox{$x \circ u =$} $f$ and \mbox{$x \circ v =$} \mbox{${0}_{A , B}$}.
Therefore \mbox{$[ f \ {0}_{A , B} ] =$} \mbox{$f \circ {u}^{\star}$}.
By Lemma~\ref{le:fgA2},
\[
f + {0}_{A , B} \ = \ [ f \ {0}_{A , B} ] \circ \Delta_{A} \ = \
f \circ {u}^{\star} \circ \Delta_{A} \ = \ f \circ {id}_{A} \ = \ f.
\]
The second equality is proved similarly.
\end{proof}
\begin{theorem} \label{the:ab-monoid}
The operation ${+}_{A , B}$ makes an abelian monoid out of Hom($A$, $B$):
addition on Hom(A , B) is associative and commutative, and \mbox{${0}_{A , B}$}
is a neutral element. 
\end{theorem}
\begin{proof}
Let \mbox{$w , x , y , z : A \rightarrow B$}. 
By Theorem~\ref{the:biproduct} and Lemma~\ref{le:fgA2}
\[
\nabla_{B} \circ \left( \begin{array}{c} w \  \ x \\ 
y \  \ z \end{array} \right) \circ \Delta_{A} \ = \ 
\nabla_{B} \circ ( [ w \ y ] \ \ [ x \ z ] ) \circ \Delta_{A} \ = \ 
\]
\[
\nabla_{B} \circ ( [ w \ y ] \circ \Delta_{A} \ \ [ x \ z ] \circ \Delta_{A} ) \ = \ 
\nabla_{B} \circ ( w + y \ \ x + z ) \ = \ 
(w + y) + (x + z)
\]
But we also have:
\[
\nabla_{B} \circ \left( \begin{array}{c} w \  \ x \\ 
y \  \ z \end{array} \right) \circ \Delta_{A} \ = \ 
\nabla_{B} \circ [ ( w \ x ) \ \ ( y \ z ) ] \circ \Delta_{A} \ = \ 
\]
\[
[ \nabla_{B} \circ ( w \ x ) \ \ \nabla_{B} \circ ( y \ z ) ] \circ \Delta_{A} \ = \ 
[ w + x \ \ y + z ] \circ \Delta_{A} \ = \ 
(w + x) + (y + z)
\]
We have shown that \mbox{$(w + y) + (x + z) =$} \mbox{$(w + x) + (y + z)$}.
That \mbox{${0}_{A , B}$} is a neutral element is Lemma~\ref{le:zero+}.
Using this and taking \mbox{$x =$} \mbox{${0}_{A , B}$} we obtain
\mbox{$(w + y) + z =$} \mbox{$w + (y + z)$}, which is associativity.
By taking \mbox{$w =$} \mbox{$z =$} \mbox{${0}_{A , B}$} we obtain
\mbox{$y + x =$} \mbox{$x + y$}, i.e., commutativity.
\end{proof}
\begin{lemma} \label{le:2}
In a B-category, for any object $A$, one has 
\mbox{$\nabla_{A} \circ \Delta_{A} = {id}_{A} + {id}_{A}$}. 
\end{lemma}
\begin{proof}
\[
{id}_{A} + {id}_{A} \ = \  \nabla_{A} \circ ({id}_{A} \oplus {id}_{A}) \circ \Delta_{A} \ = \ 
\nabla_{A} \circ {id}_{A \oplus A} \circ \Delta_{A} \ = \ 
\nabla_{A} \circ \Delta_{A}.
\]
\end{proof}
Addition behaves as expected with respect to composition: composition distributes over
addition, making B-categories
semi-abelian, see~\cite{Mitchell:Categories}.
\begin{lemma} \label{le:+comp}
Let \mbox{$h : A \rightarrow B$}, \mbox{$k : C \rightarrow D$} and
\mbox{$f , g : B \rightarrow C$} be arrows in a B-category.
One has:
\[
(f + g) \circ h \ = \  (f \circ h) + (g \circ h) \ \ \   ,  \ \ \ k \circ (f + g) \ = \  (k \circ f) + (k \circ g). 
\]
\end{lemma}
\begin{proof}
By Lemma~\ref{le:delta+}, one has:
\[
(f + g) \circ h \ = \nabla_{C} \circ (f \oplus g) \circ \Delta_{B} \circ h \ = \ 
\nabla_{C} \circ (f \oplus g) \circ (h \oplus h) \circ \Delta_{A} \ = \ 
\]
\[
\nabla_{C} \circ ((f \circ h) \oplus (g \circ h)) \circ \Delta_{A} \ = \ 
(f \circ h) + (g \circ h). 
\]
\end{proof}
Addition also behaves as expected with respect to the adjoint structure.
\begin{lemma} \label{le:+adj}
Let \mbox{$f , g : A \rightarrow B$} be arrows in a B-category. We have
\mbox{${(f + g)}^{\star} =$} \mbox{${f}^{\star} + {g}^{\star}$}.
\end{lemma}
\begin{proof}
By Lemma~\ref{le:delta} and Theorem~\ref{the:products}, item~\ref{star-prod}.
\end{proof}

\subsection{A characterization of u-coproducts} \label{sec:character}
\begin{theorem} \label{the:character}
Assume a B-category. Let \mbox{$u : A \rightarrow X$} and 
\mbox{$v : B \rightarrow X$} be right-unitary orthogonal arrows. 
Then, they are a u-coproduct iff \mbox{$u \circ {u}^{\star} + v \circ {v}^{\star} =$}
\mbox{${id}_{X}$}.
\end{theorem}
\begin{proof}
Assume $u$ and $v$ are a u-coproduct.
There is a unique arrow \mbox{$x : X \rightarrow X$} such that
\mbox{$x \circ u =$} $u$ and \mbox{$x \circ v =$} $v$.
But ${id}_{X}$ is one such arrow and \mbox{$u \circ {u}^{\star} + v \circ {v}^{\star}$} is another one
since, by Lemmas~\ref{le:+comp} and~\ref{le:zero+}: we have:
\[
(u \circ {u}^{\star} \, + \, v \circ {v}^{\star} ) \circ u \ = \  
u \circ {u}^{\star} \circ u \, + \, v \circ {v}^{\star} \circ u \ = \ 
u + v \circ {0}_{A , B} \ = \ 
u + {0}_{A , X} \ = \ 
u
\]
and similarly \mbox{$(u \circ {u}^{\star} \, + \, v \circ {v}^{\star} ) \circ v =$} $v$.

Suppose now that  \mbox{$u \circ {u}^{\star} + v \circ {v}^{\star} =$}
\mbox{${id}_{X}$} and let \mbox{$f : A \rightarrow Y$} and \mbox{$g : B \rightarrow Y$}.
If \mbox{$y : X \rightarrow Y$} is such that \mbox{$y \circ u =$} $f$ and
\mbox{$y \circ v =$} $g$ then we have:
\[
y \ = \ y \circ (u \circ {u}^{\star} + v \circ {v}^{\star}) \ = \ 
y \circ u \circ {u}^{\star} + y \circ v \circ {v}^{\star} \ = \ 
f \circ {u}^{\star} + g \circ {v}^{\star}.
\]
We have proved the uniqueness of such an arrow $y$.
We are left to prove that \mbox{$f \circ {u}^{\star} + g \circ {v}^{\star}$} is a suitable $y$.
But 
\[
(f \circ {u}^{\star} + g \circ {v}^{\star}) \circ u \ = \ 
f \circ {u}^{\star} \circ u + g \circ {v}^{\star} \circ u \ = \ 
f + {0}_{A , Y} \ = \ f
\]
and similarly
 \mbox{$(f \circ {u}^{\star} + g \circ {v}^{\star}) \circ v =$} $g$.
\end{proof}
As mentioned in Section~\ref{sec:ucoprod}, a u-coproduct (i.e., a biproduct)
 \mbox{$u : A \rightarrow A \oplus B$}, \mbox{$v : B \rightarrow A \oplus B$}
 presents \mbox{$A \oplus B$} as the type of systems for which an observable may have one
 of two definite values. Any preparation $x$ of a system of type \mbox{$A \oplus B$} can therefore
 be seen as a superposition of two orthogonal preparations, one of type $A$ and one of type $B$.
 The preparation $x$ is partly  $y$ of type $A$ and partly $z$ of type $B$, and the scalars 
 \mbox{$sqnorm(y)$} and \mbox{$sqnorm(z)$} determine the ratio of the $A$ aspect to the $B$
 aspect.
\begin{corollary} [Born's rule] \label{le:Born}
Let \mbox{$u : A \rightarrow A \oplus B$} and \mbox{$v : B \rightarrow A \oplus B$} be
a u-coproduct.
For any preparation \mbox{$x : I \rightarrow A \oplus B$}, one has:
\begin{equation} \label{eq:Born}
x \ = \  u \circ ( {u}^{\star} \circ x ) \ + \ 
v \circ ( {v}^{\star} \circ x). 
\end{equation}
Letting \mbox{$y =$} \mbox{$u \circ {u}^{\star} \circ x$}
and \mbox{$z =$} \mbox{$v \circ {v}^{\star} \circ x$} one notes 
that $y$ and $z$ are orthogonal preparations such that \mbox{$x =$} \mbox{$y + z$}.
Conversely if \mbox{$w : I \rightarrow A$} and \mbox{$w' : I \rightarrow B$} and
\mbox{$x =$} \mbox{$u \circ w \: + \: v \circ w'$} then
\mbox{$w =$} \mbox{${u}^{\star} \circ x$} and \mbox{$w' =$} \mbox{${v}^{\star} \circ x$}.
Moreover one has:
\[
sqnorm(y) \ = \ \langle y \mid x \rangle \ = \ \langle x \mid y \rangle, \ \  
sqnorm(z) \ = \ \langle z \mid x \rangle \ = \ \langle x \mid z \rangle
\]
and
\[
sqnorm(y) + sqnorm(z) \ = \ sqnorm(x)
\]
which our version of Born's rule.
\end{corollary}
\begin{proof}
By Theorem~\ref{the:character} and Lemma~\ref{le:+comp}
\[
x \ = \ ( u \circ {u}^{\star} \: + \: v \circ {v}^{\star} ) \circ x \ = \ u \circ {u}^{\star} \circ x \: + \:
v \circ {v}^{\star} \circ x.
\]
The first term is equal to $u$ up to a scalar and the second one is $v$ up to a scalar and
therefore they are orthogonal.
If \mbox{$x =$} \mbox{$u \circ w \: + \: v \circ w'$}, by Lemma~\ref{le:+comp} one has:
\mbox{${u}^{\star} \circ x =$} \mbox{$w + {0}_{I , A} -$} $w$ and similarly for $w'$.
We have
\[
sqnorm(y) \ = \ {y}^{\star} \circ y \ = \ {x}^{\star} \circ u \circ {u}^{\star} \circ u \circ {u}^{\star} \circ x
\ = \  {x}^{\star} \circ u \circ {u}^{\star} \circ x \ = \langle y \mid x \rangle \ = \langle x \mid y \rangle
\]
and similarly for \mbox{$sqnorm(z)$}.
Therefore \mbox{$sqnorm(y) + sqnorm(z) =$}
\mbox{$\langle y + z \mid x \rangle =$} \mbox{$sqnorm(x)$}.
\end{proof}

\subsection{Orthonormal bases}  \label{sec:bases}
The notion of an orthonormal basis is an important part of all traditional expositions of QM.
A basis for an object $A$ is a set of orthogonal normalized preparations that are such that
they represent all the possible values of an observable: any preparation of $A$ is a superposition
of the basis preparations.
\begin{definition} \label{def:basis}
In a B-category with unit object, a basis for an object $A$ is a set \mbox{$\{ a_{i} \mid i \in J \}$} 
of {\em normalized, pairwise orthogonal} preparations  \mbox{$a_{i} : I \rightarrow A$} such that
for any \mbox{$b : I \rightarrow A$}, if $b$ is orthogonal to every \mbox{$a_{i} , i \in J$}, then
\mbox{$b =$} \mbox{${0}_{I , A}$}. 
\end{definition}
The direct image of a basis under a unitary arrow is a basis.
\begin{theorem} \label{the:base-unitay}
In a B-category with a unit object, if \mbox{$u : A \rightarrow B$} is unitary and
\mbox{$\{a_{j} \mid j \in J \}$} is a basis for $A$, then 
\mbox{$\{u \circ a_{j} \mid j \in J \}$} is a basis for $B$.
\end{theorem}
\begin{proof}
By Theorem~\ref{the:unitary-scalar}, the preparations \mbox{$u \circ a_{j}$} are normalized
and pairwise orthogonal.
If \mbox{$b : I \rightarrow B$} is orthogonal to \mbox{$u \circ a_{j}$}, since ${u}^{\star}$ is
unitary, the preparation \mbox{${u}^{\star} \circ b$} is orthogonal to 
\mbox{${u}^{\star} \circ u \circ a_{j} =$} \mbox{$a_{j}$}.
If $b$ is orthogonal to all $a_{j}$s, then ${u}^{\star} \circ b$ is orthogonal to all elements of 
the basis and therefore null. Then \mbox{$u \circ {u}^{\star} \circ b =$} $b$ is null.
\end{proof}
The union of bases for $A$ and for $B$ form a basis for $A \oplus B$>
\begin{theorem} \label{the:bases-union}
In a B-category with unit object, suppose
\mbox{$u : A \rightarrow A \oplus B$} and \mbox{$v : B \rightarrow A \oplus B$} form a
u-coproduct.
Let \mbox{$\{ {a}_{i} \mid i \in J \} $} and \mbox{$\{ {b}_{i} \mid i \in K \} $} 
be orthogonal bases for $A$ and $B$ respectively. 
Then the set \mbox{$ \{ u \circ {a}_{i} \mid i \in J \} \cup$}
\mbox{$\{ v \circ {b}_{i} \mid i \in K \} $} 
is an orthogonal basis for \mbox{$A \oplus B$}.
\end{theorem}
\begin{proof}
By Theorem~\ref{the:unitary-scalar} the image by $u$ (or $v$) of a normalized preparation is 
normalized and the image of two orthogonal preparations is orthogonal. We have shown that 
the elements of the set we claim is a basis are normalized and to show that they are pairwise 
orthogonal we only have to show that, for any \mbox{$j \in J$}, \mbox{$k \in K$}, the preparations
\mbox{$u \circ {a}_{j}$} and \mbox{$v \circ {b}_{k}$} are orthogonal.
But \mbox{${b}^{\star}_{k} \circ {v}^{\star} \circ u \circ {a}_{j} =$} 
\mbox{${b}^{\star}_{k} \circ {0}_{A , B} \circ {a}_{j} =$} \mbox{${0}_{I , I}$}.
Let, now, \mbox{$x : I \rightarrow A \oplus B$} be orthogonal to all preparations of the set claimed
to be a basis.
For any \mbox{$j \in J$} \mbox{$\langle x \mid u \circ {a}_{j} \rangle =$} 
\mbox{${0}_{I , I} =$}
\mbox{$\langle {u}^{\star} \circ x \mid {a}_{j} \rangle$}. Since the $a_{j}$'s form a basis of $A$,
we have \mbox{${u}^{\star} \circ x =$} \mbox{${0}_{I , A}$}.
Similarly \mbox{${v}^{\star} \circ x =$} \mbox{${0}_{I , B}$}.
Theorems~\ref{the:character} and~\ref{the:ab-monoid} imply that 
\mbox{$x =$} \mbox{${0}_{I , A \oplus B}$}.
\end{proof}
In Hilbert spaces, for any bases $a_{i}$ for $A$ and $b_{j}$ for $B$, 
the preparations of the form \mbox{$a_{i} \otimes b_{j}$} form a basis for the tensor product 
\mbox{$A \otimes B$}.  This does not seem to hold in arbitrary B-categories 
that are also T-categories.

\subsection{Normalizability and regularity} \label{sec:normalizable}
Does every B-category admit an orthonormal basis? No! One needs two additional properties.
One of those properties is an expression of our point of view, already presented in Section~\ref{sec:unit-def}, that only normalized preparations are indubitably legitimate. 
Any scalar composed with a (normalized) preparation must be accepted as a preparation and
such a preparation is not, in general, normalized. We shall require that all 
preparations be obtained from normalized preparations by composition with a scalar.
\begin{definition} \label{def:normalizable}
An A-category with unit object is said to be {\em normalizable} 
iff, for any object $A$ and any preparation
\mbox{$a : I \rightarrow A$} there is some normalized preparation \mbox{$n : I \rightarrow A$}
and some scalar \mbox{$s : I \rightarrow I$} such that \mbox{$a =$} \mbox{$n \circ s$}.
\end{definition}
Note that if there is some unit object with the property described in 
Definition~\ref{def:normalizable}, then all unit objects have this property.
The second property we shall consider requires every non-zero scalar 
to be regular for composition.
This is consistent with our view that scalars, 
except for the zero scalar, represent basic symmetries. It is reasonable to suppose that
such symmetries are invertible, and this will be required in Section~\ref{sec:nocloning}.
The definition below defines a weaker property.
\begin{definition} \label{def:regular}
An Az-category with unit object is said to be {\em regular} iff for any scalars 
\mbox{$s , t : I \rightarrow I$} such that \mbox{$s \circ t =$} \mbox{${0}_{I , I}$} one has
either \mbox{$s =$} \mbox{${0}_{I , I}$} or \mbox{$t =$}  \mbox{${0}_{I , I}$}.
\end{definition}
Note again that if a unit object satisfies the property above, all unit objects do.
Before we study the properties of normalizable, regular B-categories and their bases, let us
pause a moment to prove two results concerned with A-categories and that 
will be used in Section~\ref{sec:nocloning}.
\begin{comment}
\begin{lemma} \label{le:obvious}
In a normalizable A-category with unit object, 
for any \mbox{$f , g : A \rightarrow B$}, if, for any {\em normalized}
preparation \mbox{$a : I \rightarrow A$}, one has \mbox{$f \circ a =$} \mbox{$g \circ a$}, then
\mbox{$f =$} $g$.
\end{lemma}
\begin{proof}
Let \mbox{$a' : I \rightarrow A$} be any preparation. Since the category is normalizable,
there is a normalized preparation $a$ and a scalar $s$ such that \mbox{$a' =$}
\mbox{$a \circ s$}. By assumption \mbox{$f \circ a =$} \mbox{$g \circ a$} since $a$ is normalized.
Therefore \mbox{$f \circ a \circ s =$} \mbox{$g \circ a \circ s$}.
For any preparation $a'$ we have: \mbox{$f \circ a' =$} \mbox{$g \circ a'$}.
By Definition~\ref{def:unit-object}, we have \mbox{$f =$} $g$. 
\end{proof}
\end{comment}
\begin{lemma} \label{le:fstarfzero}
In a normalizable, regular Az-category, 
for any \mbox{$f : A \rightarrow B$}, if \mbox{${f}^{\star} \circ f =$}
\mbox{${0}_{A , A}$} then \mbox{$f =$} \mbox{${0}_{A , B}$}.
In particular, for any preparation \mbox{$a :$} 
\mbox{$I \rightarrow A$}, \mbox{$sq(a) =$}
\mbox{${0}_{I , I}$}
iff \mbox{$a =$} \mbox{${0}_{I , A}$}.
\end{lemma}
\begin{proof}
Assume \mbox{${f}^{\star} \circ f =$} \mbox{${0}_{A , A}$}.
Let \mbox{$a : I \rightarrow A$} be any preparation.
Since $I$ is a unit object, it is enough to prove that \mbox{$f \circ a =$} 
\mbox{${0}_{A , B} \circ a =$} \mbox{${0}_{I , B}$}.
Since the category is normalizable, 
there is a scalar $s$ and a normalized \mbox{$n : I \rightarrow B$} such that 
\mbox{$f \circ a = n \circ s$}.
We have:
\[
0_{I , I} \ = \ {a}^{\star} \circ {0}_{A , A} \circ a \ = \ 
{a}^{\star} \circ {f}^{\star} \circ f \circ a \ = \
{s}^{\star} \circ {n}^{\star} \circ n \circ s \ = \ {s}^{\star} \circ s.
\]
Since the category is regular, either \mbox{$s =$} \mbox{$0_{I , I}$} 
or \mbox{${s}^{\star} =$} \mbox{${0}_{I , I}$} and therefore, by  Lemma~\ref{le:zero-star},
\mbox{$s =$} \mbox{${0}_{I , I}$}.
We conclude that we have \mbox{$f \circ a =$} 
\mbox{$n \circ s =$} 
\mbox{${0}_{I , B}$}.
\end{proof}
The property expressed by Lemma~\ref{le:fstarfzero} should not be confused with the property
that arrows of the form \mbox{${f}^{\star} \circ f$} are in some way {\em positive}. This last
property correspond to the property requiring that, if \mbox{${f}^{\star} \circ f + {g}^{\star} \circ g =$}
\mbox{${0}_{A , A}$}, then \mbox{$f =$} \mbox{${0}_{A , B}$}.
Appendix~\ref{sec:examples7} discusses the normalizability and regularity 
of our running examples.

\subsection{Bases in normalizable regular B-categories} \label{sec:normalized-bases}
\begin{theorem} \label{the:basis-ex}
In a normalizable regular B-category, every object admits a basis.
\end{theorem}
\begin{proof}
Let $A$ be an object in a normalizable regular B-category.
We reason by ordinal induction. 
For any ordinal $\alpha$ we define a set of normalized pairwise orthogonal
preparations $B_{\alpha}$ such that, for every \mbox{$\beta \leq \alpha$} 
\mbox{$B_{\beta} \subseteq B_{\alpha}$}.
For every limit ordinal $\alpha$ (including $0$)  we take \mbox{$B_{\alpha} =$}
\mbox{$\bigcup_{\beta < \alpha} B_{\beta}$}. For every successor ordinal
\mbox{$\alpha =$} \mbox{$\beta + 1$}, if \mbox{$B_{\beta}$} is a basis we take 
\mbox{$B_{\alpha} =$} \mbox{$B_{\beta}$} and, if it is not a basis, there is some non-null
preparation \mbox{$a_{\alpha}$} that is orthogonal to all elements of \mbox{$B_{\beta}$}.
Since the category is normalizable, there is a normalized preparation \mbox{$n : I \rightarrow  A$}
and a scalar \mbox{$s \neq$} \mbox{${0}_{I , I}$} 
such that \mbox{$a_{\alpha} =$} \mbox{$n \circ s$}.
We then take \mbox{$B_{\alpha} =$} \mbox{$B_{\beta} \cup \{ n \}$}.
We claim that $n$ is orthogonal to every element of \mbox{$B_{\beta}$}.
Indeed, for any \mbox{$x \in B_{\beta}$}, \mbox{${0}_{I , I} =$} 
\mbox{$\langle x \mid n \circ s \rangle =$} \mbox{$\langle x \mid n \rangle \circ s$}
and, since the category is regular $x$ and $n$ are orthogonal.
The sequence $B$ is a chain and there is therefore some ordinal $\alpha$ for which
\mbox{$B_{\alpha} =$} \mbox{$B_{\beta}$} for every \mbox{$\beta > \alpha$}.
The set \mbox{$B_{\alpha}$} is a basis.
\end{proof}
Note that our treatment of bases is rudimentary: weak assumptions and few results.
To obtain more standard properties for bases an additional assumption is required: that 
every right-unitary arrow that is not unitary is the injection of some u-coproduct.
In other terms every subspace has an orthogonal complement.
This last requirement parallels a property discussed and used in~\cite{SP:IJTP}.
It is remarkable that so much can be done without this property.
\begin{comment}
***** if $A$ admits a finite basis the sum of its projections is the identity    ************
This requires the {\em complementation} property: every right-unitary that is not unitary is
the injection of some u-coproduct

\subsection{BC-categories} \label{sec:BC-cat}
******* Definition and theorem
\begin{theorem} \label{the:ortho-coprod}
Assume a non-trivial BT-category.
If \mbox{$a , b : I \rightarrow C$} are orthogonal preparations, $b$ is normalized and
\mbox{$a \neq 0$}, then 
there is an object $A$ and an
arrow \mbox{$u : A \rightarrow C$} such that the pair $u$, $b$ forms a u-coproduct
(\mbox{$C =$} \mbox{$A \oplus I$}) and a preparation \mbox{$a' : I \rightarrow A$} such that
\mbox{$a =$} \mbox{$u \circ a'$} . 
\end{theorem}
\begin{proof}
Assume the assumptions of Theorem~\ref{the:ortho-coprod} hold.
We have \mbox{${b}^{\star} \circ a =$} $0$. If $b$ was unitary, 
we would have \mbox{$b \circ {b}^{\star} =$} \mbox{${id}_{C}$} and
\mbox{$a =$} \mbox{$b \circ {b}^{\star} \circ a =$} 
\mbox{$b \circ 0 =$} $0$. We conclude that $b$ is not unitary and we can

use part~\ref{T+-decomp} of Definition~\ref{def:B-cat}.

There is an object $A$ 
and an arrow
\mbox{$u : A \rightarrow C$} such that the pair $u$, $b$ forms a u-coproduct.
By Theorem~\ref{the:character}, we have:
\mbox{$u \circ {u}^{\star} + b \circ {b}^{\star} =$} \mbox{${id}_{C}$}.
Therefore \mbox{$a =$}
\mbox{$u \circ {u}^{\star} \circ a + 0 =$} \mbox{$u \circ ({u}^{\star} \circ a)$}.
 \end{proof}

\end{comment}

\section{Tensor with biproduct} \label{sec:interaction}
\subsection{BT-categories} \label{sec:BT-cat}
We now want to study the interaction between the tensor, i.e., multiplicative, and the additive
structures.
\begin{definition}
A category that is both a T-category and a B-category will be called a BT-category.
\end{definition}
It is a small miracle that there is a deep relation between tensor products and biproducts in
BT-categories: tensor distributes over biproduct.

\subsection{Distributivity} \label{sec:distributivity}
A fundamental property of physical systems is that systems composed of an $A$ part and a part
that is either $B$ or $C$ (or, equivalently both $B$ and $C$) are exactly systems that are either
composed of an $A$ part and a $B$ part or of an $A$ part and a $C$ part (equivalently, both a 
system composed of an $A$ part and a $B$ part and a system composed of an $A$ part 
and a $B$ part). In other terms, for any objects $A$, $B$ and $C$ we expect 
\mbox{$A \otimes (B \oplus C)$} to be identical (i.e., equal up to a unitary arrow) to 
\mbox{$(A \otimes B) \oplus (A \otimes C)$}.
 
As announced we shall prove the existence of a canonical unitary arrow of sort
\mbox{$(A \otimes B) \oplus (A \otimes C) \rightarrow$} \mbox{$A \otimes (B \oplus C)$}.
First, we need a distributivity property of tensor product over the Hom-set addition.
But this follows easily from the distributivity of arrow composition over Hom-set addition.
\begin{lemma} \label{le:dist+}
In a BT-category, for any preparations \mbox{$a : I \rightarrow A$} and 
\mbox{$b_{1} , b_{2} :$} \mbox{$I \rightarrow B$}, one has:
\[
a \otimes (b_{1} + b_{2}) \ = \ a \otimes b_{1} \, + \, a \otimes b_{2}  \ , \ 
(b_{1} + b_{2}) \otimes a \ = \ b_{1} \otimes a \, + \, b_{2} \otimes a.
\]
\end{lemma}
\begin{proof}
We shall prove the first assertion. The second one is proved in a similar way.
By Lemma~\ref{le:kappa-tensor}, it is enough to prove that we have:
\mbox{$\kappa_{A , B}(a , b_{1} + b_{2}) =$}
\mbox{$\kappa_{A , B}(a , b_{1}) + \kappa_{A , B}(a , b_{2})$}.
By Definition~\ref{def:bi-arrow}, this is equivalent to proving:
\mbox{$\kappa_{A, B}^{1}(a) \circ (b_{1} + b_{2}) =$}
\mbox{$\kappa_{A , B}^{1}(a) \circ b_{1} + \kappa_{A , B}^{1} \circ b_{2}$}, 
which follows from Lemma~\ref{le:+comp}.
\end{proof}
Lemma~\ref{le:dist+} implies a more general result.
\begin{theorem} \label{the:dist+}
In a BT-category, for any arrows \mbox{$a :$} \mbox{$A' \rightarrow A$},
\mbox{$b_{1} , b_{2} :$} \mbox{$B' \rightarrow B$} one has
\mbox{$a \otimes (b_{1} + b_{2}) =$}
\mbox{$a \otimes b_{1} \, + \, a \otimes b_{2}$} and
\mbox{$(b_{1} + b_{2}) \otimes a =$}
\mbox{$b_{1} \otimes a \, + \, b_{2} \otimes a$}.
\end{theorem}
\begin{proof}
By Lemma~\ref{le:dist+} and Definition~\ref{def:unit-object}, part~\ref{prep-arrows}.
\end{proof}
We may now define, for any objects $A$, $B$, $C$ of a
BT-category, canonical arrows 
\mbox{$x_{A , B , C} :$}
\mbox{$(A \otimes B) \oplus (A \otimes C) \rightarrow$}
\mbox{$A \otimes (B \oplus C)$} and
\mbox{$y_{A , B , C} :$}
\mbox{$(B \otimes A) \oplus (C \otimes A) \rightarrow$}
\mbox{$(B \oplus C) \otimes A$}.
Theorem~\ref{the:distributivity} below will show that these arrows are unitary.
\begin{definition} \label{def:x}
Let $A$, $B$ and $C$ be objects In a BT-category.
Note that \mbox{${id}_{A} \otimes {u}_{B , C} :$}
\mbox{$A \otimes B \rightarrow A \otimes (B \oplus C)$}
and \mbox{${id}_{A} \otimes {v}_{B , C} : $}
\mbox{$A \otimes C \rightarrow A \otimes (B \oplus C)$}.
We shall define 
\mbox{${x}_{A , B , C} =$}
\mbox{$[ {id}_{A} \otimes {u}_{B , C} \ \ 
{id}_{A} \otimes {v}_{B , C} ]$}. In other words, \mbox{${x}_{A , B , C} :$}
\mbox{$(A \otimes B) \oplus (A \otimes C) \rightarrow$}
\mbox{$A \otimes (B \oplus C)$} is the only arrow such that:
\begin{equation} \label{eq:x}
x_{A , B , C} \circ {u}_{A \otimes B , A \otimes C} \ = \ {id}_{A} \otimes {u}_{B , C} 
\end{equation}
\[
x_{A , B , C} \circ {v}_{A \otimes B , A \otimes C} \ = \ {id}_{A} \otimes {v}_{B , C}.
\]
Similarly we define 
\mbox{$y_{A , B , C} : (B \otimes A) \oplus (C \otimes A) \rightarrow$}
\mbox{$(B \oplus C) \otimes A$} as
\mbox{$y_{A , B , C} =$} \mbox{$[ {u}_{B , C} \otimes {id}_{A} \ \ {v}_{B , C} \otimes {id}_{A} ]$}.
\end{definition}
\begin{theorem} \label{the:distributivity}
In a BT-category, for any objects $A$, $B$, and $C$, the arrows
\mbox{${x}_{A , B , C}$} and \mbox{${y}_{A , B , C}$} are unitary.
\end{theorem}
\begin{proof}
We prove the first claim. The second one is proved similarly.
Let \mbox{$y =$} \mbox{${x}^{\star} \circ x$}.
One sees, by Equation~\ref{eq:x}, Lemmas~\ref{le:star-tens}, \ref{le:id}, \ref{le:otimes-comp}
and Corollary~\ref{co:co-funct}
that we have:
\[
{u}_{A \otimes B , A \otimes C}^{\star} \circ y \circ {u}_{A \otimes B , A \otimes C} \ = \ 
({id}_{A} \otimes {u}_{B , C}^{\star}) \circ ({id}_{A} \circ {u}_{B , C}) \ = \ 
{id}_{A} \otimes {id}_{B} \ = \ {id}_{A \otimes B}
\]
and similarly
\[
{v}_{A \otimes B , A \otimes C}^{\star} \circ y \circ {v}_{A \otimes B , A \otimes C} \ = \ 
{id}_{A \otimes C}.
\]
We also have, using the same equations and lemmas, and also Lemma~\ref{le:zero-tens}:
\[
{u}_{A \otimes B , A \otimes C}^{\star} \circ y \circ {v}_{A \otimes B , A \otimes C} \ = \ 
({id}_{A} \otimes {u}_{B , C}^{\star}) \circ ({id}_{A} \circ {v}_{B , C}) \ = \ 
\]
\[
{id}_{A} \otimes ({u}_{B , C}^{\star} \circ {v}_{B , C}) \ = \
{id}_{A} \otimes {0}_{C , B} \ = \ 
{0}_{A \otimes C , A \otimes B}
\]
and similarly
\[
{v}_{A \otimes B , A \otimes C}^{\star} \circ y \circ {u}_{A \otimes B , A \otimes C} \ = \ 
{0}_{A \otimes B , A \otimes C}.
\]
There is a unique $y$ satisfying the four equations above and therefore we conclude
that \mbox{${x}^{\star} \circ x =$} \mbox{$y =$} \mbox{${id}_{(A \otimes B) \oplus (A \otimes C)}$}.

Let us now consider \mbox{$x \circ {x}^{\star} :$}
\mbox{$A \otimes (B \oplus C) \rightarrow A \otimes (B \oplus C)$}.
By Theorem~\ref{the:character}, we have
\mbox{${u}_{A \otimes B , A \otimes C} \circ {u}_{A \otimes B , A \otimes C}^{\star} + 
{v}_{A \otimes B , A \otimes C} \circ {v}_{A \otimes B , A \otimes C}^{\star} =$}
\mbox{${id}_{(A \otimes B) \oplus (A \otimes C)}$}.
Therefore, by Lemmas~\ref{le:+comp}, \ref{le:otimes-comp}, Theorem~\ref{the:dist+} 
and Corollary~\ref{co:co-funct}, one has:
\[
x \circ {x}^{\star} \ = x \circ ({u}_{A \otimes B , A \otimes C} \circ 
{u}_{A \otimes B , A \otimes C}^{\star} + 
{v}_{A \otimes B , A \otimes C} \circ {v}_{A \otimes B , A \otimes C}^{\star}) \circ {x}^{\star} \ = \ 
\]
\[
x \circ {u}_{A \otimes B , A \otimes C} \circ {u}_{A \otimes B , A \otimes C}^{\star} \circ {x}^{\star}
\, + \,
x \circ {v}_{A \otimes B , A \otimes C} \circ {v}_{A \otimes B , A \otimes C}^{\star} \circ {x}^{\star}
\ = \ 
\]
\[
({id}_{A} \otimes {u}_{B , C}) \circ ({id}_{A} \otimes {u}_{B , C}^{\star}) \, + \, 
({id}_{A} \otimes {v}_{B , C}) \circ ({id}_{A} \otimes {v}_{B , C}^{\star}) \ = \ 
\]
\[
({id}_{A} \otimes ({u}_{B , C} \circ {u}_{B , C}^{\star})) +
({id}_{A} \otimes ({v}_{B , C} \circ {v}_{B , C}^{\star})) \ = \
\]
\[
{id}_{A} \otimes ({u}_{B , C} \circ {u}_{B , C}^{\star} +
{v}_{B , C} \circ {v}_{B , C}^{\star}) \ = \ 
{id}_{A} \otimes {id}_{B \oplus C} \ = \ {id}_{A \otimes (B \oplus C)}.
\]
\end{proof}

\section{Q-categories} \label{sec:quantic}
\subsection{Definition} \label{sec:def-Q}
What we have done so far fits the category \cR\ of sets and relations, and therefore does not
describe any specially quantic character.
We shall now make two additional assumptions on the structure of the unit object, preparations
and scalars.
Those assumptions, and in particular, the second one, are characteristic of quantum physics.
First, some notation:

\begin{notation}
From now on, we shall denote the scalar \mbox{${0}_{I , I}$} by $0$,
the scalar \mbox{${id}_{I}$} by $1$ and the scalar \mbox{$1 + 1$} by $2$.
\end{notation}
\begin{comment}
As noticed at the end of Section~\ref{sec:unit-object}, if we take seriously the idea that scalars
are symmetries, we should require every scalar to have an inverse for composition.
In Section~\ref{sec:zero}, we assumed, in particular, a zero scalar $0$. 
The zero scalar is certainly not invertible, and it does not indeed represent any symmetry.
So, it seems we have to accept that not all scalars are invertible. We shall, nevertheless, assume
that any scalar different from zero is invertible. 
\end{comment}

We shall assume that there is a scalar \mbox{${1}^{-} : I \rightarrow I$} that is different from
$1$ and is its own inverse, i.e., such that \mbox{${1}^{-} \circ {1}^{-} =$} $1$. The scalar ${1}^{-}$, 
by property~\ref{global}  of Definition~\ref{def:unit-object} represents a fundamental symmetry: 
to every arrow \mbox{$f : A \rightarrow B$} corresponds an arrow \mbox{${f}^{-} : A \rightarrow B$} 
satisfying properties described in Theorem~\ref{the:global-}.
\begin{definition} \label{def:Q-cat}
A normalizable BT-category is said to be an {\em Q-category} iff 
\begin{enumerate}
\item \label{inverse}
every scalar $s$ that is different from
$0$ is invertible, i.e., there is some scalar ${s}^{- 1}$ such that \mbox{$s \circ {s}^{- 1} =$}
$1$, and
\item \label{minus-one}
there is some scalar \mbox{${1}^{-} \neq 1$} such that \mbox{${1}^{-} \circ {1}^{-} =$} $1$.
\end{enumerate}
\end{definition}
Note that Definition~\ref{def:Q-cat} is invariant under isomorphism and therefore 
if some unit object of normalizable BT-category satisfies properties~\ref{inverse} 
and~\ref{minus-one}, all unit objects do. 
Note also that the scalar ${1}^{-}$ (i.e., $- 1$) 
is defined by its multiplicative, not additive, properties and that the category of matrices
over any field of characteristic different from $2$ (with adjoint = transpose) is an
Q-category. 
The same over a field of characteristic $2$ is not.

Trivial categories and the category $\cR$ of relations are not Q-categories since they do
not satisfy property~\ref{minus-one} above.
Hilbert spaces form a Q-category. Those fields for which matrices form a normalizable 
category (see Appendix~\ref{sec:matrices7}) are a Q-category.

In the sequel we shall study Q-categories, but, before we do, let us think a moment
about possible structures for the scalars if one requires condition~\ref{inverse}, but not
condition~\ref{minus-one} of Definition~\ref{def:Q-cat}. It is probably in this direction that one
should look for classical (i.e., non-quantic) structures.
If $1$ is the only solution to the equation \mbox{$x \circ x = 1$}, and if the scalars form a field,
then the field has characteristic $2$. If the scalars do not form a field, it may be that
\mbox{$0 = 1$} and, in this case, the category is a trivial category as in Section~\ref{sec:trivial}.
If \mbox{$0 \neq 1$}, and the additive structure of the scalars is cancellative, we may have
structures such as matrices over nonnegative rationals, where all scalars are self-adjoint, or
such as matrices of complex numbers whose real part is non-negative with adjoints being either
transpose or conjugate-transpose. If the additive structure of the scalars is not cancellative
we may have a category such as Relations.

\subsection{Properties of Q-categories} \label{sec:quantic-prop}
First, in a Q-category two eigenvectors of a self-adjoint arrow for different
eigenvalues are orthogonal.
\begin{theorem} \label{the:eigen}
In a Q-category, for any self-adjoint arrow \mbox{$f : A \rightarrow A$},
if $a$ and $b$ are eigenvectors of $f$ (i.e. : \mbox{$f \circ a =$} \mbox{$a \circ s$} and
\mbox{$f \circ b =$} \mbox{$b \circ t$} with $s$, $t$ scalars) corresponding to different eigenvalues (i.e., \mbox{$s \neq t$}), then they are orthogonal.
\end{theorem}
\begin{proof}
Assume \mbox{${f}^{\star} =$} $f$, \mbox{$f \circ x =$} \mbox{$x \circ s$} and
\mbox{$f \circ y =$} \mbox{$y \circ t$}.
On one hand we have:
\mbox{${b}^{\star} \circ f \circ a =$} \mbox{${b}^{\star} \circ a \circ s$}.
On the other hand, using successively Lemma~\ref{le:eigen1} and 
Theorem~\ref{the:scalar-com} we have:
\mbox{${b}^{\star} \circ f \circ a =$} \mbox{${b}^{\star} \circ {f}^{\star} \circ a =$}
\mbox{${(f \circ b)}^{\star} \circ a =$}
\mbox{${(b \circ t)}^{\star} \circ a =$} \mbox{${t}^{\star} \circ {b}^{\star} \circ a =$}
\mbox{$t \circ {b}^{\star} \circ a =$} \mbox{${b}^{\star} \circ a \circ t$}. 
We see that \mbox{$({b}^{\star} \circ a) \circ s =$} \mbox{$({b}^{\star} \circ a) \circ t$}.
By Definition~\ref{def:Q-cat}, property~\ref{inverse} if the scalar 
\mbox{${b}^{\star} \circ a$} is different from $0$, then it is invertible and $s = t$, 
contrary to assumptions.
We conclude that $a$ and $b$ are orthogonal.
\end{proof}
Our next result is that in a Q-category, the scalars form a (commutative) field
of characteristic different from $2$.
\begin{theorem} \label{the:field}
If \cC\ is a Q-category, then  
\begin{enumerate}
\item \label{r+1}
\mbox{$1 + {1}^{-}  =$} $0$,
\item \label{field}
the scalars 
\mbox{$\langle Hom(I , I) , + , \circ , 0 , 1 \rangle$} form a field of characteristic different from $2$, and
\item \label{starr}
\mbox{${{1}^{-}}^{\star} = {1}^{-}$}.
\end{enumerate}
\end{theorem}
\begin{proof}
By item~\ref{minus-one} of Definition~\ref{def:Q-cat}, we have 
\begin{equation} \label{eq:r}
1 + {1}^{-} \ = \ ({1}^{-} \circ {1}^{-}) + {1}^{-} \ = \ ({1}^{-} \circ {1}^{-}) + ({1}^{-} \circ 1) \ = \ 
\end{equation}
\[
{1}^{-} \circ ({1}^{-} + 1) \ = \ {1}^{-} \circ (1 + {1}^{-}).
\]
By item~\ref{inverse}, if \mbox{$1 + {1}^{-} \neq 0$},
\mbox{$1 + {1}^{-}$} is invertible and, by multiplying both sides of Equation~\ref{eq:r} on the right by
this inverse we get \mbox{$1 = {1}^{-}$}, a contradiction. 
We conclude that \mbox{$1 + {1}^{-} = 0$}.

The scalars, in any BT-category satisfy all properties defining a field except, perhaps,
three: the scalars do not always have an additive inverse, they do not always have a multiplicative
inverse and $0$ and $1$ are not always different. The existence of a multiplicative inverse is 
explicitly provided by item~\ref{inverse} of Definition~\ref{def:Q-cat}.
The fact that \mbox{$1 \neq 0$} in a Q-category follows from Lemma~\ref{le:trivial}
and item~\ref{minus-one} of Definition~\ref{def:Q-cat} that implies that there are at least two different scalars. We are left to prove that, in a Q-category, every scalar has an 
additive inverse. But, for any scalar $s$ one has:
\mbox{$s + ( s \circ {1}^{-} ) =$} \mbox{$s \circ (1 + {1}^{-}) =$} \mbox{$s \circ 0 =$} $0$ and
\mbox{$s \circ {1}^{-}$} is an additive inverse for $s$. 
Since \mbox{${1}^{-} \neq 1$}, we have \mbox{$0 =$} \mbox{$1 + {1}^{-} \neq$} \mbox{$1 + 1$}.

We have \mbox{${{1}^{-}}^{\star} \circ {{1}^{-}}^{\star} =$}
\mbox{${({1}^{-} \circ {1}^{-})}^{\star} =$} \mbox{${1}^{\star} =$} $1$.
We see that ${{1}^{-}}^{\star}$ is a solution to the equation \mbox{$s \circ s = 1$}.
But, in any field, $1$ and ${1}^{-} = - 1$ are the only solutions to this equation.
But \mbox{${{1}^{-}}^{\star} =$} $1$ implies \mbox{${1}^{-} =$} \mbox{${1}^{\star} =$} $1$ which is excluded.
We see that \mbox{${{1}^{-}}^{\star} = {1}^{-}$}.
\end{proof}
We shall now globalize the properties of the scalar ${1}^{-}$ in a Q-category into 
a global symmetry.
\begin{definition} \label{def:global-}
In a Q-category, for every arrow \mbox{$f : A \rightarrow B$} we define the
arrow \mbox{${f}^{-} : A \rightarrow B$} to be the arrow 
\mbox{$f \circ {{1}^{-}}_{A} =$} \mbox{${{1}^{-}}_{B} \circ f$}. The equality follows from 
Lemma~\ref{le:forgotten}. 
\end{definition}
We see that the notation \mbox{${1}^{-}$} chosen is consistent.
\begin{theorem} \label{the:global-}
In a Q-category, for any arrows \mbox{$f , h :$} \mbox{$A \rightarrow B$}, 
\mbox{$f' :$} \mbox{$A' \rightarrow B'$} and
\mbox{$g :$} \mbox{$B \rightarrow C$} one has:
\begin{enumerate}
\item \label{-involution}
\mbox{${({f}^{-})}^{-} = f$},
\item \label{-comp}
\mbox{${(g \circ f)}^{-} =$} \mbox{${g}^{-} \circ f =$} \mbox{$g \circ {f}^{-}$},
\item \label{-star}
\mbox{${({f}^{\star})}^{-} =$} \mbox{${({f}^{-})}^{\star}$},
\item \label{-+}
\mbox{${( f + h)}^{-} =$} \mbox{${f}^{-} + {g}^{-}$},
\item \label{--}
\mbox{$f + {f}^{-} =$} \mbox{${0}_{A , B}$},
\item \label{self}
${f}^{-}$ is self-adjoint iff $f$ is,
\item \label{-unitary}
${f}^{-}$ is right (resp. left) unitary iff $f$ is
\item \label{tens-minus}
\mbox{${(f \otimes f')}^{-} =$} \mbox{${f}^{-} \otimes f' =$} \mbox{$f \otimes {f'}^{-}$}.
\end{enumerate}
\end{theorem}
Note that part~\ref{--} expresses the fact that $f$ and ${f}^{-}$ cannot be superposed.
\begin{proof}
All claims are easy consequences of Definition~\ref{def:global-} 
and Lemmas~\ref{le:global-cat-adj} and Theorem~\ref{the:field}. 
We shall prove property~\ref{tens-minus}.
We have: \mbox{$(f \otimes f') \circ {1}_{A}^{-} =$}
\mbox{$(f \otimes f') \circ ({1}_{A}^{-} \otimes {1}_{A'}) =$} 
\mbox{$( f \circ {1}_{A}^{-}) \otimes f' =$} \mbox{${f}^{-} \otimes f'$}.
\end{proof}
We can now fulfill the promise made at the end of Section~\ref{sec:identical}.

\subsection{Special tensor products} \label{sec:tensor-sym}
In QM the case of systems composed of indistinguishable parts is of fundamental importance.
We shall define now symmetric and anti-symmetric bi-arrows 
and a uniform construction of special
tensor products that produces both the symmetric and anti-symmetric tensor products needed for
considering bosons and fermions respectively.
\begin{definition} \label{def:symmetric}
Assume \cC\ is a Q-category and let \mbox{$A , X$} be objects of \cC.
A bi-arrow \mbox{$\alpha : A , A \rightarrow X$} is said to be {\em symmetric} (resp. {\em anti-symmetric}) iff, 
for any preparations \mbox{$a , b : I \rightarrow A$} one has:
\mbox{$\alpha(a , b) =$} \mbox{$\alpha(b , a)$} (resp. \mbox{$\alpha(a , b) =$}
\mbox{${\alpha(b , a)}^{-} $}).
\end{definition}
Both the symmetric and the anti-symmetric restrictions express 
in a straightforward manner the fact that one cannot distinguish 
between the two systems \mbox{$(a , b)$} and \mbox{$(b , a)$},
i.e., acting on a system composed of a part in state $a$ and a part in state $b$ 
is indistinguishable
from acting on a system composed of a part in state $b$ and a part in state $a$. 
The fundamental symmetry revealed by ${ }^{-}$ gives this indistinguishability its dual flavor. 
\begin{lemma} \label{le:sym-bi}
In a Q-category, a bi-arrow \mbox{$\alpha : A , A \rightarrow X$} is symmetric iff there is, 
for any preparation \mbox{$a : I \rightarrow A$}, an arrow 
\mbox{$\alpha'(a) : A \rightarrow X$} such that, for any preparations
\mbox{$a , b : I \rightarrow A$}, one has \mbox{$\alpha(a , b) =$}
\mbox{$\alpha'(a) \circ b =$} \mbox{$\alpha'(b) \circ a$}.
\end{lemma}
\begin{proof}
For the {\em only if} part, assume $\alpha$ is a symmetric bi-arrow.
Then, for any preparations $a$, $b$ one has \mbox{$\alpha(a , b) =$}
\mbox{$\alpha^{1}(a) \circ b$}  and 
\mbox{$\alpha(b , a) =$} \mbox{$\alpha^{1}(b) \circ a$}.
Since \mbox{$\alpha(a , b) =$} \mbox{$\alpha(b , a)$} we can take 
\mbox{$\alpha' = \alpha^{1}$}.
The {\em if} part is obvious.
\end{proof}
\begin{lemma} \label{le:antisym-bi}
In a Q-category a bi-arrow \mbox{$\alpha : A , A \rightarrow X$} is anti-symmetric iff there is, 
for any preparation \mbox{$a : I \rightarrow A$}, an arrow 
\mbox{$\alpha'(a) : A \rightarrow X$} such that for any preparations
\mbox{$a , b : I \rightarrow A$} one has \mbox{$\alpha(a , b) =$}
\mbox{$\alpha'(a) \circ b =$} \mbox{${(\alpha'(b) \circ a)}^{-}$}.
\end{lemma}
\begin{proof}
For the {\em only if} part, assume $\alpha$ is an anti-symmetric bi-arrow.
Then, for any preparations $a$, $b$ one has \mbox{$\alpha(a , b) =$}
\mbox{$\alpha^{1}(a) \circ b$}  and 
\mbox{$\alpha(b , a) =$} \mbox{$\alpha^{1}(b) \circ a$}.
Since \mbox{$\alpha(a , b) =$} \mbox{${(\alpha(b , a))}^{-}$} we can take 
\mbox{$\alpha' = \alpha^{1}$}.
For the {\em if} part, we have, by Theorem~\ref{the:global-}  \mbox{${(\alpha'(b) \circ a)}^{-} =$}
\mbox{${(\alpha'(b))}^{-} \circ a$}. One may take 
\mbox{${\alpha}^{1} = \alpha'$} and \mbox{${\alpha}^{2}(a) =$} \mbox{${(\alpha'(a))}^{-}$}.
\end{proof}
Note that, if \mbox{$\alpha : A , A \rightarrow X$} is a symmetric (resp. anti-symmetric) bi-arrow, 
then, for any \mbox{$x : X \rightarrow Y$} and \mbox{$f : B \rightarrow A$},
 \mbox{$x \circ \alpha$} and \mbox{$\alpha \circ (f , f)$} are symmetric (resp. anti-symmetric) 
 bi-arrows.
\begin{definition} \label{def:stensor-prod}
We assume a Q-category.
A symmetric (resp. anti-symmetric) bi-arrow \mbox{$\sigma : A , A \rightarrow X$} is said to be an {\em s-tensor (resp. an a-tensor) product} (for $A$) iff:
\begin{enumerate}
\item \label{stensorprod-unique}
for any object $Y$ and any symmetric (resp. ant-symmetric) bi-arrow 
\mbox{$\alpha : A , A \rightarrow Y$} there
is a unique arrow \mbox{$x : X \rightarrow Y$} such that
\mbox{$\alpha =$} \mbox{$x \circ \sigma$}, and
\item \label{stensorprod-star}
for any preparations \mbox{$a , b , a' , b' : I \rightarrow A$} 
one has 
\[
\sigma^{\star}(a , b) \circ \sigma(a' , b') \circ 2 \ = \ 
{a}^{\star} \circ a' \circ {b}^{\star} \circ b' + {a}^{\star} \circ b' \circ {b}^{\star} \circ a
\]
\[
(resp. \ \ 
\sigma^{\star}(a , b) \circ \sigma(a' , b') \circ 2 \ = \ 
{a}^{\star} \circ a' \circ {b}^{\star} \circ b' + {({a}^{\star} \circ b' \circ {b}^{\star} \circ a)}^{-}).
\]
\end{enumerate}
In such a case we shall write \mbox{$A {\otimes}_{s} A$} 
(resp. \mbox{$A {\otimes}_{a} A$}) for $X$.
\end{definition}
Note that condition~\ref{stensorprod-star} is different from the corresponding
condition in Definition~\ref{def:tensor-product}. It expresses the fact that the scalar product of two
product preparations is the average of the two possible interpretations. In the anti-symmetric 
case proper care of the behavior under exchange has to be given.
Note also that if $H$ is a Hilbert space,
the projection of \mbox{$H \otimes H$} onto its symmetric subspace
provides an s-tensor product for $H$ and its projection onto its anti-symmetric subspace
provides an a-tensor product for $H$.
Our next result parallels Theorem~\ref{the:tensor-products}: symmetric and anti-symmetric
tensor products are defined up to a unitary arrow.
\begin{theorem} \label{the:s-tensor-products}
In a Q-category, if \mbox{$\sigma : A , A \rightarrow A \otimes_{S} A$}
is an s-tensor (resp. a-tensor) product, then \mbox{$\lambda : A , A \rightarrow X$} is an s-tensor (resp. a-tensor) product iff
there is some unitary arrow \mbox{$u : A \otimes_{S} A \rightarrow X$} such that
\mbox{$\lambda =$} \mbox{$u \circ \sigma$}. 
\end{theorem}
\begin{proof}
We treat the s-tensor case, the a-tensor is treated in the same way.
We leave the proof of the {\em if} part to the reader.
Assume, now, that both $\sigma$ and $\lambda$ are s-tensor products.
Since $\sigma$ is an s-tensor product, there exists a unique 
\mbox{$i : A \otimes_{S} A\rightarrow X$}
such that \mbox{$\lambda =$} \mbox{$i \circ \sigma$}.
Since both $\sigma$ and $\lambda$ are s-tensor products, 
by condition~\ref{stensorprod-star} of
Definition~\ref{def:stensor-prod},
for any preparations 
\mbox{$a , b, a' , b' : I \rightarrow A$}, we have
\[
{\lambda}^{\star}(a , b) \circ \lambda(a' , b') \circ 2 \ =  \
{\sigma}^{\star}(a , b) \circ \sigma(a' , b') \circ 2.
\]
The scalars form a field of characteristic different from $2$, $2$ is therefore invertible and we have:
\[
{\lambda}^{\star}(a , b) \circ \lambda(a' , b') \ =  \
{\sigma}^{\star}(a , b) \circ \sigma(a' , b').
\]
Since \mbox{$\lambda =$} \mbox{$i \circ \sigma$} we see that
\[
{\sigma}^{\star}(a , b) \circ {i}^{\star} \circ i \circ \sigma(a' , b') \ =  \
{\sigma}^{\star}(a , b) \circ \sigma(a' , b').
\]
We conclude that
\[
{\sigma}^{\star}(a , b) \circ {i}^{\star} \circ i \circ \sigma \ =  \
{\sigma}^{\star}(a , b) \circ \sigma \ , \ 
{\sigma}^{\star}(a , b) \circ {i}^{\star} \circ i \ = \ 
{\sigma}^{\star}(a , b) \ , 
\]
\[
{i}^{\star} \circ i \circ \sigma(a ,b) \ = \  \sigma(a ,b) \  , \ 
{i}^{\star} \circ i \circ \sigma \ = \  \sigma , \ {\rm and} \ 
{i}^{\star} \circ i \ = \ {id}_{A \otimes_{S} A} \  ,
\]
\[
{i}^{\star} \circ \lambda \ = \ {i}^{\star} \circ i \circ \sigma \ = \ \sigma \ , \ 
i \circ {i}^{\star} \circ \lambda \ = \ i \circ \sigma \ = \ \lambda.
\]
But $\lambda$ is an s-tensor product and
\mbox{$i \circ {i}^{\star} =$} \mbox{${id}_{X}$} and $i$ is unitary.
\end{proof}
Our next result is that a tensor product is the orthogonal sum of the symmetric 
and the anti-symmetric tensor products. 
\begin{theorem} \label{the:unitary-sym}
\begin{sloppypar}
In a Q-category,
let \mbox{$\kappa : A , A \rightarrow A \otimes A$},
\mbox{$\sigma : A , A \rightarrow A {\otimes}_{s} A$} 
and \mbox{$\tau : A , A \rightarrow A {\otimes}_{a} A$}
be, respectively, tensor product, s-tensor product and a-tensor product.
Let \mbox{$p : A \otimes A \rightarrow A {\otimes}_{s} A$} be the unique arrow such that
\mbox{$p \circ \kappa =$} \mbox{$\sigma$}
and \mbox{$q : A \otimes A \rightarrow A {\otimes}_{a} A$} the unique arrow such that 
\mbox{$q \circ \kappa =$} \mbox{$\tau$}.
\end{sloppypar}
Then
\begin{equation} \label{eq:kappa-sigma-tau}
\kappa \ = \ {p}^{\star} \circ \sigma \, + \, {q}^{\star} \circ \tau,
\end{equation}
${p}^{\star}$ and ${q}^{\star}$ provide a u-coproduct and 
\mbox{$A \otimes A =$} \mbox{$A {\otimes}_{s} A \, \oplus \, A {\otimes}_{a} A$}.
\end{theorem}
\begin{proof}
We have, for every
\mbox{$a , b , a' , b' : I \rightarrow A$}:
\[
{\kappa}^{\star}(a , b) \circ {p}^{\star}  \circ \sigma(a' , b') \circ 2 \ = \ 
{\sigma}^{\star}(a , b) \circ \sigma(a' , b') \circ 2 \ = \ 
{a}^{\star} \circ a' \circ {b}^{\star} \circ b' \, + \, {a}^{\star} \circ b' \circ {b}^{\star} \circ a' \ = \ 
\]
\[
{\kappa}^{\star}(a , b) \circ \kappa(a' , b') \, + \, {\kappa}^{\star}(a , b) \circ \kappa(b' , a') \ = \ 
{\kappa}^{\star}(a , b) \circ ( \kappa(a' , b') \, + \, \kappa(b' , a')).
\]
By Lemmas~\ref{le:kappa-tensor} and~\ref{le:dense}, we conclude that
\begin{equation} \label{eq:p-sigma}
{p}^{\star}  \circ \sigma(a' , b') \circ 2 \ = \ 
\kappa(a' , b') \, + \, \kappa(b' , a').
\end{equation}
Similarly one shows that
\begin{equation} \label{eq:q-tau}
{q}^{\star} \circ \tau(a , b') \circ 2 \ = \ 
\kappa(a' , b') + {\kappa(b' , a')}^{-}.
\end{equation}
Adding Equations~\ref{eq:p-sigma} and~\ref{eq:q-tau}, using Lemma~\ref{le:zero+}, 
Theorems~\ref{the:ab-monoid}, \ref{the:field} and~\ref{the:global-} one shows that
Equation~\ref{eq:kappa-sigma-tau} holds.
Considering Equation~\ref{eq:p-sigma}, one sees that:
\[
p \circ {p}^{\star}  \circ \sigma(a' , b') \circ 2 \ = \ 
p \circ \kappa(a' , b') \, + \, p \circ \kappa(b' , a') \ = \ 
\]
\[
\sigma(a' , b') \, + \, \sigma(b' , a') \ = \ 
\sigma(a' , b') \, + \, \sigma(a' , b') \ = \ 
\sigma(a' , b') \circ 2.
\]
Therefore we have:
\[
p \circ {p}^{\star} \circ \sigma \ = \ \sigma.
\]
But $\sigma$ provides an s-tensor product and we conclude that 
\mbox{$p \circ {p}^{\star} =$} \mbox{$ {id}_{A {\otimes}_{s} A}$}, i.e., $p$ is left-unitary and
${p}^{\star}$ is right-unitary.
Similarly one shows that $q$ is left-unitary.

Considering Equation~\ref{eq:p-sigma} once more, one sees we have, 
for any \mbox{$a , b : I \rightarrow A$}:
\[
q \circ {p}^{\star} \circ \sigma(a , b) \circ 2 \ = \ 
q \circ \kappa(a , b) \, + \, q \circ \kappa(b , a) \ = \ 
\tau(a , b) + \tau(b , a) \ = \ {0}_{I , A {\otimes}_{a} A}
\]
We see that
\[
q \circ {p}^{\star} \circ \sigma(a , b) \ = \ 
{0}_{A {\otimes}_{s} A , A {\otimes}_{a} A} \circ \sigma(a , b)
\]
and therefore \mbox{$q \circ {p}^{\star} =$} \mbox{${0}_{A {\otimes}_{s} A , A {\otimes}_{a} A}$}
and \mbox{$p \circ {q}^{\star} =$} \mbox{${0}_{A {\otimes}_{a} A , A {\otimes}_{s} A}$}.

We are left to show that ${p}^{\star}$ and ${q}^{\star}$ provide a coproduct.
Let \mbox{$x : A {\otimes}_{s} A \rightarrow X$} and \mbox{$y : A {\otimes}_{a} A \rightarrow X$}.
The arrow \mbox{$z =$} \mbox{$x \circ p \, + \, y \circ q :$} \mbox{$A \otimes A \rightarrow X$}
satisfies \mbox{$z \circ {p}^{\star} =$} \mbox{$x + 0 =$} $x$ and 
\mbox{$z \circ {q}^{\star} =$} $y$.
Any arrow satisfying \mbox{$z \circ {p}^{\star} =$} $x$ and \mbox{$z \circ {q}^{\star} =$} $y$
satisfies, by Equation~\ref{eq:kappa-sigma-tau}
\mbox{$z \circ \kappa =$} \mbox{$x \circ \sigma \, + \, y \circ \tau$}. But, since $\kappa$ is a tensor
product such an arrow is unique.
\end{proof}

\subsection{Mixed states, II} \label{sec:mixed2}
The following concerns mixed states and was announced in Section~\ref{sec:mixed}.
It clears up the background of the study of two-party entanglement~\cite{Nielsen:LOCC}.
The notion of an eigenvector used below is the obvious generalization of 
Definition~\ref{def:eigen} for the case $f$ is not necessarily an arrow but may be a generalized 
quantic operation.
We assume a Q-category, but, in fact, one does not need the existence of the scalar ${1}^{-}$,
but one needs part~\ref{inverse} of Definition~\ref{def:Q-cat}.
\begin{theorem} \label{the:equal-eigen}
In a Q-category, for any 
\mbox{$c : I \rightarrow A \otimes B$}, if \mbox{$s : I \rightarrow I$}, \mbox{$s \neq 0$} 
is an eigenvalue of multiplicity $n$ for the mixed state \mbox{$d_{c}^{A}$}, 
then $s$ is also an eigenvalue of multiplicity $n$
for the mixed state \mbox{$d_{c}^{B}$}. 
\end{theorem}
\begin{proof}
Suppose \mbox{$a : I \rightarrow A$} is normalized and \mbox{$d_{c}^{A}(a) =$} 
\mbox{$a \circ s$}.
We have:
\[
{d}_{c}^{B}(x_{c}(a)) \ = \ x_{c}(y_{c}(x_{c}(a))) \ = \ x_{c}(d_{c}^{A}(a)) \ = \ 
x_{c}(a \circ s) \ = \ 
({(a \circ s)}^{\star} \otimes {id}_{B}) \circ c \ = \ 
\]
\[
({s}^{\star} \circ {a}^{\star}) \otimes ({id}_{B} \circ {id}_{B}) \circ c \ = \ 
({s}^{\star} \otimes {id}_{B}) \circ x_{c}(a) \ = \ {s}^{\star}_{B} \circ x_{c}(a) \ = \ 
x_{c}(a) \circ {s}^{\star}.
\]
But, using the ideas of the proof of Lemma~\ref{le:eigen1} and 
Corollary~\ref{co:complete-positive}, we see that:
\[
{s}^{\star} \ = \ {s}^{\star} \circ {a}^{\star} \circ a \ = \ {(a \circ s)}^{\star} \circ a \ = \ 
{({d}_{c}^{A}(a))}^{\star} \circ a \ = \ \langle {d}_{c}^{A}(a) \mid a \rangle \ = \ 
\]
\[
\langle a \mid {d}_{c}^{A}(a) \rangle \ = \ {a}^{\star} \circ a \circ s \ = \ s.
\]
But the category is normalizable, and there is a normalized \mbox{$b : I \rightarrow B$}
and a scalar $t$ such that \mbox{$x_{c}(a) =$} \mbox{$b \circ t$}.
If \mbox{$t =$} $0$, \mbox{$x_{c}(a) =$} \mbox{${0}_{I , B}$}, \mbox{$d_{c}^{A}(a) =$}
\mbox{${0}_{I , A}$}. But the category is regular and this would imply \mbox{$s =$} $0$,
a contradiction. We see that \mbox{$t \neq 0$} and therefore invertible.
One has: 
\[
{d}_{c}^{B}(b) \ = \ {d}_{c}^{B}(x_{c}(a) \circ {t}^{- 1}) \ = \ d_{c}^{B}(x_{c}(a)) \circ {t}^{- 1} \ = \ 
\]
\[
x_{c}(a) \circ s \circ t^{- 1} \ = \ x_{c}(a) \circ t^{- 1} \circ s \ = \ b \circ s.
\]
To prove our claim on multiplicity of the eigenvalue $s$, notice that, if $a_{i}$, for
\mbox{$i = 1 , 2$}, \mbox{$a_{1} \perp a_{2}$} are orthogonal eigenvectors for $d_{c}^{A}$,
then, \mbox{$d_{c}^{A}(a_{1}) \perp a_{2}$} and, by Corollary~\ref{co:complete-positive},
\mbox{$0 =$} \mbox{$\langle a_{2} \mid d_{c}^{A}(a_{1}) =$}
\mbox{$\langle x_{c}(a_{2}) \mid x_{c}(a_{1} \rangle$}.
\end{proof}
Note that the partial order among preparations of type \mbox{$I \rightarrow A \otimes B$} 
defined by LOCC operations in~\cite{Nielsen:LOCC} cannot be defined in the present 
framework since the {\em nonnegative scalars}, i.e., the scalars of the form 
\mbox{${s}^{\star} \circ s$}, need not be closed under addition or ordered.

\section{No-cloning} \label{sec:nocloning}
In~\cite{Abramsky-Coecke:catsem}, the authors show that their categorical formalism implies
{\em no-cloning}. But the no-cloning property of their formalism is a global property: i.e., there is
no global uniform cloning method. The no-cloning property considered by, for example,
\cite{Wootters-Zurek:82} or~\cite{Dieks:82}, is a local property. We shall prove here that 
this last (local) property holds in Q-categories.
We assume a Q-category, and study tensor products of arrows.
We need to study equations such as: 
\mbox{$(a_{1} \otimes b_{1}) + (a_{2} \otimes b_{2}) \ = \  a_{3} \otimes b_{3}$}.
This will be done in Theorem~\ref{the:special-super}.
For this we need to study equations such as
\mbox{$a_{1} \otimes b_{1}  = a_{2} \otimes b_{2}$}, and, before that equations such as:
\mbox{$a \otimes b  = {0}_{I , A}$}.
Our first result is that, if the tensor product of two arrows is null, one of the arrows must be null.
\begin{lemma} \label{le:zero-tens-iff}
In a Q-category, 
for any \mbox{$f : A' \rightarrow A$}, \mbox{$g : B' \rightarrow B$}, one has
\mbox{$f \otimes g =$} \mbox{${0}_{A' \otimes B' , A \otimes B}$}
 iff either \mbox{$f =$} \mbox{${0}_{A' , A}$} or \mbox{$g =$} \mbox{${0}_{B' , B}$}.
\end{lemma}
\begin{proof}
For the {\em if} part, see Lemma~\ref{le:zero-tens}.
For the {\em only if} part, assume \mbox{$f \otimes g =$}
\mbox{${0}_{A' \otimes B' , A \otimes B}$}.
For any \mbox{$a : I \rightarrow A'$} and \mbox{$b : I \rightarrow B'$} one has, 
by Lemma~\ref{le:otimes-comp}:
\[
(f \circ a) \otimes (g \circ b) \ = \ 
(f \otimes g) \circ (a \otimes b) \ = \ {0}_{I , A \otimes B}.
\]
Therefore
\[
({(f \circ a)}^{\star} \otimes {(g \circ b)}^{\star}) \circ ((f \circ a) \otimes (g \circ b)) \ = \ 0
\]
and
\[
({(f \circ a)}^{\star} \circ (f \circ a)) \otimes ({(g \circ b)}^{\star} \circ (g \circ b)) \ = \ 0.
\]
But this is a tensor product of scalars and, by Corollary~\ref{co:tensor-scalar}
\[
{(f \circ a)}^{\star} \circ (f \circ a) \circ {(g \circ b)}^{\star} \circ (g \circ b) \ = \ 0.
\]
If \mbox{$f \neq {0}_{A' , A}$} there is, by property~\ref{prep-arrows} of 
Definition~\ref{def:unit-object}, some arrow $a$ such that \mbox{$f \circ a \neq {0}_{I , A}$}.
By Lemma~\ref{le:fstarfzero}, 
\mbox{${(f \circ a)}^{\star} \circ (f \circ a) \neq 0$} and therefore it is invertible.
Therefore we have
\mbox{${(g \circ b)}^{\star} \circ (g \circ b) = 0$} and, by Lemma~\ref{le:fstarfzero},
\mbox{$g \circ b = {0}_{I , B}$} for any preparation $b$.
We conclude that \mbox{$g = {0}_{B' , B}$}.
\end{proof}
The arrows \mbox{$(f \circ s) \otimes g$} and \mbox{$f \otimes (g \circ s)$} are the same. 
We shall now prove that an arrow that is the tensor product of two arrows may be uniquely 
decomposed in this manner up to a scalar.
\begin{lemma} \label{le:Itens=}
In a Q-category,
let \mbox{$a, a' : I \rightarrow A$} and \mbox{$b , b' : I \rightarrow B$} be such that
\mbox{$a \otimes b \neq$} \mbox{$0_{I , A \otimes B}$}.
If \mbox{$a' \otimes b' =$} \mbox{$a \otimes b$}, there exists some scalar \mbox{$s \neq 0$}
such that \mbox{$a' = a \circ s$} and \mbox{$b = b' \circ s$}.
\end{lemma}
\begin{proof}
Assuming the assumptions of the claim,
by Lemma~\ref{le:zero-tens-iff} none of $a$, $b$, $a'$ and $b'$ is null.
We have:
\[
({a}^{\star} \otimes {id}_{B}) \circ (a' \otimes b') \ = \ ({a}^{\star} \otimes {id}_{B}) \circ (a \otimes b)
\]
and therefore
\[
({a}^{\star} \circ a') \otimes b' \ = \ ({a}^{\star} \circ a) \otimes b.
\]
Let \mbox{$r = {a}^{\star} \circ a'$} and \mbox{$t = {a}^{\star} \circ a$}.
By Lemma~\ref{le:fstarfzero}, \mbox{$t \neq 0$} and therefore invertible.
By Lemma~\ref{le:global}, \mbox{$b' \circ r =$} \mbox{$b \circ t$} and
\mbox{$b =$} \mbox{$b' \circ r \circ {t}^{-1}$}.
Let \mbox{$s =$} \mbox{$r \circ {t}^{-1}$}. We see that \mbox{$s \neq 0$} since \mbox{$b \neq 0$}.
We have \mbox{$a' \otimes b' =$} \mbox{$a \otimes b =$} \mbox{$(a \otimes b') \circ s =$}
\mbox{$(a \circ s) \otimes b'$}.
Therefore we have
\[
a' \circ {b'}^{\star} \circ b' \ = \ a' \otimes ({b'}^{\star} \circ b') \ = \ 
({id}_{A} \otimes {b'}^{\star}) \circ (a' \otimes b') \ = \ 
({id}_{A} \otimes {b'}^{\star}) \circ (a \otimes b) \ = \ 
\]
\[
({id}_{A} \otimes {b'}^{\star}) \circ (a \otimes (b' \circ s)) \ = \ 
a \circ {b'}^{\star} \circ b' \circ s \ = \ 
a \circ s \circ {b'}^{\star} \circ b'.
\]
But \mbox{$b' \neq 0$} and, by Lemma~\ref{le:fstarfzero}, 
\mbox{${b'}^{\star} \circ b' \neq 0$} and therefore is invertible. We conclude that
\mbox{$a' = a \circ s$}.
\end{proof}
We can now generalize Lemma~\ref{le:Itens=}.
\begin{theorem} \label{the:tens=}
In a Q-category,
let \mbox{$f, f' : A' \rightarrow A$} and \mbox{$g , g' : B' \rightarrow B$} such that
\mbox{$f \otimes g \neq$} \mbox{$0_{A' \otimes B' , A \otimes B}$}.
Then, \mbox{$f' \otimes g' =$} \mbox{$f \otimes g$} iff there is some scalar \mbox{$s \neq 0$}
such that \mbox{ $f' =$} \mbox{$f \circ {s}_{A'}$} and \mbox{$g =$} \mbox{$g' \circ {s}_{B'}$}.
\end{theorem}
\begin{proof}
For the {\em if} part: assume that 
\mbox{$f' =$} \mbox{$f \circ s$} and \mbox{$g =$} \mbox{$g' \circ s$}.
We have \mbox{$f' \otimes g' =$} \mbox{$(f \otimes g') \circ s$} and
\mbox{$f \otimes g =$} \mbox{$(f \circ g') \circ s$}.

For the {\em only if} part: assume  \mbox{$f' \otimes g' =$} \mbox{$f \otimes g$} .
For any preparations \mbox{$a : I \rightarrow A'$} and \mbox{$b : I \rightarrow B'$} we have:
\mbox{$(f' \circ a) \otimes (g' \circ b) =$}
\mbox{$(f' \otimes g') \circ (a \otimes b) =$}
\mbox{$(f \otimes g) \circ (a \otimes b) =$}
\mbox{$(f \circ a) \otimes (g \circ a)$}.
By Lemma~\ref{le:Itens=}, there is some scalar \mbox{$s \neq 0$} such that
\mbox{$f' \circ a =$} \mbox{$f \circ a \circ s$} and \mbox{$g \circ b =$} \mbox{$g' \circ b \circ s$}.
Therefore, by Definition~\ref{def:unit-object},
\mbox{$f' \circ a =$} \mbox{$f \circ {s}_{A'} \circ a$}, \mbox{$g \circ b =$} 
\mbox{$g' \circ {s}_{B'} \circ b$}, \mbox{$f' =$} \mbox{$f \circ {s}_{A'}$} and
\mbox{$g =$} \mbox{$g' \circ {s}_{B'}$}.
\end{proof}
Our next result is fundamental: in a Q-category, a superposition of product preparations
can itself be a product preparation only in a trivial way. 
\begin{theorem} \label{the:special-super}
In a Q-category,
let, for \mbox{$i = 1 , 2 , 3$}, \mbox{$a_{i} : $} \mbox{$I \rightarrow A$} and 
\mbox{$b_{i} :$} \mbox{$I \rightarrow B$} be preparations such that
\mbox{$a_{1} \otimes b_{1} \neq$} \mbox{$0_{I , A \otimes B}$} and
\begin{equation} \label{eq:sum}
(a_{1} \otimes b_{1}) + (a_{2} \otimes b_{2}) \ = \  a_{3} \otimes b_{3}. 
\end{equation}
Then, there is some scalar $s$ such that 
\begin{itemize}
\item either \mbox{$a_{2} = a_{1} \circ s$} and \mbox{$a_{3} \otimes b_{3} =$}
\mbox{$a_{1} \otimes (b_{1} + b_{2} \circ s)$}
\item or \mbox{$b_{2} =$} \mbox{$b_{1} \circ s$} and \mbox{$a_{3} \otimes b_{3} =$}
\mbox{$(a_{1} + a_{2} \circ s) \otimes b_{1}$}.
\end{itemize}
\end{theorem}
\begin{proof}
We have:
\[
{(a_{1} \otimes {id}_{B})}^{\star} \circ (a_{1} \otimes b_{1} + a_{2} \otimes b_{2}) \ = \ 
{(a_{1} \otimes {id}_{B})}^{\star} \circ (a_{3} \otimes b_{3})
\]
and, by Lemmas~\ref{le:+comp}, \ref{le:star-tens} and~\ref{le:otimes-comp}
\[
(a_{1}^{\star} \circ a_{1}) \otimes b_{1} + (a_{1}^{\star} \circ a_{2})  \otimes b_{2} \ = \ 
(a_{1}^{\star} \circ a_{3}) \otimes b_{3} 
\]
and, by Lemma~\ref{le:global}
\[
b_{1} \circ (a_{1}^{\star} \circ a_{1}) + b_{2} \circ (a_{1}^{\star} \circ a_{2})  \ = \ 
b_{3} \circ (a_{1}^{\star} \circ a_{3}).
\]
But \mbox{$a_{1} \otimes b_{1} \neq$} \mbox{$0_{I , A \otimes B}$} and, 
by Lemmas~\ref{le:zero-star} and~\ref{le:fstarfzero}, 
\mbox{$a_{1}^{\star} \circ a_{1}$} is not null and invertible.
Therefore, we have:
\begin{equation} \label{eq:b1}
b_{1} \ = \ (b_{3} \circ (a_{1}^{\star} \circ a_{3})) + {(b_{2} \circ (a_{1}^{\star} \circ a_{2}))}^{-})
\circ {(a_{1}^{\star} \circ a_{1})}^{- 1}.
\end{equation}
Let \mbox{$s_{i} =$} \mbox{$a_{1}^{\star} \circ a_{i}$}, for \mbox{$i = 1 , 2 , 3$} and 
\mbox{$t =$} \mbox{$a_{1}^{\star} \circ a_{1}$}.
Substituting for $b_{1}$ in Equation~\ref{eq:sum}, 
using Theorem~\ref{the:global-} and reorganizing, one obtains:
\[
(a_{1} \circ s_{1}) \otimes b_{3} +  {((a_{1} \circ s_{2} \circ t^{- 1}) \otimes b_{2})}^{-} 
+ a_{2} \otimes b_{2} \ = \ a_{3} \otimes b_{3}
\]
and
\begin{equation} \label{eq:ab2}
(({a_{1} \circ s_{2} \circ t^{- 1})}^{-} + a_{2}) \otimes b_{2} \ = \ 
(a_{3} + {(a_{1} \circ s_{1})}^{-}) \otimes b_{3}
\end{equation}
We may now apply Lemma~\ref{le:Itens=}.

Assume, first, that both sides of Equation~\ref{eq:ab2} are not null.
There is a scalar \mbox{$r \neq 0$} such that 
\mbox{$b_{3} =$} \mbox{$b_{2} \circ r$}. 
Letting \mbox{$w =$} \mbox{$r \circ s_{3} + s_{2}^{-}$} and using Equation~\ref{eq:b1} 
we see that \mbox{$b_{1} =$} \mbox{$b_{2} \circ w \circ s_{1}^{- 1}$}.
But \mbox{$b_{1} \neq 0$} and therefore \mbox{$w \neq 0$} and we have, as claimed:
\mbox{$b_{2} =$} \mbox{$b_{1} \circ w^{- 1} \circ s_{1}$}.
Assume, now that both sides of Equation~\ref{eq:ab2} are null. By Lemma~\ref{le:zero-tens-iff},
either \mbox{$a_{2} =$} \mbox{$a_{1} \circ s_{2} \circ t^{- 1}$} or \mbox{$b_{2} =$}
\mbox{$0 =$} \mbox{$b_{1} \circ 0$}.

For the last claim, suppose, without loss of generality, 
that \mbox{$a_{2} =$} \mbox{$a_{1} \circ s$}.
Then: 
\[
a_{3} \otimes b_{3} \ = \ 
a_{1} \otimes b_{1} + (a_{1} \circ s) \otimes b_{2} \ = \ 
a_{1} \otimes b_{1} + a_{1} \otimes (b_{2} \circ s) \ = \ 
a_{1} \otimes (b_{1} + b_{2} \circ s).
\] 
\end{proof}
\begin{theorem} \label{the:recap}
In a Q-category, let \mbox{$a , b , c , d :$} \mbox{$I \rightarrow A$} and 
\mbox{$w_{1}, w_{2} , w_{3} :$} \mbox{$I \rightarrow W$}.
Assume \mbox{$a , c , d \neq 0$}, \mbox{$w_{1} , w_{2} , w_{3} \neq 0$} and
\mbox{$(a \otimes a) \otimes w_{1} + (b \otimes b) \otimes w_{2} =$}
\mbox{$(c \otimes d) \otimes w_{3}$}. 
Then, there is some scalar $s$ such that \mbox{$b =$} \mbox{$a \circ s$}.
\end{theorem} 
\begin{proof}
By Theorem~\ref{the:special-super}, there is some scalar $s$ such as either
\begin{itemize}
\item \mbox{$b \otimes b =$} \mbox{$(a \otimes a) \circ s$} or 
\item \mbox{$w_{2} =$} \mbox{$w_{1} \circ s$} and 
\mbox{$((a \otimes a) + (b \otimes b) \circ s) \otimes w_{1}  =$}
\mbox{$(c \otimes c) \otimes w_{3}$}.
\end{itemize}
In the first case, we have \mbox{$a \otimes a =$} \mbox{$b \otimes (b \circ s)$} and we conclude immediately, by Lemma~\ref{le:Itens=} that there is some scalar $t$ such that 
\mbox{$b =$} \mbox{$a \circ t$}.
In the second case, note that \mbox{$s \neq 0$} since \mbox{$w_{2} \neq 0$} and
apply Lemma~\ref{le:Itens=} to conclude that there is some scalar 
\mbox{$t \neq 0$} such that 
\[
(a \otimes a) \circ t + (b \otimes b) \circ s \circ t \ = \ 
a \otimes (a \circ t) + b \otimes (b \circ s \circ t) \ = \ 
c \otimes c.
\]
By Theorem~\ref{the:special-super}, now, there is some scalar $w$ such that either
\mbox{$b =$} \mbox{$a \circ w$} or \mbox{$b \circ s \circ t =$} \mbox{$a \circ t \circ w$}.
In the first case, we are through and in the second case, note that $s$ and $t$ are not null and therefore invertible and we have \mbox{$b =$} \mbox{$a \circ w \circ {s}^{- 1}$}.
\end{proof}
We shall now define the notion of cloning and prove that, in a Q-category,
only unitary preparations can be cloned, i.e., only states of a unit object can be cloned.
The property called cloning here differs from~\cite{Abramsky-Coecke:catsem}'s since they
consider cloning to be a global property and we define it as a local property.
Our definition is essentially the definition found in~\cite{Wootters-Zurek:82} 
or~\cite{Dieks:82}, but the proof is significantly different and more general.
\begin{definition} \label{def:cloning}
Assume a Q-category.
An object $A$ can be cloned iff there is an object $W$, 
an arrow \mbox{$c : A \otimes W \rightarrow (A \otimes A) \otimes W$} and
a preparation \mbox{$w :$} \mbox{$I \rightarrow W$}
such that for any normalized preparation \mbox{$a : I \rightarrow A$} there is a preparation 
\mbox{$w_{a} : I \rightarrow W$}, \mbox{${w}_{a} \neq 0$} such that
\mbox{$c \circ (a \otimes w) =$} \mbox{$(a \otimes a) \otimes w_{a}$}.
\end{definition}
The no-cloning theorem is the following.
\begin{theorem}[No cloning] \label{the:nocloning}
Assume a Q-category.
An object $A$ can be cloned iff it is a unit object.
\end{theorem}
\begin{proof}
Assume $A$ can be cloned.
For any normalized preparation \mbox{$a : I \rightarrow A$} we have
\mbox{$c \circ (a \otimes w) =$} \mbox{$(a \otimes a) \otimes {w}_{a}$}.
Let \mbox{$a , b : I \rightarrow A$} be any two normalized preparations.
We have
\[
(a \otimes a) \otimes {w}_{a} + (b \otimes b) \otimes {w}_{b} \ = \ 
c \circ (a \otimes w) + c \circ (b \otimes w) \ = \ 
\]
\[
c \circ (a \otimes w \: + \: b \otimes w) \ = \ 
c \circ ((a + b) \otimes w).
\]
Since there is a scalar $t$ and
a normalized preparation \mbox{$n : I \rightarrow A$} such that
\mbox{$a + b =$} \mbox{$n \circ t$}.
We have:
\[
c \circ ((a + b) \otimes w) \ = \
c \circ (n \otimes w) \circ (t \otimes {id}_{W}) \ = \ 
\]
\[
((n \otimes n) \otimes w_{n}) \circ (t \otimes {id}_{W}) \ = \ 
(n \otimes (n \circ t)) \otimes w_{n}.
\]
Assume, for the moment, that \mbox{$t \neq 0$}.
We have \mbox{$n \otimes (n \circ t) \neq 0$} and we can apply Theorem~\ref{the:recap}:
there is some scalar $s$ such that \mbox{$b = a \circ s$}.
If, on the contrary, \mbox{$t = 0$}, we have \mbox{$a + b =$} $0$ and
\mbox{$b =$} \mbox{${a}^{-} =$} \mbox{$a \circ {1}^{-}$}.
We have shown that for any two normalized preparations \mbox{$a , b : I \rightarrow A$}
there is some scalar $s$ such that \mbox{$b =$} \mbox{$a \circ s$}.
By Theorem~\ref{the:characterization} it follows that $A$ is a unit object (and any such preparations are unitary).

We must now show that any unit object is clonable.
Let us show, first, that $I$ is clonable. Take any object $W$, 
we have \mbox{$W =$} \mbox{$I \otimes W =$} \mbox{$(I \otimes I) \otimes W$}.
Let \mbox{$c =$} ${id}_W$, let \mbox{$w : I \rightarrow W$} be any normalized preparation and
\mbox{$w_{s} =$} \mbox{$w \circ {s}^{- 1}$} for any scalar \mbox{$s \neq 0$} and take
\mbox{$w_{0} =$} $0$.
Indeed one has \mbox{$s \otimes w =$} \mbox{$w \circ s =$}
\mbox{$w_{s} \circ s \circ s =$}
\mbox{$(s \circ s) \otimes w_{s} =$} \mbox{$(s \otimes s) \otimes w_{s}$}.
If now $A$ is any unit object, then any normalized preparation \mbox{$a : I \rightarrow A$} 
is unitary by Theorem~\ref{the:characterization}.
Take  \mbox{$c =$} ${id}_W$, let \mbox{$w :$} \mbox{$I \rightarrow W$} 
be any normalized preparation.
We define \mbox{$w_{a} =$} \mbox{$w \circ {a}^{\star}$}.
Indeed we have: \mbox{$a \otimes w =$}
\mbox{$w \circ a =$} \mbox{$w \circ a \circ a \circ {a}^{\star} =$}
\mbox{$(a \otimes a) \otimes (w \circ {a}^{\star})$}.
\end{proof}

\section{Summary and future work} \label{sec:further}
This paper has developed a framework for categorical QM that is more concrete, i.e.,
less abstract, than Abramsky and Coecke's. Its main technical innovations are the consideration
of categories equipped with an adjoint structure that defines unitary arrows and of
universal characterizations up to a unitary arrow, and a description 
of tensor product by such a universal characterization. It does not require compact closure.
For theoretical physicists, it provides descriptions of basic notions, such as, adjoints, normalized
states, unitary and self-adjoint transformations, scalars, scalar products, tensor products and
mixed states that are purely multiplicative and base-free. The presentation of mixed states
as a composition of two antilinear transformations seems original and requires more study.
The assumptions made in this paper on the structure of scalars are weak enough to enable
the consideration of models with scalars significantly different from the complex or real numbers.
Those assumptions are nevertheless sufficient to ensure no-cloning.

We conclude by a list of remarks and questions that may provide food for further work.
Hilbert spaces possess many properties that are not enjoyed by the categories presented in
this work. It is not clear which of those, if any, are necessary for QM. In particular, the categories
presented here may not enjoy the following {\em complementation} property: for any right-unitary
arrow \mbox{$u : A \rightarrow X$} that is not unitary, there is an object $B$ and a right-unitary
\mbox{$v : B \rightarrow X$} such that \mbox{$u , v$} form a u-coproduct.
\begin{itemize}
\item Can classical systems be described in the framework of A-categories? Do categories in which self-adjoint arrows commute describe classical systems?
\item Do the universal characterizations of symmetric and anti-symmetric tensor products of 
Section~\ref{sec:tensor-sym} shed any light on the fermion-bosons distinction?
\item Can the present characterization of tensor products be leveraged to gain information about
the structure of multi-party entanglement?
\end{itemize}

\section*{Acknowledgements}
During the elaboration of this work I had helpful conversations or exchanges 
with S. Abramsky, D. Aharonov, B. Coecke, C. Heunen, J. Vicary and P. Selinger for which I am very grateful.
\bibliographystyle{plain}

\appendix
\section{Examples of A-categories} \label{sec:examples1}
\subsection{Commutative monoids} \label{sec:monoid1}
Let \mbox{$\langle M , + , 0 \rangle$} be any commutative monoid. The elements of
the monoid can be considered to be the arrows of a category with only one object.
Composition is the monoid operation and $0$ is the identity.
One may always define an adjoint structure by defining \mbox{${a}^{\star} =$} $a$ for
every element $a$ of the monoid. Every arrow is self-adjoint. 
Commutativity is required by the second part of 
Equation~\ref{eq:adj}. Unitary arrows are the elements $a$ such that 
\mbox{$a + a =$} $0$.

\subsection{Groups} \label{sec:groups1}
Let \mbox{$\langle G , \circ , 1 \rangle$} be any (not necessarily commutative) group.
It defines a one-object category on which one may define an adjoint structure by
\mbox{${a}^{\star} =$} \mbox{${a}^{-1}$}. An arrow $a$ is self-adjoint iff 
\mbox{$a \circ a =$} $1$. Every arrow is unitary.
Considering Section~\ref{sec:monoid1}, one sees that, on a {\em commutative} group, 
one may define two different adjoint structures. They define different A-categories.
In the sequel, when we mention groups we shall mean the adjoint structure defined by inverses.

\subsection{Relations} \label{sec:relations1}
The category $\cal R$ has, for objects, all {\em non-empty} sets. If $A$ and $B$ are non-empty
sets, any binary relation \mbox{$r \subseteq A \times B$} is an arrow 
\mbox{$r : A \rightarrow B$}.
The adjoint of a relation is its inverse: \mbox{${r}^{\star}(b , a)$} iff
\mbox{$r(a , b)$}. Right-unitary arrows are those relations satisfying 
\mbox{$r(a , b)$} and \mbox{$r(a' , b)$} imply \mbox{$a = a'$}.
Left-unitary arrows are those relations satisfying \mbox{$r(a , b)$} and
\mbox{$r(a , b')$} imply \mbox{$b = b'$} and unitary arrows are partial injective functions.
Self-adjoint arrows are symmetric relations.
The category $\cal R$ is subsumed by the categories described below 
in Section~\ref{sec:matrices}.
\subsection{Hilbert spaces} \label{sec:Hilbert1}
The category $\cal H$ has, for objects, 
all non-zero dimensional Hilbert spaces on the complex field, and,
for arrows, linear operators. The adjoint of an arrow is defined as the adjoint operator.
The terms {\em self-adjoint} and {\em unitary} have their usual meaning.

\subsection{Matrices on a commutative semiring} \label{sec:matrices}
\subsubsection{The categorical structure} \label{sec:matrices-cat}
Let \mbox{$R =$} \mbox{$\langle M , + , 0 , \cdot , 1 \rangle$} be a {\em commutative} semiring.
Important examples of commutative semirings are:
\begin{itemize} 
\item commutative rings and, in particular, fields, and 
\item lattices where $+$ is l.u.b., $0$ is $\bot$, $\cdot$ is g.l.b. and $1$ is $\top$, and, 
in particular, complete lattices.
\end{itemize}
We define the category $\cC(R)$:
\begin{itemize}
\item
its objects are the {\em finite, non-empty} sets,
\item
for any finite sets $A$, $B$, the arrows of \mbox{$Hom(A , B)$} are the functions
from the product set \mbox{$A \times B$} to the set $M$.
\end{itemize}
If \mbox{$f : A \rightarrow B$} and \mbox{$g : B \rightarrow C$}, the arrow
\mbox{$g \circ f : A \rightarrow C$} is defined, for any \mbox{$a \in A$}, \mbox{$c \in C$}, by:
\begin{equation}
(g \circ f)(a , c) \ = \ \sum_{b \in B} g(b , c) \cdot f(a , b)
\end{equation}
The identity on a finite set $A$, \mbox{${id}_{A}$} is defined, for any \mbox{$a , a' \in A$}, by:
\begin{equation}
{id}_{A}(a , a') \ = \ \left\{ \begin{array}{l} 
1 \ {\rm if} \ a = a' \\ 0 \ {\rm otherwise}
\end{array} \right.
\end{equation}
We have restricted the class of objects to finite sets because we need sums of the type
\mbox{$\sum_{a \in A} m_{a}$}. In case there is a satisfactory notion of infinite sums of elements
of $M$, for example if $R$ is a complete lattice, one may extend the class of objects to arbitrary
sets. For example, if one takes $R$ to be the two elements lattice, then one obtains the category
of sets and relations, i.e., the category {\bf {\cal R}el} considered in Section~\ref{sec:relations1}.

One easily verifies that composition is associative.
Let, indeed, \mbox{$f : A \rightarrow B$}, \mbox{$g : B \rightarrow C$} and 
\mbox{$h : C \rightarrow D$}. For any \mbox{$a \in A$} and \mbox{$d \in D$} one has:
\[
((h \circ g) \circ f)(a , d) \ = \ \sum_{b \in B} (h \circ g)(b , d) \cdot f(a , b) \ = \ 
\]
\[
\sum_{b \in B} (\sum_{c \in C} h(c , d) \cdot g(b , c)) \cdot f(a , b) \ = \ 
\sum_{b \in B , c \in C} h(c , d) \cdot g(b , c) \cdot f(a , b)
\]
and similarly:
\[
(h \circ (g \circ f))(a , d) \ = \ 
\sum_{c \in C , b \in B} h(c , d) \cdot g(b , c) \cdot f(a , b).
\]
Note that the commutativity of $\cdot$ is not needed here.

One also verifies easily that ${id}_{A}$ has the properties needed.
For any \mbox{$f : A \rightarrow B$}, any \mbox{$a \in A$} and \mbox{$b \in B$}:
\[
({id}_{B} \circ f)(a , b) \ = \ 
\sum_{x \in B} {id}_{B}(x , b) \cdot f(a , x) \ = \ f(a , b)
\]
and similarly
\[
(f \circ {id}_{A})(a , b) \ = \ 
\sum_{x \in A} f(x , b) \cdot {id}_{A}(a , x) \ = \ f(a , b).
\]

\subsubsection{The adjoint structure} \label{sec:matrices-adj}
We shall now make $\cC(R)$ an A-category.
For any \mbox{$f : A \rightarrow B$} we define \mbox{${f}^{\star} : B \rightarrow A$}
by \mbox{${f}^{\star}(b , a) =$} \mbox{$f(a , b)$} for any \mbox{$a \in A$}, \mbox{$b \in B$}.
One easily checks that \mbox{${({f}^{\star})}^{\star} =$} $f$.
If the commutative semiring $R$ has an involution 
\mbox{$ \overline{\ } : M \rightarrow M$} such that, for any \mbox{$m , n \in M$} one has:
\mbox{${\overline{\overline{m}}} =$} $m$, \mbox{$\overline{m + n} =$} 
\mbox{$\overline{m} + \overline{n}$} and \mbox{$\overline{m \cdot n} =$}
\mbox{$\overline{m} \cdot \overline{n}$}
\begin{comment}
One cannot require \mbox{$\overline{m \cdot n} =$}
\mbox{$\overline{n} \cdot \overline{m}$} and do away with the requirement the semiring is commutative
\end{comment}
one could define \mbox{${f}^{\star} : B \rightarrow A$}
by \mbox{${f}^{\star}(b , a) =$} \mbox{$\overline{f(a , b)}$}. In the sequel we shall assume the
involution is the identity, but, mutatis mutandis, the results hold for any involution defining
the adjoint structure.

We want to show that, for any \mbox{$f : A \rightarrow B$}, \mbox{$g : B \rightarrow C$}
one has: \mbox{${(g \circ f)}^{\star} =$} \mbox{${f}^{\star} \circ {g}^{\star}$}.
Indeed, for any \mbox{$a \in B$}, \mbox{$c \in C$}, one has:
\[
{(g \circ f)}^{\star}(c , a) \ = \ (g \circ f)(a , c) \ = \ 
\sum_{b \in B} g(b , c) \cdot f(a , b)
\]
and
\[
({f}^{\star} \circ {g}^{\star})(c , a) \ = \ 
\sum_{b \in B} {f}^{\star}(b , a) \cdot {g}^{\star}(c , b) \ = \ 
\sum_{b \in B} f(a , b) \cdot g(b , c) \ = \ 
\sum_{b \in B} g(b , c) \cdot f(a , c).
\]
Note that, here, it is essential that $\cdot$ is commutative.

\subsubsection{Unitary arrows} \label{sec:matrices-unitary}
Left-unitary, right-unitary and unitary arrows are hard to characterize in general, as will be obvious
from the following example.

Suppose $R$ is the set of real numbers with addition and multiplication.
Let \mbox{$A =$} \mbox{$\{a , b\}$} be a set of two elements. We shall define an arrow
\mbox{$u : A \rightarrow A$} by
\[
u(a , a) \ = \ u(b , b) \ = \ u(a , b) \ = \ {1 \over \sqrt{2}} \ \ \ \ u(b , a) \ = \ - {1\over \sqrt{2}}.
\] 
 By inspection, one shows that the arrow $u$ is unitary.
 But we can point at a family of arrows for which the characterization of unitary arrows is easier. 
The structure $R$ is now any commutative semiring.
Let $x$ be a function ({\em not} an arrow) from $A$ to $B$.
We can define the arrow \mbox{${\chi}_{x} : A \rightarrow B$} by
\begin{equation} \label{eq:chi}
{\chi}_{x}(a , b) \ = \ \left\{ \begin{array}{l}
1 \ {\rm if} \ b = x(a) \\ 0 \ {\rm otherwise}
\end{array} \right. 
\end{equation}
Note that, if $i$ is the identity function on $A$, \mbox{${id}_{A} =$} \mbox{${\chi}_{i}$}.
\begin{lemma} \label{le:unitary}
The arrow \mbox{${\chi}_{x} : A \rightarrow B$} is
\begin{enumerate}
\item \label{runitary} right-unitary iff $x$ is injective
\item \label{bi-unitary} unitary iff $x$ is bijective.
\end{enumerate}
\end{lemma}
\begin{proof}
\begin{enumerate}
\item 
Note that we have, by Equation~\ref{eq:chi}, for any \mbox{$a , a' \in A$}:
\[
({\chi}^{\star}_{x} \circ {\chi}_{x})(a , a') \ = \  \sum_{b \in B} {\chi}^{\star}_{x}(b , a') \cdot {\chi}_{x}(a , b) \ = \ 
\sum_{b \in B} {\chi}^{\star}_{x}(x(a) , a') \ = \ {\chi}_{x}(a' , x(a)).
\]
First, notice that \mbox{$({\chi}^{\star}_{x} \circ {\chi}_{x})(a , a) =$} \mbox{$1=$} \mbox{${id}_{A}(a , a)$}.
Suppose, now, that $x$ is injective and \mbox{$a' \neq a$}. We have \mbox{$x(a') \neq x(a)$} and 
\mbox{$({\chi}^{\star}_{x} \circ {\chi}_{x})(a , a') =$} \mbox{${\chi}_{x}(a' , x(a)) =$} \mbox{$0 =$}
\mbox{${id}_{A}(a , a')$}. Using the previous remark, we conclude that ${\chi}_{x}$ is right-unitary.
 Conversely, if $x$ is not injective, there are \mbox{$a , a' \in A$}, \mbox{$a' \neq a$} such that 
 \mbox{$x(a') =$} $x(a)$ and \mbox{$({\chi}^{\star}_{x} \circ {\chi}_{x})(a , a') =$}
 \mbox{${\chi}_{x}(a' , x(a)) =$} \mbox{$1 \neq 0 =$} \mbox{${id}_{A}(a , a')$} and we conclude
 that  ${\chi}_{x}$ is not right-unitary.
\item
We have, by Equation~\ref{eq:chi}:
\[
({\chi}_{x} \circ {\chi}^{\star}_{x})(b , b') \ = \ \sum_{a \in A} {\chi}_{x}(a , b') \cdot {\chi}^{\star}_{x}(b , a) \ = \
\]
\[
\sum_{a \in A} {\chi}_{x}(a , b') \cdot {\chi}_{x}(a , b) \ = \
\sum_{a \in A , x(a) = b} {\chi}_{x}(a , b').
\]
If \mbox{$b' \neq b$}, we have \mbox{$({k}_{x} \circ {k}^{\star}_{x})(b , b') =$} 
\mbox{$\sum_{a \in A , x(a) = b} 0 =$} \mbox{$0 =$}
\mbox{${id}_{B}(b , b')$}.

If $x$ is bijective, for any \mbox{$b \in B$} there is a unique element \mbox{$a_{b} \in A$} such that
\mbox{$x(a_{b}) =$} $b$ and we have \mbox{$({k}_{x} \circ {k}^{\star}_{x})(b , b) =$} 
\mbox{$\sum_{a \in A , a = a_{b}} \chi_{x}(a , b) =$} \mbox{${\chi}_{x}(a_{b} , b) =$} \mbox{$1 =$}
\mbox{${id}_{B}(b , b)$} and we conclude that \mbox{${\chi}_{x} \circ {\chi}^{\star}_{x} =$}
\mbox{${id}_{B}$} and therefore ${\chi}_{x}$ is left-unitary. By item~\ref{runitary}, since $x$ is injective, $\chi_{x}$ is right-unitary and therefore unitary.

Conversely assume $\chi_{x}$ is unitary. Then $x$ is injective by item~\ref{runitary}.
If $x$ were not bijective there would be some \mbox{$b_{0} \in B$} that is the image
of no element of $A$ by $x$. We would have
\mbox{$({\chi}_{x} \circ {\chi}^{\star}_{x})(b_{0} , b_{0}) =$}
\mbox{$\sum_{a \in A , x(a) = b_{0}} \chi_{x}(a , b_{0} =$} $0$.
This is a contradiction to the left-unitary character of $\chi_{x}$.
We conclude that $x$ is bijective.
\end{enumerate}
\end{proof}

\subsubsection{Self-adjoint arrows}
Self-adjoint arrows \mbox{$f : A \rightarrow A$} are exactly those functions 
\mbox{$x : A \times A \rightarrow M$} such that
\mbox{$x(a , a') =$} \mbox{$x(a', a)$} for any \mbox{$a , a' \in A$}.

\section{Examples of unit objects} \label{sec:examples2}
\subsection{Commutative monoids} \label{sec:monoid2}
In any category with a unique object $I$, the object $I$ is easily seen to satisfy 
conditions~\ref{atleast} and~\ref{prep-arrows} of Definition~\ref{def:unit-object}
(Hint: consider the identity $0$).
The unique object $I$, of an A-category defined by a commutative monoid as in 
Section~\ref{sec:monoid1} is therefore a unit object since,
taking \mbox{$s_{I} = s$}, one has \mbox{$t + s =$} \mbox{${s}_{I} + t$} by commutativity.
\subsection{Groups}
The unique object $I$, of an A-category defined by a group as in 
Section~\ref{sec:groups1} is a unit object only if the group is commutative,
by Theorem~\ref{the:scalar-com}. The unique object $I$ of a commutative group is easily seen
to be a unit object.
\subsection{Relations}
In $\cal R$ any singleton is a unit object, but note it is crucial here that we exclude the
empty set from the objects of $\cal R$, in order that condition~\ref{atleast} of 
Definition~\ref{def:unit-object} be satisfied. 
There are exactly two scalars: the empty relation and the identity relation.
A preparation \mbox{$I \rightarrow A$} is a binary relation on \mbox{$I \times A$},
i.e., a subset of $A$. Any preparation is normalized.

\subsection{Hilbert spaces}
In $\cal H$ any one-dimensional space (isomorphic to the complex field) is a unit
object. Again it is crucial that we have removed the zero-dimensional space from
the category. Note that a one-dimensional space is isomorphic to the field of complex numbers 
and that a preparation \mbox{$a : I \rightarrow A$},
i.e., a linear operator from the complex field into $A$ is fully determined by
the image of the number $1$ and preparations \mbox{$I \rightarrow A$} are exactly
vectors of $A$. For any complex number $c$ the map $c_{A}$ is multiplication by $c$ in $A$.

\subsection{Matrices on a commutative semiring}
Any singleton is a unit object in $\cC(R)$. Let \mbox{$I =$} \mbox{$\{ \ast \}$}.
Note that preparations \mbox{$a : I \rightarrow A$} are arbitrary functions from $A$ to $M$ and
the set $M$ is the set of scalars.
We show now that the three conditions of Definition~\ref{def:unit-object} is satisfied.
\begin{itemize}
\item
For any \mbox{$a \in A$} define an arrow \mbox{$\psi_{a} :$} \mbox{$I \rightarrow A$} by
\[
\psi_{a}(\ast , b) \ = \ \left\{ \begin{array}{l} 1 \ {\rm if} \ b = a \\ 0 \ {\rm otherwise} \end{array} \right.
\]
for any \mbox{$b \in A$}. This is an injective functional arrow defined by \mbox{$x(\ast) = a$} and,
by Lemma~\ref{le:unitary}, it is left-unitary, i.e., normalized. Since $A$ is not empty, there is some
such normalized arrow.
\item
Let \mbox{$f : A \rightarrow B$}. Note that, for any \mbox{$a , x \in A$}, \mbox{$y \in B$} one has:
\[
(f \circ \psi_{a})(\ast , y) \ = \ \sum_{b \in A} f(b , y) \cdot \psi_{a}(\ast , b) \ = \ 
f(a , y).
\]
Let now \mbox{$f , g : A \rightarrow B$}
if \mbox{$f \circ \psi_{a} =$} \mbox{$g \circ \psi_{a}$}, then, for any \mbox{$b \in B$}, one has
\mbox{$f(a , b) =$} \mbox{$g(a , b)$}. If  \mbox{$f \circ \psi_{a} =$} \mbox{$g \circ \psi_{a}$} 
for any \mbox{$a \in A$}, then \mbox{$f = g$}.
\item
The set $M$ is the set of scalars and for any \mbox{$s , t \in M$}
one has: \mbox{$s \circ t =$} \mbox{$s \cdot t$}.
For any object $A$ and \mbox{$s \in M$}, one may define \mbox{$s_{A} :$}
\mbox{$A \rightarrow A$} by 
\[
s_{A}(a , b) \ = \ \left\{ \begin{array}{l} s \ {\rm if} \ a = b \\ 0 \ {\rm otherwise} \end{array} \right.
\]
for any \mbox{$a , b \in A$}.
Let \mbox{$f : I \rightarrow A$} be a preparation. One has, for any scalar $s$ and any 
\mbox{$a \in A$}
\[
(f \circ s)(\ast , a) \ = \ f(\ast , a) \cdot s(\ast , \ast) \ = \ 
f(\ast , a) \cdot s
\]
and
\[
(s_{A} \circ f)(\ast , a) \ = \ \sum_{b \in A} s_{A}(b , a) \cdot f(\ast , b) \ = \
s \cdot f(\ast , a).
\]
\end{itemize}

\section{Examples of bi-arrows} \label{sec:examples3}
\subsection{Commutative monoids} \label{sec:monoids3}
We do not know how to characterize bi-arrows in an arbitrary commutative monoid, considered
as an A-category with unit object, as in Sections~\ref{sec:monoid1} and~\ref{sec:monoid2},
but we know how to characterize them in a commutative monoid that is embedded in a group.

Assume $G$ is a group (not necessarily commutative). The group operation is denoted $\oplus$
and the neutral element is $0$. Let $M$ be a subset of $G$ that contains $0$, 
is closed under $\oplus$ and such that $\oplus$ is commutative on elements of $M$.
We shall describe bi-arrows in the one-object category defined by $M$. 
A bi-arrow \mbox{$I , I \rightarrow I$} (there are no others) 
is any function \mbox{$b : M \times M \rightarrow M$} 
such that there exist functions
\mbox{$f , g : M \rightarrow M$} such that for any \mbox{$m , n \in M$}
one has:
\[
b(m , n) \ = \ f(m) \oplus n \ = \ g(n) \oplus m.
\]
Considering \mbox{$m = n$} one sees that, due to the existence of an inverse $- m$ in $G$ one must have \mbox{$f = g$}.
Now for any \mbox{$m , n \in M$} one must have \mbox{$f(m) =$}
\mbox{$ m \oplus f(n) \times (- n)$}, in $G$. Therefore 
\mbox{$f(n)  \oplus (- n)$}  cannot depend on $n$ and must be some element $c$ of $G$.
But \mbox{$c = f(1)$} and $c$ is therefore an element of $M$.
A bi-morphism in $M$ is therefore described by an element $p$ of $M$, any such element, and
\mbox{$b(m , n) = p \oplus m \oplus n$}.
We see that bi-morphisms \mbox{$I , I \rightarrow I$} are in one-to-one correspondence with scalars.

\subsection{Commutative groups} \label{sec:groups3}
The characterization obtained in Section~\ref{sec:monoids3} holds for groups, the difference 
in the adjoint structure notwithstanding.

\subsection{Relations}
In ${\cal R}$, a bi-arrow \mbox{$f : A , B \rightarrow X$} is a function of two
arguments, one a subset $p$ of $A$ and the other a subset $q$ of $B$, into the subsets of $X$.
There is, for every $p$ a relation $R(p)$ on $B \times X$ and for every $q$ a relation $S(q)$ on
$A \times X$ such that \mbox{$f(p , q) =$}
\mbox{$R(p) \circ q =$} \mbox{$S(q) \circ p$}.
From the first equality follows that \mbox{$f(p_{1} \cup p_{2} , q) =$}
\mbox{$f(p_{1} , q) \cup f(p_{2} , q)$} and from the second equality follows 
\mbox{$f(p , q_{1} \cup q_{2}) =$}
\mbox{$f(p , q_{1}) \cup f(p , q_{2})$}. 
We conclude that 
\mbox{$f(p , q) =$} \mbox{$\{ x \mid x \in X , \exists a \in p, b \in q {\rm \ such \ that \ } x \in f(\{a\} , \{b\})$}.
We see that bi-arrows of sort \mbox{$A , B \rightarrow X$} are in one-to-one correspondence 
with binary relations from the cartesian product \mbox{$A \times B$} to $X$.

\subsection{Hilbert spaces}
In ${\cal H}$ a bi-arrow is a bilinear map. In the sub-category of ${\cal H}$ whose arrows are
bounded linear maps, a bi-arrow is a bilinear map that is separately continuous. It need
not be continuous.

\subsection{Matrices on a commutative semiring}
Let \mbox{$\alpha : A , B \rightarrow X$} be a bi-arrow.
For any \mbox{$i : I \rightarrow A$} and \mbox{$j : I \rightarrow B$}:
\[
\alpha(i , j) \ = \ \alpha^{1}(i) \circ j \ = \ \alpha^{2}(j) \circ i.
\]
Equivalently, for any \mbox{$x \in X$}:
\[
\alpha(i , j)(\ast , x) \ = \ \sum_{b \in B} \alpha^{1}(i)(b , x) \cdot j(\ast , b) \ = \
\sum_{a \in A} \alpha^{2}(j)(a , x) \cdot i(\ast , a).
\]
Therefore, for any \mbox{$c \in A$} we have:
\[
\alpha(\psi_{c} , j)(\ast , x) \ = \ \sum_{a \in A} \alpha^{2}(j)(a , x) \cdot \psi_{c}(\ast , a) \ = \ 
\alpha^{2}(j)(c , x)
\]
and for any \mbox{$c \in B$} we have:
\[
\alpha(i , \psi_{c})(\ast , x) \ = \ \sum_{b \in B} \alpha^{1}(i)(b , x) \cdot \psi_{c}(\ast , b) \ = \ 
\alpha^{1}(i)(c , x).
\]
We conclude that $\alpha$ is a bi-arrow iff 
\begin{equation} \label{eq:bi-arrow-mat}
\alpha(i , j)(\ast , x) \ = \ \sum_{a \in A} \alpha(\psi_{a} , j)(\ast , x) \cdot i(\ast , a) \ = \ 
\end{equation}
\[
\sum_{b \in B} \alpha(i , \psi_{b})(\ast , x) \cdot j(\ast , b).
\]

\section{Examples of T-categories} \label{sec:examples4}
\subsection{Commutative monoids embedded in a group} \label{sec:monoids4}
Section~\ref{sec:monoids3} shows that a commutative monoid embedded in
a group is a T-category.
The bi-arrow \mbox{$\kappa : I , I \rightarrow I$} defined by 
\mbox{$\kappa(m , n) =$} \mbox{$m \oplus n$} is a tensor product by Section~\ref{sec:monoids3}
and since Equation~\ref{eq:cond2} is easily seen to hold, 
by commutativity of $\oplus$ on $M$.

\subsection{Commutative groups} \label{sec:groups4}
Similarly to what has been proposed in Section~\ref{sec:monoids4}, the arrow defined by 
\mbox{$\kappa(m , n) =$} \mbox{$m \circ n$} is a tensor product by Section~\ref{sec:groups3}
and since Equation~\ref{eq:cond2} is easily seen to hold.

\subsection{Relations} \label{sec:relations4}
The category ${\cal R}$ is a T-category: 
the bi-arrow \mbox{$\kappa : A , B \rightarrow A \times B$}
is defined by \mbox{$\kappa(p , q) =$} 
\mbox{$\{\langle a , b \rangle \mid a \in p , b \in q \}$}.
Equation~\ref{eq:cond2} holds because 
\mbox{${\kappa}^{\star}(q , q) \circ \kappa(p' , q')$} is the identity if both intersections 
\mbox{$p' \cap p$} and \mbox{$q' \cap q$} are non-empty and is empty otherwise.
The same holds for the right-hand side of the equation.

\subsection{Hilbert spaces} \label{sec:Hilbert4}
The category of Hilbert spaces and linear maps is a T-category: tensor products 
are obtained in the usual way from bases. The category of Hilbert spaces and bounded
linear maps is not a T-category, since the arrow \mbox{$h : A \otimes B \rightarrow X$}
corresponding to a bounded linear bi-arrow (a separately continuous bilinear map) is not
always continuous, i.e., bounded.

\subsection{Matrices on a commutative semiring} \label{sec:matrices4}
For any objects $A$ , $B$, we shall define the canonical bi-arrow \mbox{$\kappa_{A , B} :$}
\mbox{$A , B \rightarrow A \times B$}, where $A \times B$ is the cartesian product of the
sets $A$ and $B$, i.e., the set of ordered pairs \mbox{$\langle a , b \rangle$} for
\mbox{$a \in A$}, \mbox{$b \in B$}.
We let 
\[
\kappa_{A , B}(i , j)(\ast , \langle a , b \rangle) \ = \ i(\ast , a) \cdot j(\ast , b)
\]
for any \mbox{$i : I \rightarrow A$}, \mbox{$j : I \rightarrow B$}, \mbox{$a \in A$} and
\mbox{$b \in B$}.
We must now check that $\kappa_{A , B}$ is indeed a bi-arrow, i.e., satisfies Equation~\ref{eq:bi-arrow-mat}.
Indeed:
\[
\sum_{c \in A} \kappa_{A , B}(\psi_{c}, j)(\ast , \langle a , b \rangle) \cdot i(\ast , c) \ = \ 
\sum_{c \in A} \psi_{c}(\ast , a) \cdot j(\ast , b) \cdot i(\ast , c) \ = \
j(\ast , b) \cdot i(\ast , a) 
\]
and
\[
\sum_{c \in B} \kappa_{A , B}(i , \psi_{c})(\ast , \langle a , b \rangle) \cdot j(\ast , c) \ = \ 
\sum_{c \in B} i(\ast , a) \cdot \psi_{c}(\ast , b) \cdot j(\ast , c) \ = \
i(\ast , a) \cdot j(\ast , b).
\]

We now want to show that the bi-arrow $\kappa_{A , B}$ provides a tensor product for $A$ 
and $B$.
Let \mbox{$\alpha : A , B \rightarrow X$} be any bi-arrow.
Assume, first that \mbox{$f : A \times B \rightarrow X$} is such that
\mbox{$f \circ \kappa_{A , B} =$} $\alpha$.
We have, for any \mbox{$i : I \rightarrow A$}, 
\mbox{$j : I \rightarrow B$} and \mbox{$x \in X$}:
\[
\alpha(i, j)(\ast , x) \ = \ 
(f \circ \kappa_{A , B})(i , j)(\ast , x) \ = \
\sum_{c \in A \times B} f(c , x) \cdot \kappa_{A , B}(i , j)(\ast , c) \ = \ 
\]
\[
\sum_{a \in A , b \in B} f(\langle a , b \rangle , x) \cdot i(\ast , a) \cdot j(\ast , b).
\]
By taking \mbox{$i = \psi_{a}$} and \mbox{$j = \psi_{b}$} one sees that,
for any \mbox{$a \in A$}, \mbox{$b \in B$} and \mbox{$x \in X$} one has:
\begin{equation} \label{eq:unicity}
\alpha(\psi_{a} , \psi_{b})(\ast , x) \ = \ 
f(\langle a , b \rangle , x).
\end{equation}
We have proved unicity of arrow $f$.
Let us now check that Equation~\ref{eq:unicity} provides an arrow that satisfies 
\mbox{$f \circ \kappa_{A , B} =$} $\alpha$.
Indeed, using Equation~\ref{eq:bi-arrow-mat}:
\[
\sum_{a \in A , b \in B} \alpha(\psi_{a} , \psi_{b})(\ast , x) \cdot i(\ast , a) \cdot j(\ast , b) \ = \ 
\sum_{b \in B} \alpha(i , \psi_{b})(\ast , x) \cdot j(\ast , b) \ = \ 
\alpha(i , j)(\ast , x).
\]

We must now show that, for any preparations \mbox{$a , a' : I \rightarrow A$} and
\mbox{$b , b' : I \rightarrow B$} we have
\mbox{$\kappa^{\star}_{A , B}(a , b) \circ \kappa_{A , B}(a' , b') =$}
\mbox{${a}^{\star} \circ a' \circ {b}^{\star} \circ b'$}.
Indeed:
\[
(\kappa^{\star}_{A , B}(a , b) \circ \kappa_{A , B}(a' , b'))(\ast , \ast) \ = \
\sum_{c \in A \times B} \kappa^{\star}_{A , B}(a , b)(c , \ast) \cdot 
\kappa_{A , B}(a' , b')(\ast , c) \ = \ 
\]
\[
\sum_{x \in A , y \in B} \kappa_{A , B}(a , b)(\ast , \langle x , y \rangle) \cdot 
\kappa_{A , B}(a' , b')(\ast , \langle x , y \rangle) \ = \ 
\]
\[
a(\ast , x) \cdot b(\ast , y) \cdot a'(\ast , x) \cdot b'(\ast , y) \ = \ 
{a}^{\star}(x , \ast) \cdot a'(\ast , x) \cdot {b}^{\star}(y , \ast) \cdot b'(\ast , y) \ = \ 
\]
\[
({a}^{\star} \circ a')(\ast , \ast) \cdot ({b}^{\star} \circ b')(\ast , \ast) \ = \ 
({a}^{\star} \circ a' \circ {b}^{\star} \circ b')(\ast , \ast).
\]
We have shown that, for any commutative semiring $R$, the category $\cC(R)$ is a T-category.

\section{Examples of families of zero arrows} \label{sec:examples5}
\subsection{Commutative monoids} \label{monoids5}
In a commutative monoid there is a family of zero arrows iff there is an element \mbox{$w \in M$}
such that, for any \mbox{$m \in M$} one has \mbox{$m + w =$} $w$. This may be the case and
may not be the case.
A commutative monoid embedded in a group has no family of zero arrows, 
unless it is a singleton.

\subsection{Commutative groups} \label{groups5}
Similarly, a commutative group has no family of zero arrows unless it is a singleton.

\subsection{Relations} \label{relations5}
The category $\cR$ has a family of zero arrows: the zero arrows are defined by 
\mbox{${0}_{A , B} =$} $\emptyset$.

\subsection{Hilbert spaces} \label{Hilbert5}
The category of Hilbert spaces and linear maps has a family of zero arrows:
\mbox{${0}_{A , B}(\vec{x}) =$} 
\mbox{$\vec{0} \in B$} for every \mbox{$\vec{a} \in A$}.

\subsection{Matrices on a commutative semiring} \label{matrices5} 
The category $\cC(R)$ has a family of zero arrows defined by:
\mbox{${0}_{A , B}(a , b) =$} $0$ for every \mbox{$a \in A$}, \mbox{$b \in B$}.

\section{Examples of B-categories} \label{sec:examples6}
\subsection{Commutative monoids} \label{sec:monoids6}
Inspection shows that the only Az commutative monoid that has u-coproducts is the singleton
monoid, that generates a trivial category.

\subsection{Commutative groups} \label{sec:groups6}
The only commutative group that is a B-category is the singleton group.

\subsection{Relations} \label{sec:relations6}
The category $\cR$ is a B-category. For any sets $A$ and $B$, 
let \mbox{$A + B$} denote the disjoint union of sets $A$ and $B$.  
The arrows
\mbox{$u : A \rightarrow A + B$} and \mbox{$v : B \rightarrow A + B$} defined by:
\mbox{$u(a' , \langle c$} iff \mbox{$a' =$} $c$ and 
\mbox{$v(b' , c)$} iff \mbox{$b' =$} $c$ are right-unitary, and orthogonal and they provide 
a coproduct.

\subsection{Hilbert spaces} \label{sec:Hilbert6}
The category of Hilbert spaces and linear maps is a B-category. The orthogonal sums provide
a u-coproduct.

\subsection{Matrices on a commutative semiring} \label{matrices6}
Let $A$, $B$ be objects. 
Define \mbox{$i_{1} : A \rightarrow A \oplus B$} and \mbox{$i_{2}: B \rightarrow A \oplus B$}
to be a co-product {\em in the category of sets and functions}, i.e., $A \oplus B$ is the disjoint union
of a copy of $A$ and a copy of $B$ and $i_{1}$ and $i_{2}$ are the canonical injections.
The set $A \oplus B$ is an object in $\cC(R)$. We claim that, in $\cC(R)$,
\mbox{$\chi_{{i}_{1}} :$} \mbox{$A \rightarrow A \oplus B$} and 
\mbox{$\chi_{{i}_{2}} :$} \mbox{$B \rightarrow A \oplus B$} provide a u-coproduct.

First, let us show that they provide a coproduct.
Let \mbox{$f : A \rightarrow X$} and \mbox{$g : B \rightarrow X$}. 
Let \mbox{$[f \ g] :$} \mbox{$A \oplus B \rightarrow X$} be defined by
\begin{equation} \label{eq:fg}
[f \ g](c , x) \ = \ \sum_{a \in A} f(a , x) \cdot \psi_{c}(\ast , i_{1}(a)) + 
\sum_{b \in B} g(b , x) \cdot \psi_{c}(\ast , i_{2}(b)).
\end{equation}
We have:
\[
([ f \ g ] \circ \chi_{i_{1}})(a , x) \ = \ \sum_{c \in A \oplus B} [f \ g ] (c , x) \circ \chi_{i_{1}}(a , c) \ = \ 
[f \ g ] (i_{1}(a) , x) \ = \ f(a , x)
\]
and
\[
([ f \ g ] \circ \chi_{i_{2}})(b , x) \ = \ \sum_{c \in A \oplus B} [f \ g ] (c , x) \circ \chi_{i_{2}}(b , c) \ = \ 
[f \ g ] (i_{2}(b) , x) \ = \ g(b , x).
\]
To show unicity, assume that, \mbox{$h : A \oplus B \rightarrow X$} such that
\mbox{$h \circ \chi_{i_{1}} =$} $f$ and \mbox{$h \circ \chi_{i_{2}} =$} $g$.
By considering in turn the cases \mbox{$c =$} \mbox{$i_{1}(a)$} and 
\mbox{$c =$} \mbox{$i_{2}(b)$} we show that
\[
h(c , x) \ = \ \sum_{a \in A} f(a , x) \cdot \psi_{c}(\ast , i_{1}(a)) + 
\sum_{b \in B} g(b , x) \cdot \psi_{c}(\ast , i_{2}(b)).
\]

Secondly, by Lemma~\ref{le:unitary}, since the functions $i_{1}$ and $i_{2}$ are injective, we see that
$\chi_{{i}_{1}}$ and $\chi_{{i}_{2}}$ are right-unitary.

Thirdly, $\chi_{{i}_{1}}$ and $\chi_{{i}_{2}}$ are orthogonal since
\[
(\chi^{\star}_{i_{1}} \circ \chi_{i_{2}})(b , a) \ = \ 
\sum_{c \in A \oplus B} \chi^{\star}_{{i}_{1}}(c , a) \cdot \chi_{i_{2}}(b , c) \ = \ 
\sum_{c \in A \oplus B} \chi_{{i}_{1}}(a , c) \cdot \chi_{i_{2}}(b , c) \ = \ 0
\]

\section{Examples of normalizable and regular categories} \label{sec:examples7}
\subsection{Commutative monoids}
Every commutative monoid is normalizable: $0$ is normalized and
\mbox{$m =$} \mbox{$0 + m$}.
Some commutative monoids with a zero element are regular, others are not.
The singleton monoid, which is the only commutative monoid embedded in a group with a 
zero element is regular.

\subsection{Commutative groups}
As above, all commutative groups are normalizable. The only commutative group with a zero,
the singleton group, is regular.

\subsection{Relations}
The category $\cR$ is normalizable since every preparation is normalized.
It is also regular.

\subsection{Hilbert spaces}
Hilbert spaces form a regular category since the scalars are a field.
They form a normalizable category since, for every vector $\vec{x}$ one has:
\mbox{$\vec{x} =$} \mbox{$\frac {\vec{x}} {\| \vec{x} \|} \: \| \vec{x} \|$}.
Note that, for this, we need the fact that norms are non-negative reals, the existence of multiplicative inverses and of square roots for such numbers.

\subsection{Matrices on a commutative semiring} \label{sec:matrices7}
Let \mbox{$f : I \rightarrow A$} be a preparation, i.e., a function from $A$ to $R$. 
A normalized preparation \mbox{$n : I \rightarrow A$} is such that 
\mbox{$\sum_{a \in A} ({n(a))}^{\star} \cdot n(a) =$} $1$.
If \mbox{$f =$} \mbox{$n \circ s$} then 
\mbox{$\sum_{a \in A} {(f(a))}^{\star} \cdot f(a) =$}
\mbox{${f}^{\star} \circ f =$}
\mbox{${s}^{\star} \circ s =$} \mbox{$\sum_{a \in A} {(s(a))}^{\star} \cdot s(a)$}.
If we call {\em nonnegative} an element of $M$ that is of the form 
\mbox{${m}^{\star} \cdot m$}
we see that any sum of nonnegative elements must be nonnegative.
If this is the case, and, in addition, in any sum of nonnegative elements 
that is equal to zero all terms are equal to zero and every element of $M$ 
different from $0$ has 
a multiplicative inverse, then matrices on such a semiring form a normalizable category:
take \mbox{$n(a) =$} \mbox{$f(a) \cdot {s}^{- 1}$} if \mbox{$s \neq 0$} and 
\mbox{$n(a) =$} $1$ otherwise.

Matrices on a commutative semiring form a regular category iff all elements of the semiring
are regular for multiplication.
\end{document}